\documentclass[a4paper,fleqn]{scrartcl}
\usepackage{amsmath}
\usepackage{amssymb}
\usepackage[pdfborder={0 0 0},breaklinks=true]{hyperref}
\usepackage{breakurl}
\usepackage{array}
\usepackage{calc}
\usepackage{longtable}
\usepackage{fullpage}
\usepackage{pstricks}
\usepackage{pst-text}
\usepackage{graphicx}
\usepackage{xspace}
\usepackage{ifthen}
\usepackage{booktabs}
\usepackage{feynmp}  
\usepackage{cite}
\usepackage{mcite}

\numberwithin{equation}{section}

\newcommand{\Sherpa}{S\protect\scalebox{0.8}{HERPA}\xspace}
\newcommand{\Hadrons}{H\protect\scalebox{0.8}{ADRONS++}\xspace}
\newcommand{\Photons}{P\protect\scalebox{0.8}{HOTONS++}\xspace}
\newcommand{\PHOTOS}{P\protect\scalebox{0.8}{HOTOS}\xspace}

\newcommand{\BLOR}{B\scalebox{0.8}{LOR}\xspace}
\newcommand{\KLOR}{K\scalebox{0.8}{LOR}\xspace}
\newcommand{\Babar}{B\scalebox{0.8}{A}B\scalebox{0.8}{AR}\xspace}
\newcommand{\Belle}{B\scalebox{0.8}{ELLE}\xspace}

\newcommand{\KTeV}{KTeV\xspace}
\newrgbcolor{mred}{.8 .1 .1}
\newrgbcolor{brick}{.6 .1 .1}
\newrgbcolor{olive}{.1 .6 .1}
\newrgbcolor{dgreen}{.1 .4 .1}
\newrgbcolor{ogray}{.85 .85 .85}
\newrgbcolor{lgray}{.6 .6 .6}
\newrgbcolor{dgray}{.4 .4 .4}
\newrgbcolor{dyellow}{.8 .7 .01}
\newrgbcolor{dblue}{.05 .05 .3}
\newrgbcolor{lblue}{.1 .1 .6}
\long\def\symbolfootnote[#1]#2{\begingroup%
\def\thefootnote{\fnsymbol{footnote}}\footnote[#1]{#2}\endgroup}

\newcommand{\done}{{\rm d}}

\newcommand{\dthree}{{\rm d}^3}

\newcommand{\dpar}{\partial}
\newcommand{\order}{\mathcal{O}}
\newcommand{\QED}{{\mbox{\scalebox{0.6}{QED}}}}
\newcommand{\QCD}{{\mbox{\scalebox{0.6}{QCD}}}}
\newcommand{\EW}{{\mbox{\scalebox{0.6}{EW}}}}
\newcommand{\GeV}{{\rm GeV}}

\newcolumntype{a}[2]{>{\raggedleft}p{#1}@{}p{#2}}
\newcolumntype{B}[1]{>{\centering}p{#1}}

\newcommand{\vp}{\vphantom{\frac{|}{|}}}

\newcommand{\g}{\gamma}
\newcommand{\G}{\Gamma}

\newcommand{\GF}{G_{\text{F}}}
\newcommand{\GFhat}{\hat{G}_{\text{F}}}
\newcommand{\f}{\text{f}}

\newcommand{\ud}{\text{d}}
\newcommand{\bve}[1]{\big({#1}\big)}
\newcommand{\ds}[1]{\dash{#1}}
\newcommand{\ph}[1]{\phantom{#1}}

\newcommand{\pr}{P_{\text{R}}}
\newcommand{\half}{\frac{1}{2}}

\newcommand{\bra}{\langle}
\newcommand{\ket}{\rangle}

\newcommand{\dash}[1]{{#1}\!\!\!/}

\newcommand{\myfigure}[3]{
  \begin{figure}[#1]
    \begin{center}
      #2\\\myfigcaption{\textwidth}{#3}
    \end{center}
  \end{figure}
}

\newcommand{\mytable}[3]{
  \begin{table}[#1]
    \begin{center}
      #2\\\mytabcaption{\textwidth}{#3}
    \end{center}
  \end{table}
}

\newcommand{\bt}{\begin{tabular}}
\newcommand{\et}{\end{tabular}}
\def\be{\begin{equation}}
\def\ee{\end{equation}}
\def\bc{\begin{center}}
\def\ec{\end{center}}
\newcommand{\nnb}{\nonumber}
\newcommand{\bea}{\begin{eqnarray}}
\newcommand{\eea}{\end{eqnarray}}
\newcommand{\bi}{\begin{itemize}}
\newcommand{\ei}{\end{itemize}}

\newcommand{\eps}{\varepsilon}

\makeatletter
\newif\if@preliminary
\@preliminaryfalse
\def\preliminary{\@preliminarytrue}
\def\preprintno#1{\def\@preprintno{#1}}
\def\address#1{\def\@address{#1}}

\def\abstract#1{\def\@abstract{#1}}
\renewcommand\abstractname{ABSTRACT}
\newlength\preprintnoskip
\newlength\abstractwidth
\renewcommand\maketitle{\begin{titlepage}%
  \let\footnotesize\small
  \hfill\parbox{\preprintnoskip}{%
  \begin{flushright}\@preprintno\end{flushright}}\hspace*{1cm}
  \vskip 60\p@
  \begin{center}%
    {\Large\bf\boldmath \@title \par}\vskip 1cm%
    {\sc\@author \par}\vskip 3mm%
    {\@address \par}%
    \if@preliminary
      \vskip 2cm {\large\sf PRELIMINARY DRAFT \par \@date}%
    \fi
  \end{center}\par
  \@thanks
  \vfill
  \begin{center}%
    \parbox{\abstractwidth}{\centerline{\abstractname}%
    \vskip 3mm%
    \@abstract}
  \end{center}
  \end{titlepage}%
  \setcounter{footnote}{0}%
  \let\thanks\relax\let\maketitle\relax
  \gdef\@thanks{}\gdef\@author{}\gdef\@address{}%
  \gdef\@title{}\gdef\@abstract{}\gdef\@preprintno{}
}%
\makeatother

\hypersetup{
  pdfauthor={Florian U.~Bernlochner, Marek Schoenherr},
  pdftitle={Comparing different ansatzes to describe electroweak radiative 
            corrections to exclusive semileptonic B meson decays into 
            (pseudo)scalar final state mesons using Monte-Carlo techniques},
  pdfkeywords={B-physics, NLO, EW Corrections}
}
\begin{document}
\begin{fmffile}{graphs/graphs}
\fmfcmd{
 vardef cross_bar (expr p, len, ang) =
  ((-len/2,0)--(len/2,0))
    rotated (ang + angle direction length(p)/2 of p)
    shifted point length(p)/2 of p
 enddef;
 vardef circle (expr p, len) =
  ((-len/2,0)..(0,len/2)..(len/2,0)..(0,-len/2)..cycle)
    rotated (angle direction length(p)/2 of p)
    shifted point length(p)/2 of p
 enddef;
 style_def crossed expr p =
  cdraw p;
  ccutdraw cross_bar (p, 5mm,  45);
  ccutdraw cross_bar (p, 5mm, -45);
  ccutdraw circle (p, 5mm)
 enddef;
 style_def vcrossed expr p =
  ccutdraw cross_bar (p, 5mm,  45);
  ccutdraw cross_bar (p, 5mm, -45);
  ccutdraw circle (p, 5mm)
 enddef;
}

\title{Comparing different ansatzes to describe electroweak radiative 
       corrections to exclusive semileptonic $B$ meson decays into 
       (pseudo)scalar final state mesons using Monte-Carlo techniques}
\preprintno{MCNET-10-19 \\ HU-EP-10/62}
\author{
 Florian U.~Bernlochner$^{1}$,
 Marek Sch\"onherr$^{2}$
}
\address{\it%
$^{1}$ Institut f\"ur Physik, Humboldt Universit\"at Berlin, Germany\\
$^{2}$ Institut f\"ur Kern- und Teilchenphysik, Technische Universit\"at Dresden, Germany
}

\abstract{
  In this publication electroweak next-to-leading order corrections to 
  semileptonic $B$-meson decays into (pseudo)scalar final states are 
  presented. To this end, these corrections of $\order(\alpha\,\GF)$ have been 
  calculated in the {\sc Qed}-enhanced phenomenological model, incorporating 
  the bound-state mesons as its degrees of freedom, and matched to a similar 
  calculation on the level of constituent partons in the full Standard Model. 
  Consequently, the effects arising due to corrections of the partial decay 
  widths on the extraction of the CKM matrix elements $|V_\text{cb}|$ and 
  $|V_\text{ub}|$ are detailed.  Further, the results of two independent
  Monte-Carlo implementations are presented: One is the dedicated, strict
  fixed-order generator \BLOR, and the other is embedded into the 
  generic Yennie-Frautschi-Suura-type resummation of \Photons, which is part of 
  the multi-purpose event generator \Sherpa. The resulting distributions are 
  compared against the standard tool used in many experimental analyses, 
  \PHOTOS, showing improvements on the shapes of kinematic distributions of 
  both the lepton and the final state meson.
}

\maketitle
\tableofcontents

\section{Introduction}\label{Sec:Intro}
 In the Standard Model, the Cabibbo-Kobayashi-Maskawa (CKM) matrix 
\cite{Cabibbo:1963yz,Kobayashi:1973fv} governs the charged current 
weak interactions between the up- and down-type quarks of the three fermion
generations. The precision determination of its matrix 
elements and its CP-violating complex phase in the $B$ meson 
sector has been the focus of intense research over the past decade. The 
combination of various measurements to test the unitarity of the CKM matrix 
is considered a strong instrument in the search for physics 
beyond the Standard Model \cite{Charles:2004jd,Bona:2005eu}.

In the present paper, a calculation of the electroweak next-to-leading order 
corrections in exclusive semileptonic $B$ meson 
decays into (pseudo)scalar mesons, $B \to D \, \ell \, \nu_\ell$, 
$B \to D_0^* \, \ell \, \nu_\ell$ and $B \to \pi \, \ell \, \nu_\ell$, $\ell$ 
denoting either an electron or a muon, is presented. 
Next-to-leading order corrections to such 
decays are an important aspect in the extraction of the CKM matrix elements 
$|V_{\text{cb}}|$ and $|V_{\text{ub}}|$ at $B$-factory experiments. Virtual 
electroweak bosons running in the loop as well as real photon emissions off 
all charged particles present in the decay alter the resulting decay 
dynamics and enhance the weak decay rate. To correct for the 
changed decay dynamics, experimentalists use approximative all-purpose 
next-to-leading order algorithms. These exploit universal factorisation 
theorems in the soft and/or collinear photon energy limit 
\cite{Low:1958sn,Yennie:1961ad,Altarelli:1977zs}. In addition, the total 
hadronic decay rate of semileptonic decays are corrected by the known leading 
logarithm of the virtual corrections of the partonic decay 
\cite{Sirlin:1977sv,Sirlin:1981ie}.

Experience from exclusive semileptonic $K$ meson decays illustrate the 
importance of having a good understanding of such radiative effects: 
until 2004 the global average of the extracted value of $|V_{\text{us}}|$ 
from $K_{l3}^+$ and $K_{l3}^0$ decays implied the violation of CKM unitarity 
by two standard deviations \cite{Eidelman:2004wy}. Further measurements 
proved dissonant with these findings \cite{Sher:2003fb, Alexopoulos:2004sw, 
Alexopoulos:2004sx, Alexopoulos:2004sy}, indicating that the achieved 
experimental precision needed an improved understanding of electroweak 
corrections. Since for many decays next-to-leading 
order calculations do not exist, experiments often use the approximative 
all-purpose algorithm \PHOTOS \cite{Barberio:1990ms,Barberio:1993qi} 
to study the reconstruction efficiency and acceptance. The accuracy of this 
approach was tested by the \KTeV collaboration, using the measured photon 
spectra from radiative $K_{l3}^0$ decays: the angular distribution of the 
simulated photons did not agree well with the predicted spectrum 
\cite{Andre:2004fs}. This lead to the development of the next-to-leading 
order Monte Carlo generator \KLOR (see \cite{Andre:2004fs}), whose 
next-to-leading order calculation is based on a phenomenological model. Its 
predicted angular photon distribution agreed satisfactorily with the measured 
spectra. Although this approach is very precise, it is also the most 
complicated one to adopt for an experiment: electroweak next-to-leading 
order calculations only exist for a few decay modes, sometimes only valid 
in a limited region of phase-space. Most of these calculations are 
evaluated numerically and rely on customised Monte Carlo generators. 

Over the last 10 years, an increasing amount of data and a better 
understanding of detector effects lead to a very accurate picture of
physics at the $B$-factory experiments. This increased precision then lead 
to the demand of knowledge of next-to-leading order electroweak 
effects beyond the precision of approximative all-purpose algorithms. 
The present paper aims at improving the status quo by providing a prediction 
for both total decay rates and differential distributions of a few 
representative kinematic variables.
For the latter the predictions of a dedicated Monte-Carlo generator, \BLOR 
\cite{Bernlochner:2010yd}, is compared to two all-purpose generators, 
\Sherpa/\Photons \cite{Gleisberg:2008ta,Schonherr:2008av} and \PHOTOS 
\cite{Barberio:1990ms,Barberio:1993qi}. While the latter is a 
{\sc Qed}-parton-shower Monte Carlo program intended to supplement generic 
leading logarithmic corrections to pure leading order decay generators, the 
former is a full-fledged hadron-level Monte Carlo generator for collider physics 
whose internal leading order (hadronic) decays are supplemented by a universal 
soft-photon-resummation systematically improved, where possible, by known exact 
next-to-leading order matrix elements.
While the improved description of inclusive decay rates directly 
gives small corrections to the extracted values of $|V_{\text{cb}}|$ and 
$|V_{\text{ub}}|$ from semileptonic decays, the improved description of the 
decay kinematics influence extrapolation to corners of the phase space and, 
therefore, leads to both direct and indirect corrections.

The considerations of the present paper proceed as follows: Sec.~\ref{treelevel} 
briefly reviews exclusive $B \to X \, \ell \, \nu_\ell$ decays at tree-level. 
Thereafter, Sec.~\ref{Sec:NLO} develops the next-to-leading order formalism, 
reviewing both the partonic short-distance results of 
\cite{Sirlin:1977sv,Sirlin:1981ie}, the hadronic long-distance 
{\sc Qed}-improved effective decay and their matching to one-another, including 
also a detailed discussion on non-universal structure-dependent terms in 
Sec.~\ref{Sec:theory_SD_terms}. This model is then embedded into the 
resummation in the soft limit of Yennie, Frautschi and Suura 
\cite{Yennie:1961ad} in Sec.~\ref{Sec:theory_resummation}. 
Sec.~\ref{Sec:Methods} then shortly reviews the basic 
principles of both \BLOR and \Sherpa/\Photons where the calculations of 
Sec.~\ref{Sec:Theory} have been implemented, and of \PHOTOS. The total 
inclusive decay rates obtained are shown in Sec.~\ref{Sec:res_rates} while 
differential distributions are shown in Sec.~\ref{Sec:res_diff}, also 
detailing the improvement over the current estimates. The influence of the 
structure-dependent terms, where known, on the results is presented in 
Sec.~\ref{Sec:result_SD_terms}. Sec.~\ref{Sec:Conclusions} finally summarises 
the results.

Note that the charge-conjugated modes are implied throughout the present paper.

\section{Phenomenological model}\label{Sec:Theory}
 \subsection{Tree-level decay revised}\label{treelevel}

\myfigure{t}{
  \begin{fmfgraph*}(120,80)
   \fmfleft{a}
   \fmfright{b,c,d}
   \fmf{phantom}{a,x,c}
   \fmffreeze
   \fmf{dashes,lab=$B$}{a,x}
   \fmf{dashes,lab.sid=left,lab=$\bar X$}{x,d}
   \fmf{fermion,lab.sid=left,lab=$\ell^+$}{c,x}
   \fmf{fermion,lab=$\nu_\ell$}{x,b}
   \fmfblob{0.1w}{x}
  \end{fmfgraph*}
  \hspace{2cm}
  \begin{fmfgraph*}(120,80)
   \fmfleft{a}
   \fmfright{b,c,d}
   \fmf{phantom}{a,x,c}
   \fmffreeze
   \fmf{fermion,lab=$\bar b$,lab.sid=left}{x,a}
   \fmf{fermion,lab=$\bar x$}{d,x}
   \fmf{fermion,lab.sid=left,lab=$\ell^+$}{c,x}
   \fmf{fermion,lab=$\nu_\ell$}{x,b}
   \fmfdot{x}
  \end{fmfgraph*}\vspace*{2mm}}
  {The tree-level weak $B \to \bar X\, \ell^+\, \nu_\ell$ decay is shown both in the 
   phenomenological picture (left) and, at parton level, in Fermi's theory as 
   low energy approximation of the Standard Model (right). The shaded circle 
   represents the effective vertex parametrised by form factors $\f_\pm$, 
   $x\in\{u,c\}$. \label{Fig:tree_graphs}}

The phenomenological interaction Lagrangian of the weak $B \to X\, \ell\, \nu$ 
decay to a (pseudo)scalar final state in Fermi's theory, with constant form 
factors of the hadronic current, $\f_\pm$, is given by
\bea\label{weaktreelevelinteraction}
  \mathcal{L}_{W} 
& = & \frac{\GF}{\sqrt{2}} V_{\text{xb}} 
      \left[ \bve{\f_+ + \f_-} \phi_X \, \partial^\mu \phi_B
            +\bve{\f_+ - \f_-} \phi_B \, \partial^\mu \phi_X \right]
      \; \bar\psi_\nu \pr\g_\mu \psi_\ell  
      \;\;+\;\; \text{h.c.} \,,
\eea
where $\psi_\ell$ and $\psi_\nu$ are the Dirac fields of the lepton and the 
neutrino, $\phi_B$ and $\phi_X$ are the scalar fields of the 
initial and final state mesons, $\GF$ the Fermi coupling,  $V_{\text{xb}}$ 
the CKM matrix element governing the strength of the $b \to x$ transition, 
and $\pr = 1 + \g_5$ is derived from the right-handed projection operator by 
absorbing the factor $\frac{1}{2}$ into the coupling definition. The Lagrangian 
of eq.~(\ref{weaktreelevelinteraction}) leads to the transition matrix 
element\footnote{Throughout this paper $\mathcal{M}^n_m$  denote a matrix 
                 element at $\mathcal{O}( \GF \, \alpha^n)$ with \emph{m} 
                 photons in the final state. The total decay rate at  
                 $\mathcal{O}( \GF \, \alpha^n)$ is denoted as $\G^n_m$. }
\bea \label{Eq:tree_ME}
 \mathcal{M}^0_0 
& = & -i\, \frac{\GF}{\sqrt{2}}\, V_{\text{xb}}\;H_\mu(p_B,p_X;t)\;\;
        \bar u_\nu \,\pr\gamma^\mu \, v_\ell  \,,
\eea
with the hadronic current, generalised to variable form factors,
\bea\label{Eq:hadcurtreelev}
 H_\mu(p_B,p_X;t) 
\;\;=\;\; \bra X | \bar\psi_x\pr\g_\mu\psi_b | B \ket  
\;\;=\;\; \bve{p_B + p_X}_\mu \, \f_+(t) + \bve{p_B - p_X}_\mu \, \f_-(t) \,. 
\eea
The four-momenta labels in eq.~(\ref{Eq:tree_ME}) are introduced in 
Fig.~\ref{Fig:tree_graphs}. The generalised form factors $\f_{\pm} = \f_{\pm}(t)$ 
now describe the phase-space dependent influence of the strong interaction on 
the weak decay dynamics and are functions of the squared momentum 
transfer from the hadronic to the leptonic system only, given at tree-level by
\bea\label{Eq:squaredmomentumtrans}
 t \;\;=\;\; \bve{p_B - p_X}^2 \;\;=\;\; \bve{p_\ell + p_\nu}^2 \, .
\eea
The tree-level differential decay rate in the $B$-meson rest frame is then 
given by 
\bea\label{Eq:treeleveldifferentialrate}
 \ud \Gamma^0_0 
& = & \frac{1}{64 \, \pi^3 m_B} \big| \mathcal{M}^0_0 \big|^2 \, 
      \ud E_X \, \ud E_\ell \, ,
\eea
with $E_X = p_X^0$ and $E_\ell = p_\ell^0$. The explicit expressions of 
the $\f_{\pm}(t)$ as function of the momentum transfer squared for the 
processes considered in this paper, $B \to D \, \ell \, \nu$, 
$B \to D_0^* \, \ell \, \nu$ and $B \to \pi \, \ell \, \nu$, can be found in 
App.~\ref{App:form_factors}.

 \subsection{Next-to-leading order corrections}\label{Sec:NLO}

The arising electroweak next-to-leading order corrections can be divided 
into two energy regimes: short-distance corrections at parton level, and 
long-distance corrections within the phenomenological model. 
First, Sec.~\ref{Sec:matching} will discuss how both descriptions can be 
matched and renormalised.
Sec.~\ref{Sec:SD_NLO_Corr} then reviews the calculation of the virtual 
short-distance corrections of \cite{Sirlin:1977sv,Sirlin:1981ie}. The 
long-distance corrections, following from an extension of the phenomenological 
model, are then discussed in Sec.~\ref{Sec:LD_NLO_Corr}.

\subsubsection{Matching of different energy regimes}
\label{Sec:matching}

The aim of this section is to develop a formalism to calculate the corrections 
at $\order(\alpha\,\GF)$.
The standard approach involves calculating the one-loop graphs for the 
$B \to X \, \ell \, \nu_\ell$ decay in the effective theory with counterterms 
and compare it to the renormalised Standard Model result. Fixing the 
counterterms results in the desired matching of both results. The effective 
theory itself is non-renormalisable, but the Standard Model can be renormalised 
to measured quantities, e.g. the Fermi coupling constant of the muon decay, the 
electron mass, and the fine-structure constant, in order to produce finite 
predictions. Such a matching procedure was carried out in great detail by 
\cite{DescotesGenon:2005pw} for semileptonic Kaon decays where the leading 
order phenomenological decay is described by a chiral Lagrangian.

In the present paper, however, an alternative route is pursued. Consider a 
general logarithmically divergent $N$-point tensor integral of rank $p$ with 
a single massless photon propagator. It can be cast in the form
\bea\label{Eq:npoint}
 T^{\mu_1\ldots\mu_p} (p_1,\ldots,p_{N-1}) 
& \propto & \int\done^4k \;\frac{k^{\mu_1}\cdots k^{\mu_p}}{k^2\, d_1\ldots d_{N-1}} \,,
\eea
with denominators $d_i = (p_i - k)^2 - m_i^2$. The integral can 
then be split according to 
\bea\label{Eq:npointsep}
 T^{\mu_1\ldots\mu_p} (p_1,\ldots,p_{N-1}) 
& \propto & \int\done^4k
            \left[\,\frac{k^{\mu_1}\cdots k^{\mu_p}}{k^2\,d_1\cdots d_{N-1}}\,
                 -\,\frac{k^{\mu_1}\cdots k^{\mu_p}}{[k^2-\Lambda^2]\,d_1\cdots d_{N-1}}\,
            \right]\nnb\\
&&{}       +\int\done^4k\;\frac{k^{\mu_1}\cdots k^{\mu_p}}{[k^2-\Lambda^2]\,d_1\cdots d_{N-1}}\;.
\eea
This amounts to regulating the ultraviolet behaviour of the first term using 
an unphysical photon-like vector field of mass $\Lambda$ and opposite norm, as 
proposed by Pauli and Villars in \cite{Pauli:1949zm}. Its infrared behaviour is 
left unchanged, thus, relying on the Kinoshita-Lee-Nauenberg theorem 
\cite{Kinoshita:1962ur,Lee:1964is}, these divergences are left to be canceled by 
the real corrections. The second term of eq.~(\ref{Eq:npointsep}) is the 
equivalent of eq.~(\ref{Eq:npoint}), this time only with a massive photon. 
Hence, it is infrared finite and possesses the identical ultraviolet behaviour. 

Transferring this observation to the present case of semileptonic $B$ meson 
decays where, both in the effective theory and in the Standard Model, there is 
at most one massless photon propagator in any one-loop diagram, the virtual 
emission matrix element can be decomposed as
\bea\label{Eq:virtual_decomp}
 \mathcal{M}_0^1
& = & \mathcal{M}_{0,\text{ld}}^1(\Lambda) + \mathcal{M}_{0,\text{sd}}^1(\Lambda) \,.
\eea
The term $\mathcal{M}_{0,\text{ld}}^1$ is now comprised of the Pauli-Villars 
regulated exchange of a massless photon, including its infrared divergence. The 
specific UV regulator effectively restricts the virtual photon's momentum to be 
smaller than $\Lambda$. Hence, it describes long-distance (ld) interactions 
only.

The term $\mathcal{M}_{0,\text{sd}}^1$, on the other hand, carries the full 
ultraviolet behaviour of $\mathcal{M}_0^1$. It thus can be used for 
renormalising all parameters. Consequently, because eq.~(\ref{Eq:npointsep}) is 
exact, all parameters in $\mathcal{M}_{0,\text{ld}}^1$ are then renormalised 
automatically. Through the photon mass, its virtual propagator's momentum is 
effectively restricted to be larger than $\Lambda$. Hence, this term describes 
the short-distance (sd) interactions only.

The above is exact as long as the same Lagrangian input is used to calculate 
both the short-distance and the long-distance parts. In practice, however, 
due to the confining, non-perturbative nature of {\sc Qcd} this is not 
feasible for the processes at hand. 
For scales larger than the hadron mass, its parton content 
can be resolved, electroweak corrections have to be calculated on the 
basis of (constituent) quarks. For scales smaller than the hadron mass, 
its parton content cannot be resolved, the bound-state hadrons themselves are 
the relevant degrees of freedom. Thence, supposing $\Lambda$ is set such, 
that it effectively separates those two regimes, the long-distance {\sc Qed} 
corrections $\mathcal{M}_{0,\text{ld}}^1$ can be calculated using the 
phenomenological model and for the short-distance corrections 
$\mathcal{M}_{0,\text{sd}}^1$ the full Standard Model has to be invoked.
This is justified, in principle, by the assumption, that the phenomenological 
model describes the Standard Model and its effective degrees of freedom at 
these low scales. This mere fact, however, directly leads to inconsistencies 
at the matching scale $\Lambda$, where both models should give the same answer. 
Thus, this matching is only approximate and the associated systematic 
uncertainties to this method can be estimated by varying the matching 
parameter, cf.~Sec.~\ref{Sec:res_rates}.

The above reasoning leads to an optimal value for $\Lambda$: the smallest 
hadronic mass in the decay. Then, only low-energy virtual photons, not able to 
resolve either of both mesons, are described by the effective theory, while
high-energy virtual photons are described by the short-distance picture of the 
full Standard Model, resolving the partonic content of both the charged and the 
neutral meson involved. Further, as long as $\Lambda>E_\gamma^\text{max}$, the 
kinematic limit of the photon energy in single photon emission\footnote{
    The maximum photon energy for single photon radiation in the 
    rest frame of a decaying particle is half its mass, neglecting 
    all other decay products' masses. Allowing for massive decay 
    products further reduces this kinematic limit.}, the real emission of 
photons off these charged mesons are also correctly described by the 
phenomenological model (except for structure dependent terms discussed in 
Sec.~\ref{Sec:theory_SD_terms}). 

Nonetheless, it has to be noted that there are conceptual problems if both 
hadronic scales differ significantly. Then, there is a large intermediate 
regime, where virtual photons are able to resolve one meson, but not the other. 
By the above choice of $\Lambda$ it is expected to give the best approximate 
description in this region. Further, if the third scale $E_\gamma^\text{max}$ 
exceeds $\Lambda$, real radiation, in the present ansatz always described using 
the phenomenological model, is able to resolve the final state meson, be it 
charged or neutral, as well. However, even if a considerable fraction is 
radiated at scales above $\Lambda$, this should have only negligible effects on 
the total decay rate as the bulk of the radiation is in the region $k\to 0$ and, 
therefore, adequately described.

\subsubsection{Short-distance next-to-leading order corrections}
\label{Sec:SD_NLO_Corr}

\myfigure{t}{
  \begin{fmfgraph*}(100,60)
    \fmfstraight
    \fmfleft{l}
    \fmfright{r}
    \fmftop{t0,t1,t2,t4,t5}
    \fmfbottom{b0,b1,b2,b4,b5}
    \fmf{phantom}{b1,xh,t2}
    \fmf{phantom}{b4,xl,t5}
    \fmffreeze
    \fmf{phantom}{xh,y,xl}
    \fmffreeze
    \fmf{fermion}{t1,xh,b1}
    \fmf{fermion}{t5,xl,t4}
    \fmflabel{$\bar b$}{b1}
    \fmflabel{$\bar x$}{t1}
    \fmflabel{$\nu_\ell$}{t4}
    \fmflabel{$\ell^+$}{t5}
    \fmf{dashes,lab=$W$}{xh,y}
    \fmf{dashes,lab=$W$}{y,xl}
    \fmf{dashes,lab=$Z,,\gamma,,h$,left}{xh,y}
    \fmfv{decor.shape=circle,decor.filled=empty,decor.size=0.1w}{xh}
  \end{fmfgraph*}
  \hspace*{15mm}
  \begin{fmfgraph*}(100,60)
    \fmfstraight
    \fmfleft{l}
    \fmfright{r}
    \fmftop{t0,t1,t2,t4,t5}
    \fmfbottom{b0,b1,b2,b4,b5}
    \fmf{phantom}{b1,xh,t2}
    \fmf{phantom}{b4,xl,t5}
    \fmffreeze
    \fmf{phantom}{xh,y,xl}
    \fmf{phantom}{xl,z,t5}
    \fmffreeze
    \fmf{fermion}{t1,xh,b1}
    \fmf{fermion}{t5,xl,t4}
    \fmflabel{$\bar b$}{b1}
    \fmflabel{$\bar x$}{t1}
    \fmflabel{$\nu_\ell$}{t4}
    \fmflabel{$\ell^+$}{t5}
    \fmf{dashes,lab=$W$,lab.sid=left}{xh,xl}
    \fmf{dashes,lab=$Z,,\gamma,,h$,right}{xh,z}
    \fmfv{decor.shape=circle,decor.filled=empty,decor.size=0.1w}{xh}
  \end{fmfgraph*}
  \hspace*{15mm}
  \begin{fmfgraph*}(100,60)
    \fmfstraight
    \fmfleft{l}
    \fmfright{r}
    \fmftop{t0,t1,t2,t4,t5}
    \fmfbottom{b0,b1,b2,b4,b5}
    \fmf{phantom}{b1,xh,t2}
    \fmf{phantom}{b4,xl,t5}
    \fmffreeze
    \fmf{phantom}{xh,y,xl}
    \fmf{phantom}{xl,z,t4}
    \fmffreeze
    \fmf{fermion}{t1,xh,b1}
    \fmf{fermion}{t5,xl,t4}
    \fmflabel{$\bar b$}{b1}
    \fmflabel{$\bar x$}{t1}
    \fmflabel{$\nu_\ell$}{t4}
    \fmflabel{$\ell^+$}{t5}
    \fmf{dashes,lab=$W$}{xh,xl}
    \fmf{dashes,lab=$Z$,left=0.5}{xh,z}
    \fmfv{decor.shape=circle,decor.filled=empty,decor.size=0.1w}{xh}
  \end{fmfgraph*}\vspace*{15mm}\\
  \begin{fmfgraph*}(100,60)
    \fmfstraight
    \fmfleft{l}
    \fmfright{r}
    \fmftop{t0,t1,t2,t4,t5}
    \fmfbottom{b0,b1,b2,b4,b5}
    \fmf{phantom}{b1,xh,t2}
    \fmf{phantom}{b4,xl,t5}
    \fmffreeze
    \fmf{phantom}{xh,xl}
    \fmffreeze
    \fmf{fermion}{t1,xh,b1}
    \fmf{fermion}{t5,xl,t4}
    \fmflabel{$\bar b$}{b1}
    \fmflabel{$\bar x$}{t1}
    \fmflabel{$\nu_\ell$}{t4}
    \fmflabel{$\ell^+$}{t5}
    \fmf{dashes,lab=$W$}{xh,xl}
    \fmf{dashes,lab=$\begin{array}{c}W,,Z\\\gamma,,h\end{array}$,left=900}{xh,xh}
    \fmfv{decor.shape=circle,decor.filled=empty,decor.size=0.1w}{xh}
  \end{fmfgraph*}
  \hspace*{15mm}
  \begin{fmfgraph*}(100,60)
    \fmfstraight
    \fmfleft{l}
    \fmfright{r}
    \fmftop{t0,t1,t2,t4,t5}
    \fmfbottom{b0,b1,b2,b4,b5}
    \fmf{phantom}{b1,xh,t2}
    \fmf{phantom}{b4,xl,t5}
    \fmffreeze
    \fmf{phantom}{xh,xl}
    \fmf{phantom}{xh,w,l}
    \fmffreeze
    \fmf{fermion}{t1,xh,b1}
    \fmf{fermion}{t5,xl,t4}
    \fmflabel{$\bar b$}{b1}
    \fmflabel{$\bar x$}{t1}
    \fmflabel{$\nu_\ell$}{t4}
    \fmflabel{$\ell^+$}{t5}
    \fmf{dashes,lab=$W$}{xh,xl}
    \fmf{dashes}{xh,w}
    \fmfv{decor.shape=circle,decor.filled=empty,decor.size=0.1w}{xh}
    \fmfv{decor.shape=circle,decor.filled=empty,decor.size=0.1w}{w}
  \end{fmfgraph*}
  \hspace*{15mm}
  \begin{fmfgraph*}(100,60)
    \fmfstraight
    \fmfleft{l}
    \fmfright{r}
    \fmftop{t0,t1,t2,t4,t5}
    \fmfbottom{b0,b1,b2,b4,b5}
    \fmf{phantom}{b1,xh,t2}
    \fmf{phantom}{b4,xl,t5}
    \fmffreeze
    \fmf{phantom}{xh,y,xl}
    \fmffreeze
    \fmf{fermion}{t1,xh,b1}
    \fmf{fermion}{t5,xl,t4}
    \fmflabel{$\bar b$}{b1}
    \fmflabel{$\bar x$}{t1}
    \fmflabel{$\nu_\ell$}{t4}
    \fmflabel{$\ell^+$}{t5}
    \fmf{dashes,lab=$W$}{xh,y}
    \fmf{dashes,lab=$W$}{y,xl}
    \fmfv{decor.shape=circle,decor.filled=empty,decor.size=0.1w}{xh}
    \fmfv{decor.shape=circle,decor.filled=empty,decor.size=0.1w}{y}
  \end{fmfgraph*}
  \vspace*{5mm}
  }
  {Representative Feynman diagrams of the Standard Model partonic decay 
   $\bar b\to \bar x\,\ell^+\nu_\ell$ are shown. The white circles indicate 
   hadronic contributions that are neglected in the short-distance expansion.
   \label{Fig:SD_graphs}}

The well-known Standard Model Lagrangian is used in this study to describe the 
partonic $b \to x \, \ell \, \nu$ decay in the short-distance regime. A 
representative collection of relevant next-to-leading order corrections to 
the tree-level decay, involving the exchange of virtual photons, $W$ and $Z$ 
bosons as well as Higgs scalars, is depicted in Fig.~\ref{Fig:SD_graphs}. 
Besides vertex corrections to the $b$-$x$-$W$ and $\ell$-$\nu$-$W$ vertices, 
wave function and propagator corrections, box diagrams involving the additional 
exchange of a neutral $\gamma$, $Z$ or $h$ bosons between the hadronic and 
leptonic systems are present. 
These next-to-leading order corrections read, calculated in \cite{Sirlin:1977sv}
within the current algebra framework and concentrating on the 
renormalisation of the bare Fermi coupling $\GFhat$ for this process, to 
leading logarithmic accuracy
\bea
 \hat{\mathcal{M}}_{0,\text{sd}}^1(\Lambda)
& = & \frac{\alpha \, \GFhat}{4\pi}
      \left[ 3\ln\frac{m_W}{\Lambda}
            +6\bar{Q}\ln\frac{m_W}{\Lambda}
            -3\bar{Q}\ln\frac{m_W^2}{m_Z^2}
            +\ldots
      \right]
      \tilde{\mathcal{M}}_0^0 \,,
\eea
with $\hat{\mathcal{M}}_0^0=\GFhat\tilde{\mathcal{M}}_0^0$, i.e.~the leading 
order matrix element stripped of the Fermi coupling constant, and
$\bar{Q}$ being the average charge of the quark line. The ellipsis stands for 
non-logarithmic terms. For the photonic 
contributions an infrared regulator has been introduced in the form of a photon 
mass $\Lambda\ll m_W$. Thus, the resulting matrix element has exactly the form 
required for the matching outlined in Sec.~\ref{Sec:matching}. The arising loop 
corrections are ultraviolet (UV) divergent and have been regularised in 
\cite{Sirlin:1977sv} by a UV cutoff set to $m_W$. Renormalisation of the 
parameters, again focusing on $\GFhat$, is then achieved by comparison to the 
muon decay computed in the same computational framework, yielding the relation
\bea
 \GF
& = & \GFhat \left[
                    1+\frac{3\alpha}{8\pi}\ln\frac{m_W^2}{m_Z^2}+\ldots
             \right]\,,
\eea
where $\GF$ is now the renormalised Fermi decay constant as measured in the 
muon decay. This leads to the partonic short-distance virtual matrix element in 
the renormalised 
theory
\bea
 \mathcal{M}_{0,\text{sd}}^1(\Lambda)
& = & \frac{\alpha \, \GF}{4\pi}
      \left[
             3\ln\frac{m_W}{\Lambda}
            +6\bar{Q}\ln\frac{m_W}{\Lambda}
            -\frac{3}{2}\left(1+2\bar{Q}\right)\ln\frac{m_W^2}{m_Z^2}
            +\ldots
      \right]
      \tilde{\mathcal{M}}_0^0 \nnb\\
& = & \frac{3\alpha}{4\pi}
      \left(1+2\bar{Q}\right)\ln\frac{m_Z}{\Lambda}\cdot
      \mathcal{M}_0^0 \,,
\eea
where $\GF$ has been reabsorbed into the leading order matrix element 
$\mathcal{M}_0^0$. 
In the case of semileptonic $B$ decays 
$\bar{Q}=\tfrac{1}{2}\,|Q_{\bar{b}}+Q_{\bar{x}}|=\tfrac{1}{6}$, $x\in\{u,c\}$, 
this gives
\bea\label{m10sd}
 \mathcal{M}_{0,\text{sd}}^1(\Lambda)
& = & \frac{\alpha}{\pi} \ln\frac{m_Z}{\Lambda}\cdot\mathcal{M}^0_0
      \;+\; \dots \,.
\eea
The logarithm in eq.~(\ref{m10sd}) then represents the leading logarithmic 
corrections to $\order(\alpha\,\GF)$ due to virtual particle exchange with 
(virtual) photon energies above $\Lambda$.

\subsubsection{Long-distance next-to-leading order corrections}
\label{Sec:LD_NLO_Corr}

\myfigure{t}{  
  \begin{fmfgraph*}(80,60)
    \fmfcurved
    \fmfleft{on1,on2,on3,is,on4,on5,on6}
    \fmfright{of1,of2,fs,of5,of6}
    \fmftop{t1}
    \fmf{phantom}{is,v0,v1,v2,v3,v4,v5,v6,fs}
    \fmffreeze
    \fmf{phantom}{of6,v9,v8,v7,v3}
    \fmffreeze
    \fmf{dashes}{is,v3}
    \fmf{dashes}{of6,v3}
    \fmf{fermion}{fs,v3,of1}
    \fmf{photon}{v5,of5}
    \fmfblob{0.15w}{v3}
    \fmfdot{v5}
    \fmflabel{$p_B$}{is}
    \fmflabel{$k$}{of5}
    \fmflabel{$p_X$}{of6}
    \fmflabel{$p_\ell$}{fs}
    \fmflabel{$p_\nu$}{of1}
    \fmflabel{$\text{a)}$}{on6}
  \end{fmfgraph*}
  \hspace{15mm}
  \begin{fmfgraph*}(80,60)
    \fmfcurved
    \fmfleft{on1,on2,on3,is,on4,on5,on6}
    \fmfright{of1,of2,fs,of5,of6}
    \fmftop{t1}
    \fmf{phantom}{is,v0,v1,v2,v3,v4,v5,v6,fs}
    \fmffreeze
    \fmf{phantom}{of6,v9,v8,v7,v3}
    \fmffreeze
    \fmf{dashes}{is,v3}
    \fmf{dashes}{of6,v3}
    \fmf{fermion}{fs,v3,of1}
    \fmf{photon}{v1,t1}
    \fmfblob{0.15w}{v3}
    \fmfdot{v1}
    \fmflabel{$\text{b)}$}{on6}
  \end{fmfgraph*}
  \hspace{15mm}
  \begin{fmfgraph*}(80,60)
    \fmfcurved
    \fmfleft{on1,on2,on3,is,on4,on5,on6}
    \fmfright{of1,of2,fs,of5,of6}
    \fmftop{t1}
    \fmf{phantom}{is,v0,v1,v2,v3,v4,v5,v6,fs}
    \fmffreeze
    \fmf{phantom}{of6,v9,v8,v7,v3}
    \fmffreeze
    \fmf{dashes}{is,v3}
    \fmf{dashes}{of6,v3}
    \fmf{fermion}{fs,v3,of1}
    \fmf{photon}{v3,of5}
    \fmfblob{0.15w}{v3}
    \fmflabel{$\text{c)}$}{on6}
  \end{fmfgraph*}\\
  \vspace{15mm}
  \begin{fmfgraph*}(80,60)
    \fmfcurved
    \fmfleft{on1,on2,on3,is,on4,on5,on6}
    \fmfright{of1,of2,fs,of5,of6}
    \fmftop{t1}
    \fmf{phantom}{is,v0,v1,v2,v3,v4,v5,v6,fs}
    \fmffreeze
    \fmf{phantom}{of6,v9,v8,v7,v3}
    \fmffreeze
    \fmf{dashes}{is,v3}
    \fmf{dashes}{of6,v3}
    \fmf{fermion}{fs,v3,of1}
    \fmf{photon,left}{v4,v6}
    \fmfblob{0.15w}{v3}
    \fmfdot{v4,v6}
    \fmflabel{$\text{d)}$}{on6}
  \end{fmfgraph*}
  \hspace{15mm}
  \begin{fmfgraph*}(80,60)
    \fmfcurved
    \fmfleft{on1,on2,on3,is,on4,on5,on6}
    \fmfright{of1,of2,fs,of5,of6}
    \fmftop{t1}
    \fmf{phantom}{is,v0,v1,v2,v3,v4,v5,v6,fs}
    \fmffreeze
    \fmf{phantom}{of6,v9,v8,v7,v3}
    \fmffreeze
    \fmf{dashes}{is,v3}
    \fmf{dashes}{of6,v3}
    \fmf{fermion}{fs,v3,of1}
    \fmf{photon,left}{v0,v2}
    \fmfblob{0.15w}{v3}
    \fmfdot{v0,v2}
    \fmflabel{$\text{e)}$}{on6}
  \end{fmfgraph*}
  \hspace{15mm}
  \begin{fmfgraph*}(80,60)
    \fmfcurved
    \fmfleft{on1,on2,on3,is,on4,on5,on6}
    \fmfright{of1,of2,fs,of5,of6}
    \fmftop{t1}
    \fmf{phantom}{is,v0,v1,v2,v3,v4,v5,v6,fs}
    \fmffreeze
    \fmf{phantom}{of6,v9,v8,v7,v3}
    \fmffreeze
    \fmf{dashes}{is,v3}
    \fmf{dashes}{of6,v3}
    \fmf{fermion}{fs,v3,of1}
    \fmf{photon,left=1.2}{v2,v4}
    \fmfblob{0.15w}{v3}
    \fmfdot{v2,v4}
    \fmflabel{$\text{f)}$}{on6}
  \end{fmfgraph*} \\
  \vspace{15mm}
  \begin{fmfgraph*}(80,60)
    \fmfcurved
    \fmfleft{on1,on2,on3,is,on4,on5,on6}
    \fmfright{of1,of2,fs,of5,of6}
    \fmftop{t1}
    \fmf{phantom}{is,v0,v1,v2,v3,v4,v5,v6,fs}
    \fmffreeze
    \fmf{phantom}{of6,v9,v8,v7,v3}
    \fmffreeze
    \fmf{dashes}{is,v3}
    \fmf{dashes}{of6,v3}
    \fmf{fermion}{fs,v3,of1}
    \fmf{photon,left=1.2}{v3,v5}
    \fmfblob{0.15w}{v3}
    \fmfdot{v5}
    \fmflabel{$\text{g)}$}{on6}
  \end{fmfgraph*}
  \hspace{15mm}
  \begin{fmfgraph*}(80,60)
    \fmfcurved
    \fmfleft{on1,on2,on3,is,on4,on5,on6}
    \fmfright{of1,of2,fs,of5,of6}
    \fmftop{t1}
    \fmf{phantom}{is,v0,v1,v2,v3,v4,v5,v6,fs}
    \fmffreeze
    \fmf{phantom}{of6,v9,v8,v7,v3}
    \fmffreeze
    \fmf{dashes}{is,v3}
    \fmf{dashes}{of6,v3}
    \fmf{fermion}{fs,v3,of1}
    \fmf{photon,left=1.2}{v1,v3}
    \fmfblob{0.15w}{v3}
    \fmfdot{v1}
    \fmflabel{$\text{h)}$}{on6}
  \end{fmfgraph*}\vspace*{5mm}
  }
  {The Feynman diagrams for the next-to-leading order corrections to 
   $B^+ \to \bar X^0 \, \ell^+ \, \nu$ decays are shown. \label{Fig:nlofg1}}

\myfigure{t}{  
  \begin{fmfgraph*}(80,60)
    \fmfcurved
    \fmfleft{on1,on2,on3,is,on4,on5,on6}
    \fmfright{of1,of2,fs,of5,of6}
    \fmftop{t1}
    \fmf{phantom}{is,v0,v1,v2,v3,v4,v5,v6,fs}
    \fmffreeze
    \fmf{phantom}{of6,v9,v8,v7,v3}
    \fmffreeze
    \fmf{dashes}{is,v3}
    \fmf{dashes}{of6,v3}
    \fmf{fermion}{fs,v3,of1}
    \fmf{photon}{v5,of5}
    \fmfblob{0.15w}{v3}
    \fmfdot{v5}
    \fmflabel{$p_B$}{is}
    \fmflabel{$k$}{of5}
    \fmflabel{$p_X$}{of6}
    \fmflabel{$p_\ell$}{fs}
    \fmflabel{$p_\nu$}{of1}
    \fmflabel{$\text{a)}$}{on6}
  \end{fmfgraph*}
  \hspace{15mm}
  \begin{fmfgraph*}(80,60)
    \fmfcurved
    \fmfleft{on1,on2,on3,is,on4,on5,on6}
    \fmfright{of1,of2,fs,of5,of6}
    \fmftop{t1}
    \fmf{phantom}{is,v0,v1,v2,v3,v4,v5,v6,fs}
    \fmffreeze
    \fmf{phantom}{of6,v9,v8,v7,v3}
    \fmffreeze
    \fmf{dashes}{is,v3}
    \fmf{dashes}{of6,v3}
    \fmf{fermion}{fs,v3,of1}
    \fmf{photon}{v8,of5}
    \fmfblob{0.15w}{v3}
    \fmfdot{v8}
    \fmflabel{$\text{b)}$}{on6}
  \end{fmfgraph*}
  \hspace{15mm}
  \begin{fmfgraph*}(80,60)
    \fmfcurved
    \fmfleft{on1,on2,on3,is,on4,on5,on6}
    \fmfright{of1,of2,fs,of5,of6}
    \fmftop{t1}
    \fmf{phantom}{is,v0,v1,v2,v3,v4,v5,v6,fs}
    \fmffreeze
    \fmf{phantom}{of6,v9,v8,v7,v3}
    \fmffreeze
    \fmf{dashes}{is,v3}
    \fmf{dashes}{of6,v3}
    \fmf{fermion}{fs,v3,of1}
    \fmf{photon}{v3,of5}
    \fmfblob{0.15w}{v3}
    \fmflabel{$\text{c)}$}{on6}
  \end{fmfgraph*}\\
  \vspace{15mm}
  \begin{fmfgraph*}(80,60)
    \fmfcurved
    \fmfleft{on1,on2,on3,is,on4,on5,on6}
    \fmfright{of1,of2,fs,of5,of6}
    \fmftop{t1}
    \fmf{phantom}{is,v0,v1,v2,v3,v4,v5,v6,fs}
    \fmffreeze
    \fmf{phantom}{of6,v9,v8,v7,v3}
    \fmffreeze
    \fmf{dashes}{is,v3}
    \fmf{dashes}{of6,v3}
    \fmf{fermion}{fs,v3,of1}
    \fmf{photon,left}{v4,v6}
    \fmfblob{0.15w}{v3}
    \fmfdot{v4,v6}
    \fmflabel{$\text{d)}$}{on6}
  \end{fmfgraph*}
  \hspace{15mm}
  \begin{fmfgraph*}(80,60)
    \fmfcurved
    \fmfleft{on1,on2,on3,is,on4,on5,on6}
    \fmfright{of1,of2,fs,of5,of6}
    \fmftop{t1}
    \fmf{phantom}{is,v0,v1,v2,v3,v4,v5,v6,fs}
    \fmffreeze
    \fmf{phantom}{of6,v9,v8,v7,v3}
    \fmffreeze
    \fmf{dashes}{is,v3}
    \fmf{dashes}{of6,v3}
    \fmf{fermion}{fs,v3,of1}
    \fmf{photon,left}{v7,v9}
    \fmfblob{0.15w}{v3}
    \fmfdot{v7,v9}
    \fmflabel{$\text{e)}$}{on6}
  \end{fmfgraph*}
  \hspace{15mm}
  \begin{fmfgraph*}(80,60)
    \fmfcurved
    \fmfleft{on1,on2,on3,is,on4,on5,on6}
    \fmfright{of1,of2,fs,of5,of6}
    \fmftop{t1}
    \fmf{phantom}{is,v0,v1,v2,v3,v4,v5,v6,fs}
    \fmffreeze
    \fmf{phantom}{of6,v9,v8,v7,v3}
    \fmffreeze
    \fmf{dashes}{is,v3}
    \fmf{dashes}{of6,v3}
    \fmf{fermion}{fs,v3,of1}
    \fmf{photon,left=0.2}{v8,v5}
    \fmfblob{0.15w}{v3}
    \fmfdot{v8,v5}
    \fmflabel{$\text{f)}$}{on6}
  \end{fmfgraph*} \\
  \vspace{15mm}
  \begin{fmfgraph*}(80,60)
    \fmfcurved
    \fmfleft{on1,on2,on3,is,on4,on5,on6}
    \fmfright{of1,of2,fs,of5,of6}
    \fmftop{t1}
    \fmf{phantom}{is,v0,v1,v2,v3,v4,v5,v6,fs}
    \fmffreeze
    \fmf{phantom}{of6,v9,v8,v7,v3}
    \fmffreeze
    \fmf{dashes}{is,v3}
    \fmf{dashes}{of6,v3}
    \fmf{fermion}{fs,v3,of1}
    \fmf{photon,left=1.2}{v3,v5}
    \fmfblob{0.15w}{v3}
    \fmfdot{v5}
    \fmflabel{$\text{g)}$}{on6}
  \end{fmfgraph*}
  \hspace{15mm}
  \begin{fmfgraph*}(80,60)
    \fmfcurved
    \fmfleft{on1,on2,on3,is,on4,on5,on6}
    \fmfright{of1,of2,fs,of5,of6}
    \fmftop{t1}
    \fmf{phantom}{is,v0,v1,v2,v3,v4,v5,v6,fs}
    \fmffreeze
    \fmf{phantom}{of6,v9,v8,v7,v3}
    \fmffreeze
    \fmf{dashes}{is,v3}
    \fmf{dashes}{of6,v3}
    \fmf{fermion}{fs,v3,of1}
    \fmf{photon,left=1.2}{v3,v8}
    \fmfblob{0.15w}{v3}
    \fmfdot{v8}
    \fmflabel{$\text{h)}$}{on6}
  \end{fmfgraph*}\vspace*{5mm}
  }
  {The Feynman diagrams for the next-to-leading order corrections to 
   $B^0 \to X^- \, \ell^+ \, \nu$ decays are shown. \label{Fig:nlofg2}}

The {\sc Qed} long-distance corrections to the phenomenological hadron decay 
can be calculated in an effective model that arises by requiring the 
phenomenological Lagrangian of the leading order decay to be invariant 
under local $U(1)_\text{em}$ gauge transformations. Assigning the usual charges 
the following interaction terms in the Lagrangian arise in addition to 
eq.~(\ref{weaktreelevelinteraction})
\bea\label{Eq:QED_W_Lagrangian}
  \mathcal{L}_{\text{int},\QED}
& = &{}-eQ_\ell\bar\;\psi_\ell\gamma^\mu\psi_\ell A_\mu
      - ieQ_\phi A_\mu(\phi^+\dpar^\mu\phi^--\phi^-\dpar^\mu\phi^+)
      + e^2Q_\phi^2A_\mu A^\mu \phi^+\phi^- \nnb\\
&&{}
      + ie\sqrt{2}\GF V_{xy}\mathrm{f}_\pm(Q_B\pm Q_X)\phi_B\phi_X A_\mu\;
        \bar\psi_\nu P_R \gamma^\mu\psi_\ell 
      \;\;+\;\; \text{h.c.} \quad,
\eea
wherein the summation over $\phi\in\{\phi_B,\phi_X\}$ is implied.
In addition to the point-like lepton-photon and meson-photon interactions, 
a vertex emission term arises. This term is connected to the 
bound-state nature of the meson. It is infrared finite and needed for gauge 
invariance. Further, in eq.~(\ref{Eq:QED_W_Lagrangian}) it is assumed that 
the meson-photon interaction is sufficiently described by scalar {\sc Qed}. 
Additional terms arise when moving away from this assumption, including 
intermediate lines of excited hadrons necessitating $X^*\to X\gamma$ vertices 
as well as contributions due to off-shell currents. These terms are discussed 
on general grounds in Sec.~\ref{Sec:theory_SD_terms} and will be largely 
neglected in this study. This, in most cases, roots in their unavailability or, 
where known, in their smallness.

Figs.~\ref{Fig:nlofg1} and \ref{Fig:nlofg2} depict the relevant real and 
virtual diagrams for $B^+ \to \bar X^0 \, \ell^+ \, \nu$ and 
$B^0 \to X^- \, \ell^+ \, \nu$ decays at $\order(\alpha\,\GF)$. The real 
corrections diagrams a, b and c correspond to the emission of a real photon 
from either the charged legs of the decay, or the charged vertex itself. 
The virtual corrections group into three categories: diagrams d and e concern 
the wave-function renormalisation of the charged legs while diagram f is the 
dominant inter-particle photon exchange. Diagrams g and h are again due to 
emissions off the charged effective vertex and are, thus, infrared finite. 
The corresponding subtleties involving the vertex emissions are detailed in 
App.~\ref{App:NLO_ME}. 

In the virtual amplitude $\mathcal{M}_{0,\text{ld}}^1$ the arising ultraviolet 
divergences are regularised using the Pauli-Villars prescription 
\cite{Pauli:1949zm} by introducing an unphysical heavy photon of mass $\Lambda$ 
and opposite norm. Consequently, the virtual corrections have exactly the form 
required by the matching procedure outlined in Sec.~\ref{Sec:matching} and all 
real emission processes are described in the long-distance picture.

Squaring the phenomenological real and virtual matrix elements results in the 
real and virtual next-to-leading order differential rates. In the $B$ meson 
rest frame they read
\bea \label{Eq:nlo_rate}
 \ud \Gamma^1_1 
&\!=\!& \frac{1}{(2 \pi)^{12}}
        \frac{\dthree p_X}{E_X}\frac{\dthree p_\ell}{E_\ell}
        \frac{\dthree p_\nu}{E_\nu}\frac{\dthree k}{E_k} \;
        \delta^{(4)}\bve{p_B - p_X - p_l - p_\nu - k}\,
        \big| \mathcal{M}_{1,\text{ld}}^\half \big|^2 \; 
        \,, \nnb\\
\done\Gamma_0^0 + \done\Gamma_0^1 
&\!=\!& \frac{1}{64\,\pi^3 m_B} 
        \left(\big|\mathcal{M}_0^0 \big|^2 
              +2\mathcal{R}\text{e}\!\left[\mathcal{M}_0^0 \mathcal{M}_{0,\text{ld}}^{1*}(\Lambda)\right]
              +2\,\big|\mathcal{M}^0_0\big|^2\,\frac{\alpha}{\pi}\,\ln\frac{m_Z}{\Lambda}
        \right) \, \done E_X \, \done E_\ell  \,.
\eea
Integrating eqs.~(\ref{Eq:nlo_rate}) results in the next-to-leading order total 
decay rate. Comparing with the total tree-level decay rate, the integral over 
phase-space of eq.~(\ref{Eq:treeleveldifferentialrate}), yields the long-distance 
enhancement factor  $\delta_{\text{ld}}$ due to next-to-leading order effects. 
It is 
\bea
 \Gamma
 \;\;=\;\; \left( 1 + \delta_{\text{sd}} + \delta_{\text{ld}} \right) \, \Gamma^0_0
 \;\;=\;\; \Gamma^0_0 + \Gamma^1_0 + \Gamma^1_1 + \order(\alpha^2\,\GF)\,.
\eea
with $\delta_{\text{sd}} =  \frac{2\alpha}{\pi} \ln \frac{m_Z}{\Lambda} $ from 
eq.~(\ref{m10sd}). The exact form of the next-to-leading order matrix elements 
can be found in App.~\ref{App:NLO_ME}.

 \subsection{Structure dependent terms}
\label{Sec:theory_SD_terms}

This section discusses the arising additional electromagnetic next-to-leading 
order corrections that cannot be grasped by simply replacing 
$\dpar_\mu\to D_\mu$ to arrive at a $U(1)_\text{em}$ gauge invariant 
phenomenological Lagrangian, as was described in the previous section. 
These include both deviations arising from the point-like meson-photon 
interaction assumed above and additional terms arising in the interaction of an 
off-shell hadronic current. Nonetheless, it is clear that in the relevant 
phase space region for the total inclusive decay rate, namely the region as 
$k\to 0$ near the infrared divergence, both the real and virtual next to 
leading order matrix elements are completely determined by the leading order 
decay, and the above procedure accurately reproduces the full theory in this 
region \cite{Low:1958sn,Yennie:1961ad,Burnett:1967km}. Hence, the real emission 
squared amplitude in this limit reads
\bea
 \big|\mathcal{M}_1^\half\big|^2
&\stackrel{k\to 0}{=}& -e^2\left(\frac{p_M}{k\cdot p_M}
                                -\frac{p_\ell}{k\cdot p_\ell}\right)^{\!\!2}
                        \big|\mathcal{M}_0^0\big|^2\;
                        \qquad\mbox{with}\;\;p_M\in\{p_B,p_X\}\,,
\eea
depending on whether $B$ or $X$ is charged.
Physically this roots in the fact that the wavelength of an infinitely soft 
photon is much larger than the size of any strongly bound hadron. It, thus, 
cannot resolve its substructure and effectively interacts with its summed, then 
point-like charge. Further, such soft photons cannot push the hadronic current 
significantly off-shell, such that off-shell current interactions can become 
sizeable.

Introducing non-point-like meson-photon interactions does not only lead to 
corrections due to the hadron's size and its internal charge distribution, it 
also leads to additional vertices of the type $X\to X^*\gamma$, where $X^*$ is a 
higher resonance of the $X$ meson. This necessarily also introduces additional 
terms in the interaction of the hadronic and the leptonic current, especially if 
the resonance differs in its spin. $B^*$ resonances in the initial 
state occur in the unphysical region. Hence, they are only relevant if their 
width is comparable to, or larger than, the mass separation $m_B^2-m_{B^*}^2$ to 
the initial state $B$ meson. In contrast, $D^*$ resonances, for example, 
occurring in a final state line are allowed to be on-shell for a range of photon 
energies. Thus, a considerable correction may arise. \cite{Becirevic:2009xp} 
find the $D^{*+}\to D^+\gamma$ coupling to be compatible with zero while the 
$D^{*0}\to D^0\gamma$ coupling is small, but considerable. Both are considered 
and discussed in detail in Sec.~\ref{Sec:result_SD_terms}. 

Generally, following the argumentation of \cite{Gasser:2004ds,Becirevic:2009fy} 
the electromagnetic current of the hadronic system can be split into two 
components: \emph{inner-bremsstrahlung} (IB) contributions, which account for 
photon radiation from the external charged particles and are completely 
determined by the non-radiative process, and \emph{structure-dependent} (SD) 
contributions, which describe intermediate hadronic states and represent new 
information with respect to the IB contributions. 

The amplitude of a semileptonic $B$ meson decay with full electromagnetic 
corrections reads
\bea\label{Eq:eleccoupl}
 \mathcal{A}_\nu
& = & i\,e\,\frac{\GF}{\sqrt{2}} V_{\text{xb}}\;
      \bar u_\nu\, \pr\gamma^\mu\, 
      \left({}- \frac{ H_{\mu}}{2 p_\ell \cdot k}   
                \left( \gamma_\nu \ds{k} + 2 p_{\ell,\nu} \right)
              + V_{\mu\nu} - A_{\mu\nu}
      \right) v_\ell\,,
\eea
with the hadronic current $H_\mu$, as introduced in 
eq.~(\ref{Eq:hadcurtreelev}). The hadronic vector and axial form factors of the 
photon-emitting hadronic current, incorporating among others the 
$X\to X^*\gamma$ coupling, are given by the unknown non-local operator
\bea\label{Eq:nonlocalop}
  V_{\mu\nu} - A_{\mu\nu} 
& = & \int \ud^4 x \; e^{i \, k \cdot x} \;
      \bra X | T \, [ \hat h_{\mu} (0) \, J^{\text{em}}_\nu(x) ] \, | B \ket \, .
\eea
$J^{\text{em}}_\nu$ denotes the electromagnetic current, $\hat h_{\mu}$ the 
quark-current in position-space and $k$ is the photon momentum. The vector and 
axial-vector operators of eq.~(\ref{Eq:nonlocalop}) obey the electromagnetic 
Ward-identities, obtained by contracting $k^\nu A_\nu$ of eq.~(\ref{Eq:eleccoupl}),
\bea \label{Eq:wardident}
 k^\nu \, V_{\mu\nu} & = & H_\mu \,, \nnb \\
 k^\nu  A_{\mu\nu} & = & 0 \,.
\eea
These properties of the individual pieces of the amplitude in conjunction with 
Low's theorem \cite{Low:1958sn,Burnett:1967km} lead to the fact, that the 
leading terms of next-to-leading order amplitude in powers of the photon 
four-momentum $k$, i.e.~the terms proportional to $k^{-1}$ and $k^0$, are 
completely determined by the on-shell form factors of the tree-level decay. 

Following \cite{Gasser:2004ds}, corrections beyond $\mathcal{O}(k^0)$ can be 
included by separating the non-local operator eq.~(\ref{Eq:nonlocalop}) into SD 
and IB contributions. Since the IB and SD describe different physical 
mechanisms they are separately gauge invariant. Further, the SD amplitude must 
be of $\order(k)$ or higher. This, however, does not prevent the IB amplitude from 
containing terms of $\order(k)$ and higher as well. 

Splitting the amplitude under these restrictions allows more terms to be 
collected in the IB part, still using only the knowledge of the non-radiative 
matrix element. This offers the advantage to obtain more precise predictions 
for the decay process, without formulating the (mostly unknown) SD 
contributions. The splitting of the transition matrix element requires a 
corresponding splitting of the non-local operator eq.~(\ref{Eq:nonlocalop}) 
into SD and IB parts. 

The axial contributions are strictly zero for photons emitted from the $B$- or 
$X$-meson, and therefore can be considered purely SD. They can be written in 
the form \cite{Bijnens:1992en} 
\bea
 A_{\mu\nu} 
\;\;=\;\; A_{\mu\nu}^{\text{SD}} 
&   =  &{}-i \, \epsilon_{\mu\nu\rho\sigma} 
          \Big[ A_1 \, p_X^\rho \, k^\sigma
                +A_2 \, k^\rho \left(p_\ell + p_\nu\right)^\sigma\Big] \nnb \\
&      &{}-i \, \epsilon_{\nu\lambda\rho\sigma} \, p_X^\lambda \, k^\rho \, 
          \left(p_\ell + p_\nu\right)^\sigma  
          \Big[ A_3 \, \left(p_\ell + p_\nu\right)_\mu + A_4 \, p_\mu \Big] \,.
\eea
Note that the Lorentz-invariant scalars $A_{i}$ are non-singular in the limit 
$k\to 0$ by construction and are functions of the three independent scalar 
variables that can be built with $p_B$, $p_X$ and $k$. The decomposition of the 
vector current reads
\bea\label{veccursplit}
 V_{\mu\nu} 
& = &  V_{\mu\nu}^{\text{IB}} + V_{\mu\nu}^{\text{SD}} \,, 
\eea
where the IB piece is chosen in such a way, that 
\bea \label{IBWard}
 k^\nu \, V_{\mu\nu}^{\text{IB}} & = & H_{\mu} \nnb\\
 k^\nu \, V_{\mu\nu}^{\text{SD}} & = & 0 \,.
\eea
Thus, the decay amplitude separates as
\bea \label{sdibsplit}
\mathcal{A}_\nu 
& = & i\,e\,\frac{\GF}{\sqrt{2}} V_{\text{xb}}\;
      \bar u_\nu\, \pr\gamma^\mu\, 
      \left({}- \frac{ H_{\mu}}{2 p_\ell \cdot k}   
                \left( \g^\rho \ds{k} + 2 p_\ell^\rho \right)
              + V_{\mu\nu}^{\text{IB}}
      \right) v_\ell \nnb\\
&&{} +i\,e\,\frac{\GF}{\sqrt{2}} V_{\text{xb}}\;
      \bar u_\nu\, \pr\gamma^\mu\, 
      \left( V_{\mu\nu}^{\text{SD}} - A_{\mu\nu}^{\text{SD}} \right) v_\ell \,.
\eea
Herein, $V_{\mu\nu}^{\text{IB}}$ can be constructed from leading order 
information only, cf.~App.~\ref{App:NLO_ME}. The SD vector contributions, on the 
other hand, contain additional information. They can be written as 
\cite{Poblaguev:1999ys}
\bea\label{sdveccur}
V_{\mu\nu}^{\text{SD}}
& = &  V_1 \Big[ k_\mu \, p_{X\nu} - (p\cdot k) \, g_{\mu\nu} \Big]
     + V_2 \Big[ k_\mu (p_\ell + p_\nu)_\nu 
                  - (k\cdot(p_\ell + p_\nu)) g_{\mu\nu} \Big] \nnb\\
&&{} + V_3 \Big[ (k\cdot(p_\ell + p_\nu))(p_\ell + p_\nu)_\mu \, p_{X\nu}
                  - (p_X\cdot k)(p_\ell + p_\nu)_\mu (p_\ell + p_\nu)_\nu \Big] \nnb \\
&&{} + V_4 \Big[ (k\cdot(p_\ell + p_\nu))(p_{X\mu} \, p_{X\nu}
                  - (p_X\cdot k) \, p_{X\mu} (p_\ell + p_\nu)_\nu \Big] \,,
\eea
where the Lorentz-invariant scalars $V_{i}$ are functions of the three 
independent scalar variables that can be built with $p_B$, $p_X$ and $k$. 
All IB and SD contributions are finite as $k\to 0$. 
The IB and SD parametrization of \cite{Cirigliano:2005ms} based on 
\cite{Fearing:1970zz} can be obtained by a change of basis in 
eq.~(\ref{sdveccur}), and correspondingly shifting terms of $\mathcal{O}(k)$ 
and higher into the SD contributions.
The IB coupling to the electromagnetic current can be used to construct the 
next-to-leading order matrix elements, receiving in principle additional 
corrections from the SD coupling. 

The knowledge of the full SD contributions 
for semileptonic $B$ meson decays is modest: \cite{Cirigliano:2005ms} discusses 
the matter for $B\to \pi\,\ell\,\nu\,\gamma$ decays, using the soft-collinear 
effective theory to isolate the expressions for the SD contributions in the 
soft-pion and hard-photon part of phase-space. Sec.~\ref{Sec:result_SD_terms} 
compares their findings with the pure IB result from this study. The 
differences are non sizable. The recent work of \cite{Becirevic:2009fy} 
addresses the real SD corrections to $B\to D\,\ell\,\nu\,\gamma$ decays by 
using lattice results of the $D^*\to D\,\gamma$ coupling to estimate the 
dominant SD contributions when the $D^*$ is on-shell. 
Sec.~\ref{Sec:result_SD_terms} also compares these SD contribution with the complete 
SD+IB picture. Again the differences turn out to be non sizable. The SD 
corrections to $B\to D^*_0\,\ell\,\nu\,\gamma$ are unknown, but given the large 
widths of the $D^*_0$ and $D^*_1$ states a non-negligible correction to the 
pure IB prediction can be expected.

 \subsection{Soft-resummation and inclusive exponentiation}
\label{Sec:theory_resummation}

This section discusses a systematic improvement of the fixed order results 
discussed in the previous section. Centring on the exponentiability of 
soft-radiative corrections and following the approach of Yennie, Frautschi and 
Suura~\cite{Yennie:1961ad}, the fully inclusive decay rate
\bea\label{Eq:fully_incl_dec_rate}
 \Gamma
& = & \frac{1}{2M}\!\sum_{n_R=0}^\infty\frac{1}{n_R!}\;
      \int\done\Phi_{p_f}\done\Phi_k\,
      (2\pi)^4\delta^4\left(p_B-p_X-p_\ell-p_\nu-\sum k\right)
      \left|\sum_{n_V=0}^\infty\mathcal{M}_{n_R}^{n_V+\frac{1}{2}n_R}\right|^2
\eea
can be rewritten as
\bea\label{Eq:fully_incl_dec_rate_yfs}
 \Gamma
& = & \frac{1}{2M}\!\sum_{n_R=0}^\infty\frac{1}{n_R!}\;
      \int\done\Phi_{p_f}\done\Phi_k^\prime\,
      (2\pi)^4\delta^4\left(p_B-p_X-p_\ell-p_\nu-\sum k\right) \nnb\\
&& \hspace{30mm}
      \times\,e^{Y(\Omega)}
      \prod_{i=1}^{n_R}\tilde{S}(k_i)\Theta(k_i,\Omega)
      \left(\tilde{\beta}_0^0+\tilde{\beta}_0^1
            +\sum_{i=1}^{n_R}\frac{\tilde{\beta}_1^1(k_i)}{\tilde{S}(k_i)}
            +\hspace{3mm} \mathcal{O}(\alpha^2)\hspace{3mm}\right)
\eea
by separating the universal spin-independent infrared divergent terms from the 
virtual and real emission amplitudes. $\done\Phi_{p_f}$ and $\done\Phi_k$ are 
the leading order and (multiple) extra emission phase space elements, while 
$n_R$ and $n_V$ count the additional real and virtual photons present in each 
amplitude. Therefore, using the same convention as before the sub- and 
superscripts of the (squared) matrix elements $\mathcal{M}$, $M$ and $\tilde\beta$ 
denote their real emission photon multiplicity and their order of $\alpha$ in 
the perturbative expansion relative to the leading order.

The separation of infrared divergences proceeds by splitting
\bea
 \mathcal{M}_0^1 
& = & \alpha B \mathcal{M}_0^0 + M_0^1\,, \\
  \frac{1}{2(2\pi)^3}\big|M_1^\half\big|^2 
& = & \tilde{S}(k)\big|M_0^0\big|^2 + \tilde{\beta}_{1}^{1}(k)\,,
\eea
wherein $M_0^1$ and $\tilde{\beta}_{1}^{1}(k)$ are free of any infrared 
singularities due to virtual or real photon emissions.
This separation can be continued iteratively, leading to
\bea\label{Eq:n_V_virtual_emissions}
\left|\sum_{n_V=0}^\infty\mathcal{M}_{n_R}^{n_V+\frac{1}{2}n_R}\right|^2 & = &
\exp(2\alpha B)\left|\sum_{n_V=0}^\infty 
        M_{n_R}^{n_V+\frac{1}{2}n_R}\right|^2\,.
\eea
and
\bea\label{Eq:n_R_real_emissions}
\lefteqn{\left(\frac{1}{2(2\pi)^3}\right)^{n_R}
\left|\sum_{n_V=0}^\infty M_{n_R}^{n_V+\frac{1}{2}n_R}\right|^2}\nnb\\
& = & \tilde{\beta}_0 \prod_{i=1}^{n_R}\left[\vp\tilde{S}(k_i)\right]
    +\sum_{i=1}^{n_R}\left[\frac{\tilde{\beta}_1(k_i)}{\tilde{S}(k_i)}\right] 
        \prod_{j=1}^{n_R}\left[\vp\tilde{S}(k_j)\right]
    +\sum_{\genfrac{}{}{0pt}{}{i,j=1}{i\neq j}}^{n_R}\left[
        \frac{\tilde{\beta}_2(k_i,k_j) }{\tilde{S}(k_i)\tilde{S}(k_j)}\right]
        \prod_{l=1}^{n_R}\left[\vp\tilde{S}(k_l)\right]
    +\dots \nnb\\
&& {}+\sum_{i=1}^{n_R}\left[\vp
        \tilde{\beta}_{n_R-1}(k_1,\dots,k_{i-1},k_{i+1},\dots,k_{n_R})\,
        \tilde{S}(k_i)\right]
   +\tilde{\beta}_{n_R}(k_1,\dots,k_{n_R})\,,
\eea
with $\tilde{\beta}_{n_R} = \sum_{n_V=0}^\infty\tilde{\beta}_{n_R}^{n_V+n_R}$. 
Note that for a given phase space phase space configuration $\{p_1,\ldots,p_n,
k_1,\ldots,k_{n_R}\}$ the infrared subtracted squared matrix elements 
$\tilde\beta_1(k_i)$ involve a projection onto the single emission subspace 
$\{\mathcal{P}p_1,\ldots,\mathcal{P}p_n,\mathcal{P}k_i\}$. Of course, 
momentum conservation holds for each projected subset. Thus, for every radiated 
photon the $\tilde\beta_1(k_i)$ are evaluated as if this photon was the only 
one in the event. Hence, truncating the pertubative series in the 
$\tilde\beta_{n_R}$ at the next-to-leading order leaves every single photon 
emission correct at $\order(\alpha)$.

Exponentiating the integral of the eikonal $\tilde S(k)$ upon insertion of the 
identity of eq.~(\ref{Eq:n_R_real_emissions}) over the unresolved 
phase space $\Omega$, containing the infrared singularity gives rise to the 
Yennie-Frautschi-Suura form factor
\bea
 Y(\Omega) & = & 2\alpha(B+\tilde B(\Omega)) 
\qquad\mbox{with}\qquad
 2\alpha \tilde B(\Omega)\;\;=\;\;\int_\Omega\frac{\dthree k}{k}\;\tilde S(k)
\eea
and the residual perturbative series of the infrared-subtracted squared 
amplitudes $\tilde\beta_{n_R}^{n_V+n_R}$ in 
eq.~(\ref{Eq:fully_incl_dec_rate_yfs}). 
Hence, photon emissions contained in the unresolved soft region $\Omega$ are 
assumed to have a negligible effect on differential distributions, but are 
included in the overall normalisation.
Furthermore, $Y(\Omega)$ is UV-finite and, thus, does not interfere with 
renormalisation of the $\tilde\beta_{n_R}^{n_V+n_R}$.

\section{Methods \& Implementations}\label{Sec:Methods}
 
In this section a short overview over the generators used in this study and 
their underlying principles is given.

\subsection{\protect\BLOR}
\label{Sec:methods_BLOR}

\BLOR \cite{Bernlochner:2010yd} is a fixed order Monte Carlo Event generator 
specialised on {\sc Qed} corrections in semileptonic $B$ meson decays. It 
separately generates decays for the born ($1\to 3$) and real emission 
($1\to 4$) phase space events according to their individual cross sections
\bea
 \done\Gamma_{1\to 3}
& = & \done\Gamma_0^0 + \done\Gamma_0^1 \nnb\\
 \done\Gamma_{1\to 4}
& = & \done\Gamma_1^1 \,.
\eea
The infrared divergences present in $\done\Gamma_0^1$ and $\done\Gamma_1^1$ are 
regulated introducing a small but finite photon mass $\lambda$ set to 
$10^{-7}$GeV. The expressions of Sec.~\ref{Sec:NLO} are altered accordingly.

\subsection{\protect\Sherpa/\protect\Photons}
\label{Sec:methods_SHERPA_PHOTONS}

The \Sherpa Monte Carlo \cite{Gleisberg:2008ta} is a complete event generation 
frame work for high energy physics processes. Although its traditional 
strengths lie in the perturbative aspects of lepton and hadron colliders 
it also encompasses several modules for all non-perturbative aspects. In this 
work the hadron decay module \Hadrons \cite{Krauss:2010xx} and the universal 
higher order {\sc Qed} correction tool \Photons \cite{Schonherr:2008av} are 
used.

\Hadrons generates the leading order decay events according to the respective 
form factor parametrisations of the involved hadronic and 
leptonic currents. \Photons then rewrites the differential all orders 
inclusive decay width of eq.~(\ref{Eq:fully_incl_dec_rate_yfs}) as 
\bea\label{Eq:sherpa_monte_carlo_equation}
 \Gamma 
=  \Gamma_0\;\sum_{n_\gamma}\frac{1}{n_\gamma!}\,
      \int\done\Phi_k\, J_{\text{P}}(k)\, e^{Y(\Omega)}
      \prod_{i=1}^{n_\gamma}\left[\tilde{S}(k_i)\Theta(k_i,\Omega)\right]
      \left(1+\frac{\tilde{\beta}_0^1}{\tilde{\beta}_0^0}
            +\sum_{i=1}^{n_R}\frac{\tilde{\beta}_1^1(k_i)}{\tilde{\beta}_0^0\cdot\tilde{S}(k_i)}
            +\mathcal{O}(\alpha^2)\right),\hspace*{-10mm}\nnb\\
{}
\eea
cf.~\cite{Schonherr:2008av}. Hence, the leading order input is corrected both 
for soft photon effects to all orders and hard photon emission to an arbitrary 
order. $J_{\text{P}}\le1$ represents the various Jacobians occurring when 
factoring out the leading order term. The perturbative series in the 
infrared-subtracted squared matrix elements includes only terms up to 
$\order(\alpha)$ for this study. The {\sc Nlo} real emission squared amplitudes 
can alternatively be approximated using Catani-Seymour splitting function 
\cite{Catani:1996vz,Dittmaier:1999mb,Catani:2002hc}
\bea
 \tilde\beta_{1,\text{CS}}^1(k)
& = &  -\frac{\alpha}{4\pi^2}\sum_{i<j}Z_iZ_j\theta_i\theta_j
        \left(\bar{\mbox{\sl g}}_{ij}(p_i,p_j,k)+
        \bar{\mbox{\sl g}}_{ji}(p_j,p_i,k)\right)\tilde{\beta}_0^0\,,
\eea
where in $i$ and $j$ run over all particles in the process. The exact form of 
the $\bar{\mbox{\sl g}}_{ij}$ can be found in \cite{Schonherr:2008av}. This 
approximation is also used if the exact real emission matrix element is not 
known.

The infrared cut-off was set to $10^{-6}$GeV in the rest frame of the charged 
dipole, i.e.~the $B^+\!\!-\ell^+$ or the $X^-\!\!-\ell^+$ system, respectively.

\subsection{\protect\PHOTOS}
\label{Sec:methods_PHOTOS}

The \PHOTOS Monte Carlo \cite{Barberio:1990ms} is an 
\lq\lq after-burner\rq\rq~algorithm, which adds approximate brems\-strahlung 
corrections to leading order events produced by an external code. In this study 
it is taken as a reference for the differential distributions since it is 
widely in use. \PHOTOS bases on the factorisation of the real emission matrix 
element in the collinear limit
\bea
 \big|\mathcal{M}_1^\half\big|^2
& = & \sum_i\big|\mathcal{M}_0^0\big|^2 \cdot\; f(p_i,k)\,.
\eea
The radiation function $f(p_i,k)$ is given to leading logarithmic accuracy. 
It incorporates the Altarelli-Parisi emission kernel for radiation off the 
particular final state particle. Its exact form is spin dependent and can be 
found in \cite{Barberio:1990ms,Barberio:1993qi}. In its exponentiated mode the 
number of photons follows a Poissonian distribution, while the individual 
photon's kinematics are determined by applying above equation iteratively. To 
also recover the soft limit of real photon emission matrix elements an 
additional weight was introduced \cite{Golonka:2005pn}
\bea\label{Eq:PHOTOS_soft_weight}
 W_{\text{soft}}
& = & \frac{\sum\limits_{i=1}^{n_\gamma}
            \left|\sum\limits_{j=1}^{n_C} Q_j
                  \frac{p_j\cdot\eps_i^*}{p_j\cdot k_i}\right|^2}
           {\sum\limits_{i=1}^{n_\gamma}\sum\limits_{j=1}^{n_C} Q_j^2
            \left|\frac{p_j\cdot\eps_i^*}{p_j\cdot k_i}\right|^2}\,,
\eea
wherein $n_\gamma$ and $n_C$ are the photon and the final state charged 
multiplicity of the process. To regularise the emission function in the soft 
limit an energy cut-off is imposed in the rest frame of the decaying particle. 
The collinear divergence is regularised by the emitter's mass.

Initial state radiation is not accounted for as the mass of the decaying 
particle is the largest scale in the process and there are no associated 
collinear divergences or logarithmic enhancements. It could only be accounted 
for by supplementing \PHOTOS with a matrix element correction. In case of heavy 
initial states eq.~(\ref{Eq:PHOTOS_soft_weight}) still approximately recovers 
the soft limit.

In this analysis \PHOTOS version 2.13 has been used in its exponentiated mode, 
including the soft interference terms. The infrared cut-off was set to 
$10^{-7} m_{B^{0,+}}$, respectively. \BLOR supplemented the leading order decay 
events.



\section{Results}\label{Sec:Results}
 \mytable{h!}{
  \begin{minipage}{0.45\textwidth}
    \begin{tabular}{cl}
      Parameter & Value \\ \hline
      $m_{\Upsilon(4S)}$ & 10.5794 GeV \\
      $\Gamma_{\Upsilon(4S)}$ & 20.5 MeV \\
      $m_{B^0}$ & 5.27950 GeV \\
      $m_{B^+}$ & 5.27913 GeV \\
      $m_{D^0}$ & 1.86950 GeV \\
      $m_{D^-}$ & 1.86484 GeV \\
      $m_{D^{*\,0}_0}$ & 2.403 GeV \\
      $m_{D^{-\,0}_0}$ & 2.352 GeV \\
      $m_{\pi^0}$ & 0.134976 GeV \\
      $m_{\pi^-}$ & 0.13957 GeV \\
    \end{tabular}
  \end{minipage}
  \begin{minipage}{0.03\textwidth}
  \end{minipage}
  \begin{minipage}{0.45\textwidth}
    \begin{tabular}{cl}
      Parameter & Value \\ \hline
      $m_e$ & 0.0005109989 GeV \\
      $m_\mu$ & 0.10565837 GeV \\
      $m_{\nu_e}$ & 0 \\
      $m_{\nu_\mu}$ & 0 \\
      $m_W$ & 80.419 GeV \\
      $\alpha$ & 1/137.035999 \\
      $\GF$ & $1.16637\cdot 10^{-5}\;\GeV^{-2}$
    \end{tabular}\vspace*{6.1mm}
  \end{minipage}\vspace*{3mm}
  }
  {Parameters used for all inclusive and differential decay rate calculations. 
   All particle widths, except the $\Upsilon(4S)$ width, are considered 
   negligible. 
    \label{Tab:params}}

In this section the results of the inclusion of the complete next-to-leading 
order corrections, both real and virtual short and long distance contributions, 
are reviewed for different decay channels with different form factor 
parametrisations of the hadronic current. First, results for the 
next-to-leading order inclusive decay rates and their effects on the 
extraction of $|V_\text{xb}|$ from decay
measurements will be shown in Sec.~\ref{Sec:res_rates}. Then, in 
Sec.~\ref{Sec:res_diff} the effects on differential distributions and spectra 
are investigated and compared against the standard tool used in many 
experimental analyses, \PHOTOS. The parameters used are detailed in 
Tab.~\ref{Tab:params}.

 \subsection{Next-to-leading order corrections to decay rates}
\label{Sec:res_rates}

One key prediction of this publication is the process specific correction 
factor for higher order electroweak effects. These higher order corrections 
enter measurements of the {\sc Ckm} mixing angles $V_\text{xb}$ via 
\bea
 \Gamma_\text{measured}
& = & \eta_\QCD^2\;\eta_\EW^2\;|V_\text{xb}|^2\;\tilde\Gamma_\text{LO}\,,
\eea
and, thus,
\bea
 |V_\text{xb}|
& = & \frac{1}{\eta_\EW}\;\cdot\;
      \sqrt{\frac{\Gamma_\text{measured}}{\eta_\QCD^2\tilde\Gamma_\text{LO}}}\;\;.
\eea
$\tilde\Gamma_\text{LO}$ is the leading order phenomenological decay rate 
stripped of the {\sc Ckm} mixing angle. $\eta_\QCD$ and $\eta_\EW$ incorporate 
the higher order {\sc Qcd} and electroweak corrections. Both contributions 
factorise at the {\sc Nlo} level. The electroweak correction factor is determined 
in this study at {\sc Nlo} accuracy in the {\sc Qed}-improved phenomenological 
long-distance description of the hadronic decay and at leading logarithmic 
accuracy in the underlying short-distance partonic decay in the Standard Model. 
While the leading logarithm of the short-distance correction depends only on 
the matching scale $\Lambda$ of both descriptions and is, thus, independent of 
the actual decay properties, the long-distance corrections, in contrast, are 
sensitive to exactly these specifics. 

The resulting correction factors, 
\bea
 \eta_\EW^2
\;\; = \;\; 1+\delta_\text{sd}+\delta_\text{ld} 
\;\; = \;\; 1+\frac{\Gamma_0^1+\Gamma_1^1}{\Gamma_0^0}+\order(\alpha^2)
\eea
are a central outcome of this study. They are presented in 
Tab.~\ref{Tab:SD_LD_emu_sep} for the different semileptonic decay channels of 
charged and neutral $B$ mesons into (pseudo)scalar mesons. 
As is evident, the long-distance {\sc Qed} corrections break the strong isospin 
symmetry. This originates in the different masses of the charged mesons in the 
strong isospin rotated decays. These masses both determine the amount 
of radiation from the meson line and enter the loop integrals.
For the same reasons the corrections for pions are larger than for $D$ mesons. 
On the other hand, the difference in the size of the correction between the two 
leptonic channels of each decay mode is only very small since both masses 
are insignificant compared to the hadronic mass scales. 

\mytable{tbp!}{
  \begin{tabular}{l lll} 
      & $\eta_\EW^2$ & $1/\eta_\EW$    \\ \hline 
    $B^0 \to D^- \, e^+ \, \nu_e \, (\g)$ & $1.0222(1 \pm 2 \pm 17 \pm 1)$  & $0.9891(1 \pm 1 \pm 4 \pm 1)$  \\
    $B^0 \to D^- \, \mu^+ \, \nu_\mu \, (\g)$ & $1.0222(1 \pm 2 \pm17 \pm 1)$  & $0.9891(1 \pm 1 \pm 4 \pm 1)$  \\
    $B^+ \to \bar D^0 \, e^+ \, \nu_e \, (\g)$  & $1.0146(1 \pm 1 \pm 39 \pm 16 )$  & $0.9928(1 \pm 1 \pm 10 \pm 4)$  \\
    $B^+ \to \bar D^0 \, \mu^+ \, \nu_\mu \, (\g)$ & $1.0147(1 \pm 1 \pm 39 \pm 16)$  & $0.9927(1 \pm 1 \pm 10 \pm 4)$  \\
    \hline
    $B^0 \to \pi^- \, e^+ \, \nu_e \, (\g)$     & $1.0555(1 \pm 4 \pm 148 \pm 48)$  & $0.9734(1 \pm 1 \pm 33 \pm 10)$  \\
    $B^0 \to \pi^- \, \mu^+ \, \nu_\mu \, (\g)$ & $1.0545(1 \pm 4 \pm 136 \pm 48)$  & $0.9738(1 \pm 1 \pm 31 \pm 11)$  \\
    $B^+ \to \pi^0 \, e^+ \, \nu_e \, (\g)$     & $1.0411(1 \pm  3 \pm 100)$  & $0.9801(1 \pm 1 \pm 23)$  \\
    $B^+ \to \pi^0 \, \mu^+ \, \nu_\mu \, (\g)$ & $1.0401(1 \pm  3 \pm 89)$  & $0.9805(1 \pm 1 \pm 21)$   \\
    \hline
    $B^0 \to D^{*\,-}_0 \, e^+ \, \nu_e \, (\g)$    & $1.0224(1 \pm 2 \pm 10)$  & $0.9890(1 \pm 1 \pm 2)$  \\
    $B^0 \to D^{*\,-}_0 \, \mu^+ \, \nu_\mu \, (\g)$ & $1.0226(1 \pm 2 \pm 10)$  & $0.9889(1 \pm 1 \pm 2)$  \\
    $B^+ \to \bar D^{*\,0}_0 \, e^+ \, \nu_e \, (\g)$   & $1.0142(1 \pm 1 \pm 35)$  & $0.9930(1 \pm 1 \pm 8)$  \\
    $B^+ \to \bar D^{*\,0}_0 \, \mu^+ \, \nu_\mu \, (\g)$ & $1.0144(1 \pm 1 \pm 35)$  & $0.9929(1 \pm 1 \pm 8)$  
  \end{tabular}\vspace*{3mm}}
  {Predictions for $\eta_\EW=\sqrt{1+\delta_{\text{sd}} + \delta_{\text{ld}}}$ for the summed 
   next-to-leading corrections are listed. The uncertainties in the parentheses 
   are the sum of numerical, next-to-next-to-leading order, matching and 
   missing structure dependent contributions. For 
   $B^+\to\pi^0\,\ell\,\nu\,(\g)$ and $B^{0,+} \to D_0^{*\,-,0}\,\ell\,\nu\,(\g)$ 
   decays no predictions for structure dependent contributions are known. \label{Tab:SD_LD_emu_sep}}

\mytable{tbp!}{
  \begin{tabular}{r lll} 
      & $\eta_\EW^2$ & $1/\eta_\EW$    \\ \hline 
    $B^0 \to D^- \, \ell \, \nu$ & $1.0222(17)$  & $0.9891(4)$  \\
    $B^+ \to D^0 \, \ell \, \nu$  &  $1.0146(43)$& $0.9928(10)$   \\
    \hline
    $B \to D \,\ell \, \nu$  & $1.0186(29)$ & $0.9909(7)$ \\
  \end{tabular}\vspace{2mm}\\
  \begin{tabular}{r ll} 
      & $\eta_\EW^2$ & $1/\eta_\EW$     \\ \hline 
    $B^0 \to \pi^- \, \ell \, \nu$      & $1.0550(150) $  &  $0.9736(34)  $ \\
  \end{tabular}\vspace{3mm}}
  {Averaged integration results for $\eta_\EW^2=1+\delta_{\text{sd}}+\delta_{\text{ld}}$
   and $1/\eta_\EW$: 
   The uncertainties in the parentheses are the sum of numerical, 
   next-to-next-to-leading order, matching and missing structure dependent 
   contributions.\label{Tab:LD_SD_emu_av}}

The total uncertainty $\sigma_{\text{total}}$ of the summed long- and 
short-distance correction is given by
\bea
 \sigma^2_{\text{total}} 
& = & \sigma^2_{\text{numerical}} + \sigma^2_{\text{nnlo}}
     + \sigma^2_{\Lambda} + \sigma^2_{\text{SD}}\,.
\eea
The leading uncertainty originates from the matching of the short- and 
long-distance results. Here, the mismatch of both theories, the Standard Model 
for the short-distance corrections and the {\sc Qed}-enhanced effective theory 
for the long-distance corrections, at the scale $\Lambda$, as discussed in 
Sec.~\ref{Sec:matching}, is the main source. These matching uncertainties are 
estimated choosing $\Lambda$ of the scale of the final state meson's mass as a 
central value for $\Lambda$  and then taking the 
difference to the result when using the scale $2\Lambda$. Note that half the 
charged final state meson's mass is not a sensible choice for the matching 
scale since in this case a considerable fraction of the real radiation cross 
section would occur at scales greater than $\Lambda$ and being described 
by the phenomenological model, thus ending up in the wrong picture. 

Along these lines, also again referring to Sec.~\ref{Sec:matching}, the results 
for decays into pions should be considered with care: apart from the afore 
mentioned conceptual problems introduced by the large separation of both 
hadronic mass scales, large parts of the real 
emission phase space, described in the phenomenological model with a point-like 
pion, are actually able to resolve the pion, the charged and the neutral one.
Thus, the model used in this study is not entirely accurate for such decays. 

Further, the effect of additional real short-distance contributions is studied 
by comparing the pure inner-bremsstrahlung calculation with the real emission 
results of \cite{Cirigliano:2005ms} and \cite{Becirevic:2009fy}. This, however, 
can only be done for and $B \to D \, \ell \, \nu \, (\gamma)$ and 
$B^0 \to \pi^- \, \ell^+ \, \nu \, (\gamma)$ decays. Such terms are unknown for 
the other decay modes. Consequently, no error can be associated with them.
Finally, the electroweak next-to-next-to-leading order effects are estimated as 
$\sigma_{\text{nnlo}}=\alpha\left(\delta_{\text{sd}}+\delta_{\text{ld}}\right)$. 
It is clear that the 
complete model dependence from our phenomenological treatment of the 
long-distance contributions cannot be grasped with the above error estimation. 

\mytable{t!}{
 \begin{tabular}{r ll} 
   Measurement   & $\mathcal{G}(1) \, |V_{\text{cb}}| \times 10^{-3}$ with & $\mathcal{G}(1) \, |V_{\text{cb}}| \times 10^{-3}$ with  \\
     &  sd corrections only  & sd and ld corrections   \\ \hline
   \Babar tagged$\vphantom{\frac{|}{|}}$  \cite{Aubert:2009ac} & $42.30(2.36)$ & $42.21(2.34)$ \\ \hline
     &&\\
   Measurement   & $|V_{\text{ub}}| \times 10^{-3}$ with & $|V_{\text{ub}}| \times 10^{-3}$ with  \\
     & no corrections  & sd and ld corrections   \\ \hline 
   \Babar untagged$\vphantom{\frac{|}{|}}$  \cite{Aubert:2006px} & $ 3.60({}^{+0.62}_{-0.42})$ & $3.50({}^{+0.61}_{-0.42})$ \\ \hline
  \end{tabular}\vspace{3mm}}
  {The impact of long-distance corrections on the correction factor $\eta_\EW$ 
   is shown. Both measurements estimate real next-to-leading order corrections 
   via \PHOTOS and, thus, use phase-space cuts and variables that provide a 
   reduced sensitivity on the modeling of final state radiation. $\mathcal{G}(1)$ is 
   the normalisation of the heavy quark form factor. Note that $|V_{\text{ub}}|$ 
   aimed analyses usually do not apply short-distance corrections and that the 
   form factor prediction of \cite{Ball:2004ye} was used to extract 
   $|V_{\text{ub}}|$ from the measured $B^0\to\pi^-\,\ell^+\,\nu_\ell$ partial 
   branching fraction. 
   \label{Tab:TOTAL_impact_on_Vxb}}

Tab.~\ref{Tab:LD_SD_emu_av} presents the same results for 
$B\to D\,\ell\,\nu\,(\g)$ and $B^0\to\pi^-\,\ell^+\,\nu\,(\g)$ averaged over the 
different lepton species and the isospin rotated decays. The isospin averaged 
result is corrected for the difference in the production rate of $B^0$ and 
$B^+$ mesons, i.e.
\bea
 \delta_{\text{sd}} + \delta_{\text{ld}} 
& = & \bve{ \delta_{\text{sd}} + \delta^+_{\text{ld}} } \, f_{+-}\, 
      + \bve{ \delta_{\text{sd}} + \delta^0_{\text{ld}} } \, f_{00}\,,
\eea
with $\delta^+_{\text{ld}}$ and $\delta^0_{\text{ld}}$ the isospin breaking 
contributions for charged and uncharged charmed decays extracted from 
Tab.~\ref{Tab:SD_LD_emu_sep}. The latter correction factor is the adequate choice
to correct $|V_{\text{cb}}|$ from measurements demanding isospin. 

The impact of these correction factors on two selected measurements are listed 
in Tab.~\ref{Tab:TOTAL_impact_on_Vxb}. Note that both measurements used \PHOTOS 
to estimate the effect on radiative corrections. Applying the stated factors 
only corrects the overall normalisation, differences due to changes in 
kinematic distributions, see Sec.~\ref{Sec:res_diff}, result in another 
correction for the extracted values of $|V_{\text{cb}}|$ and $|V_{\text{ub}}|$ 
that cannot be estimated here. The correction for $|V_{\text{ub}}|$ shown here 
is therefore an illustration only and should be used with care.

 \subsection{Next-to-leading order corrections to differential rates}
\label{Sec:res_diff}

In this section the results of both \BLOR and \Sherpa/\Photons are presented 
and compared against \PHOTOS. The focus lies on the absolute value of the spatial 
momentum of the produced meson and the lepton, i.e.~$|\vec p_X|$ and 
$|\vec p_\ell|$. The chosen frame for these observables is the centre-of-momentum system of the 
electron and positron beam. Thus, the decaying $B^0$ and $B^+$ mesons already 
carry momentum corresponding to the $\Upsilon(4S)\to B^0\bar B^0$ and 
$\Upsilon(4S)\to B^+B^-$ decay kinematics. All quantities are shown as bare quantities, 
i.e.~no recombination of photon and lepton/meson momenta was used. This is 
directly applicable if the charged particle momenta are extracted by measuring 
their curved tracks in a magnetic field or in photon-free calorimeters, as is 
the case in many \Babar and \Belle analyses.

The prediction of each generator is normed to its inclusive decay width and 
the ratio plots show the relative difference 
\bea
 \Delta_O
& = &
 \frac{\frac{1}{\Gamma_{\text{tot},i}}\frac{\done\Gamma_i}{\done O}-
       \frac{1}{\Gamma_{\text{tot,ref}}}\frac{\done\Gamma_\text{ref}}{\done O}}
      {\frac{1}{\Gamma_{\text{tot},i}}\frac{\done\Gamma_i}{\done O}+
       \frac{1}{\Gamma_{\text{tot,ref}}}\frac{\done\Gamma_\text{ref}}{\done O}}
 \nnb
\eea
to the \PHOTOS prediction for the given observable $O$ in the given bin. Hence, 
the short-comings of the approximations inherent in the standard tool currently 
used by most experiments are plainly visible. Further, due to the choice of 
normalisation, systematic errors, shown to be dominant in the previous section, 
are negligible here. Hence, the error bands shown are statistical errors only 
and are of comparable magnitude for all three generators predictions.

\subsubsection{Decays $B \to D \, \ell \, \nu_\ell$}
\label{Sec:BDLN}


\myfigure{tbp}{
  \includegraphics[width=0.48\textwidth]{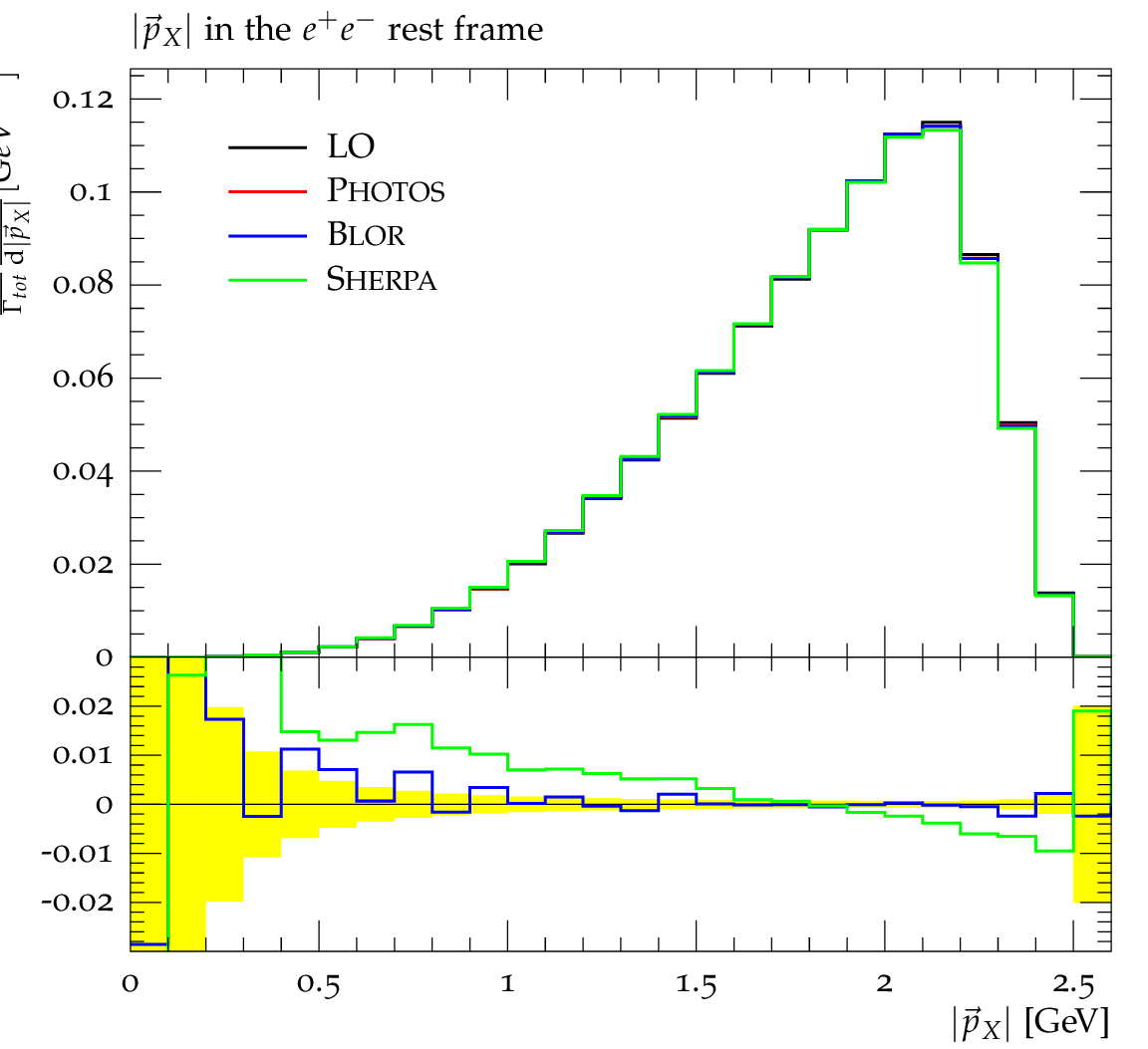} 
  \hspace*{0.02\textwidth}
  \includegraphics[width=0.48\textwidth]{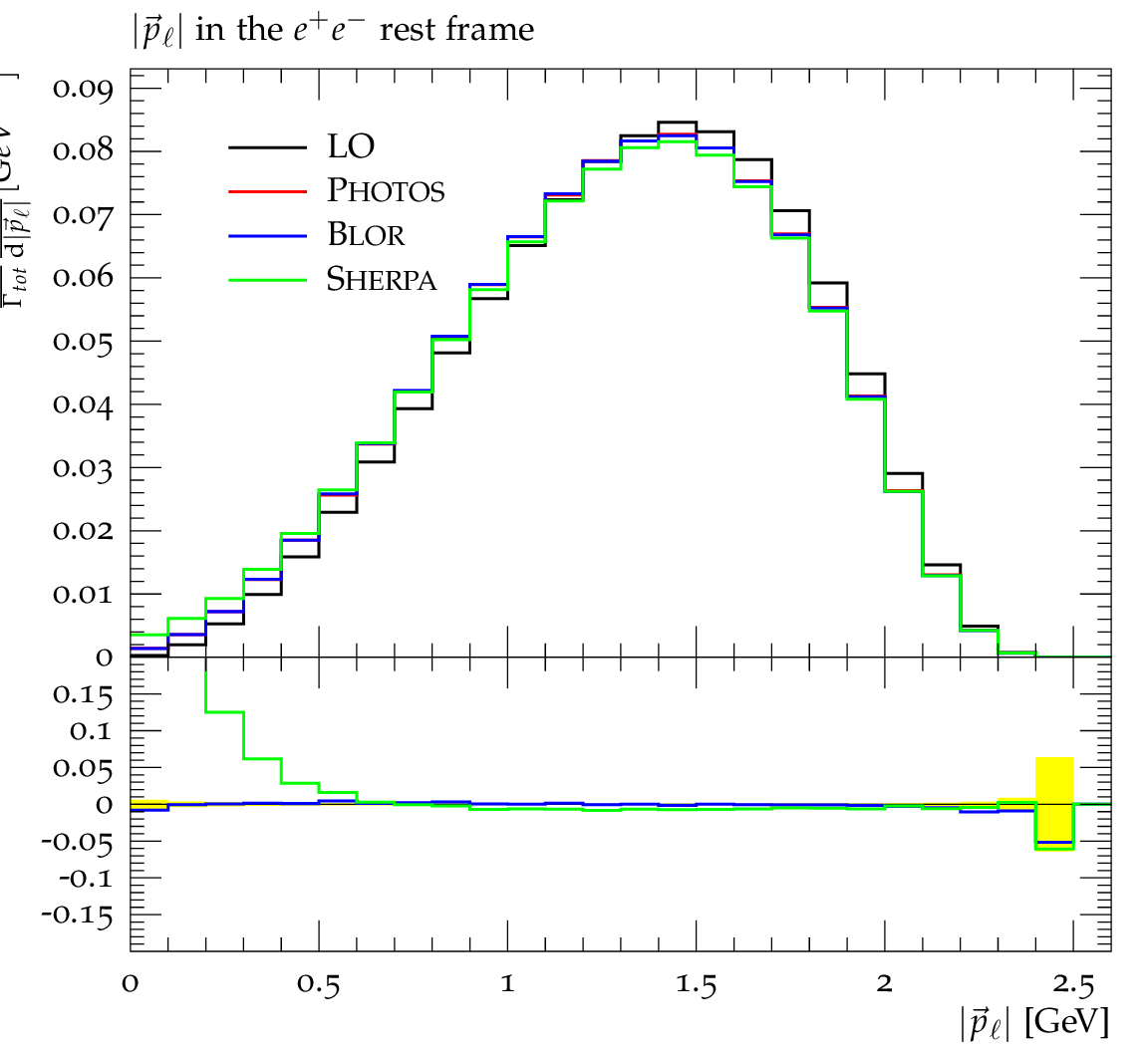}}
  {Lepton and Meson momentum spectrum in the $e^+e^-$ rest frame in the decay 
   $B^0 \to D^- \, e^+ \, \nu_e$. All spectra are normed to the total inclusive 
   decay width predicted by the respective generator. The ratio plot gives the 
   relative deviation, bin by bin, of the predicted shapes with \protect\PHOTOS 
   as reference. The shaded yellow area gives the statistical uncertainty of the 
   reference distribution.\label{Fig:B0_D-_e_nu}}

\myfigure{tbp}{
  \includegraphics[width=0.48\textwidth]{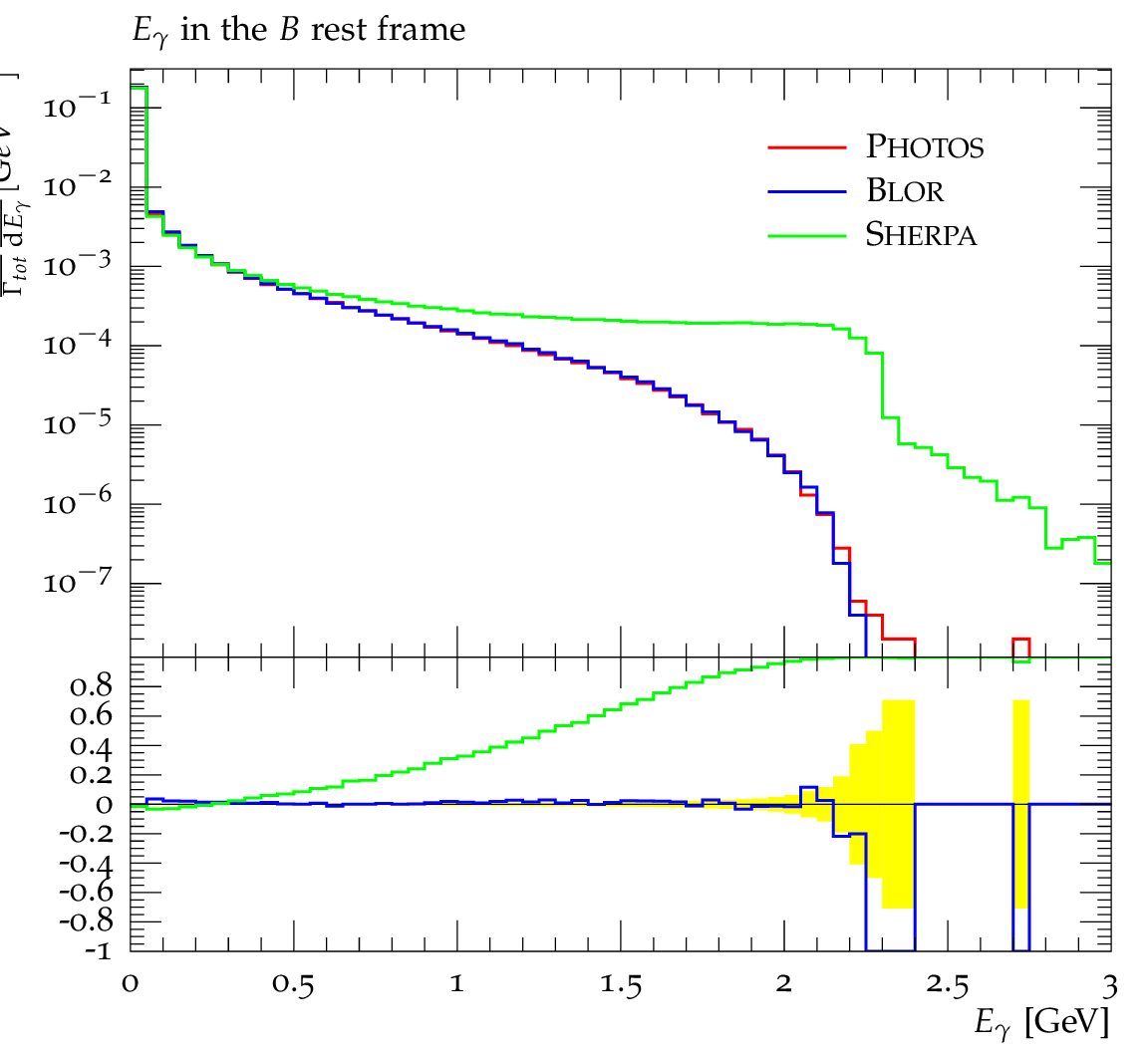} 
  \hspace*{0.02\textwidth}
  \includegraphics[width=0.48\textwidth]{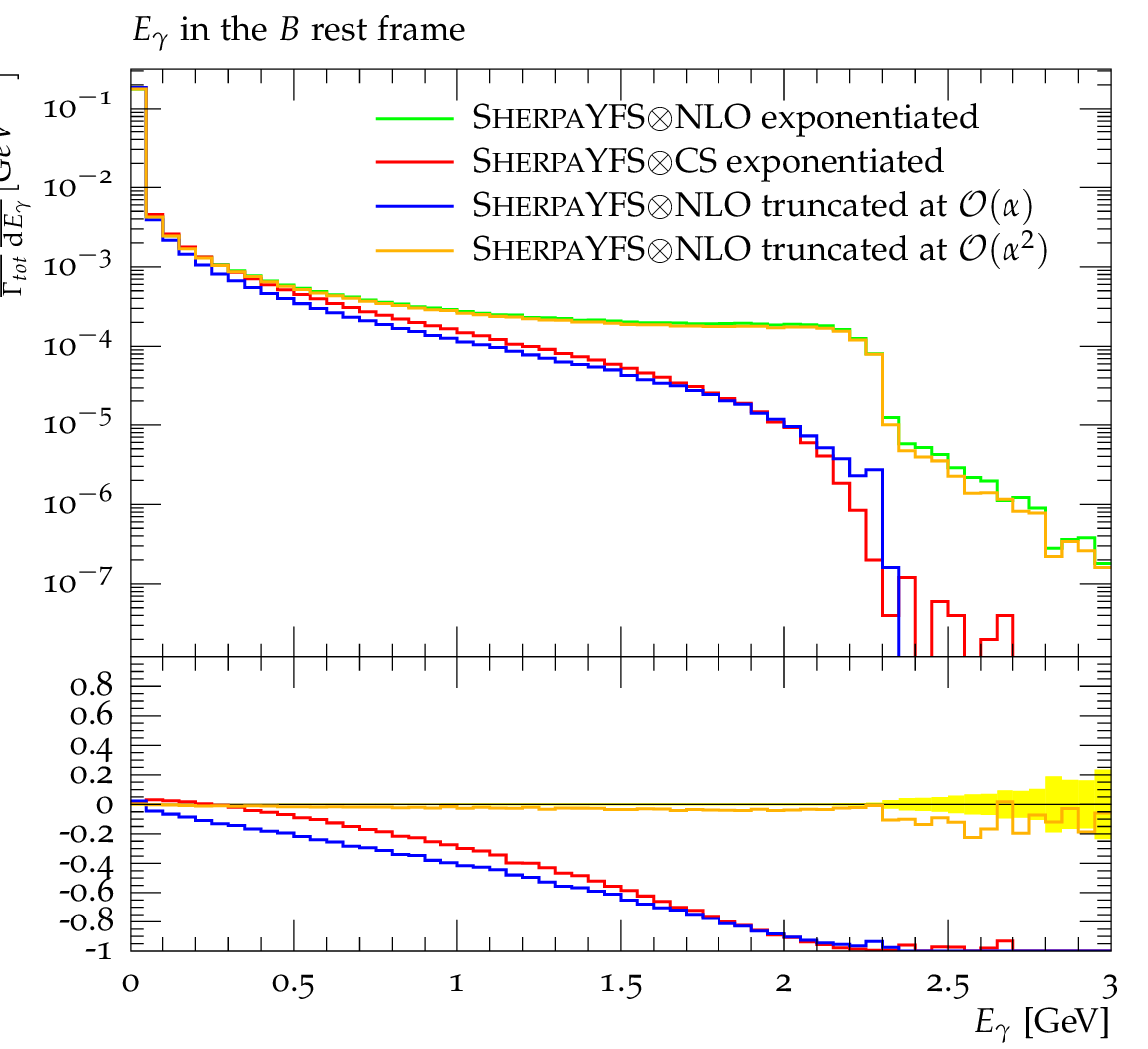}
  }
  {Total radiative energy loss, i.e.~the sum of all photons radiated, in the 
   decay $B^0 \to D^- \, e^+ \, \nu_e$ in the $B$ rest frame. All spectra are 
   normed to the total inclusive decay width predicted by the respective 
   generator. The left panel shows the predictions of all three generators and 
   the \protect\PHOTOS prediction is taken as the reference in the ratio plot.
   The right panel shows the predictions of \protect\Sherpa/\protect\Photons in 
   its full YFS$\otimes${\sc Nlo} exponentiated mode (green), a mode where the 
   exact {\sc Nlo} matrix element of the perturbative expansion is replaced by 
   universal Catani-Seymour dipole splitting kernels (red) and two modes where 
   the exact real emission matrix elements are used, but the expansion in the 
   resolved emission region is truncated at $\order(\alpha)$ (blue) and 
   $\order(\alpha^2)$ (orange), thus allowing at most one and two photons, 
   respectively. Here, the full exponentiated \protect\Sherpa/\protect\Photons 
   prediction is taken as reference in the ratio plot.
   \label{Fig:B0_D-_e_nu_different_photons}}

In Fig.~\ref{Fig:B0_D-_e_nu} the predictions of all three generators for the 
decay $B^0 \to D^- \, e^+ \, \nu_e$ using a form factor parametrisation from 
Heavy Quark Effective Theory, cf.~App.~\ref{App:hqetformfactorsintroduction}, 
are presented. For most of the phase space the agreement of the next-to-leading 
order shape is good. However, as can be seen in the absolute lepton momentum 
plot, \Sherpa/\Photons predicts a slightly different shape close to 
$|\vec p_\ell|,|\vec p_X| = 0$. This limit cannot be measured directly in most 
experiments. Nonetheless, it influences the extrapolation to the full phase 
space, and thus determinations of total decay widths. In \cite{Aubert:2008yv}, 
for example, uncertainties related to such an extrapolation constituted up to 
50\% of the total experimental error. Further, the slight slope of the order 
of 1\% exhibited in the prediction of \Sherpa/\Photons for $|\vec p_X|$ is also 
present in the $|\vec p_\ell|$ spectrum. Both differences originate in the 
different modeling of radiative energy loss between the different programs, 
as shown in the following.

Fig.~\ref{Fig:B0_D-_e_nu_different_photons} displays the radiative energy loss, 
i.e.~the sum of the energies of all photons radiated, in the rest frame of the 
decaying $B^0$. The predictions of the three different generators are shown in 
the left panel.
For single photon emission in such a final state dipole there exists a 
kinematic limit on the photon's energy: $E_\gamma^\text{max}=\frac{m_B}{2}\left(
\frac{m_B}{m_X+m_\ell}-\frac{m_X+m_\ell}{m_B}\right)$ in the dipole's rest 
frame or $E_\gamma^\text{max}=2.3083\GeV$ in the $B$ rest frame. This limit is clearly 
visible in Fig.~\ref{Fig:B0_D-_e_nu_different_photons}. All events exceeding 
it, i.e.~radiating more than $E_\gamma^\text{max}$, must exhibit multi-photon radiation. 
Here (at least) two hard photons recoil against each other. Hence, this feature 
is present in both the \Sherpa/\Photons and \PHOTOS predictions, but not in the 
fixed order {\sc Nlo} prediction of \BLOR. This tail is an $\order(\alpha^2)$ 
effect in the hard radiation and can only be described approximatively here. In 
\PHOTOS it is described by an iteration of the emission kernels while 
\Sherpa/\Photons describes this part of the spectrum by the next-to-leading 
order hard emission amplitude $\tilde\beta_1^1$ summed over all projections 
onto the single emission subspaces, cf.~Sec.~\ref{Sec:theory_resummation} and 
\ref{Sec:methods_SHERPA_PHOTONS}. 

On the other hand, multi-photon radiation also enhances the amount of 
radiation in the region $E_\gamma<E_\gamma^\text{max}$, if the probability of two 
relatively hard photons is sufficiently large. As exemplified in 
Fig.~\ref{Fig:B0_D-_l_nu_NGamma}, in $B^0\to D^-\,e^+\,\nu_e$ double photon 
emission is relatively probable and, hence, leads to such an enhancement, 
whereas due to the much larger muon mass this feature is nearly absent in the 
decay to muons, Fig.~\ref{Fig:B0_D-_mu_nu_different_photons}. Of course, this 
enhancement of the radiative energy loss is an effect of $\order(\alpha^2)$ 
and can therefore only be described approximatively here.

Of equal importance as multi-photon radiation is the presence of the exact real 
emission matrix element, as is also shown in the right panel of 
Figs.~\ref{Fig:B0_D-_e_nu_different_photons} and 
\ref{Fig:B0_D-_mu_nu_different_photons}. Approximating the real emission 
matrix elements with Catani-Seymour splitting functions, reproducing the 
Altarelli-Parisi splitting functions used in \PHOTOS in the (quasi-) collinear 
limit, leads to a mis-estimation of the radiative energy loss in the regime 
close to the kinematic boundary. It seems, however, that in the present cases 
collinearly approximated multi-photon emission mimics the exact fixed-order {\sc Nlo} 
behaviour reasonably well.

Close to the kinematic boundary on single photon emission, $E_\text{max}$, the 
vertex emission diagrams become important, as do the corrections for 
$t\neq t'$ (cf. App. \ref{App:NLO_ME}). These corrections have different sizes for the electron and muon 
channels due to their different masses and radiative properties. In principle, 
here also the structure-dependent corrections of 
Sec.~\ref{Sec:theory_SD_terms} play a role. But, as is investigated in 
Sec.~\ref{Sec:result_SD_terms} they have negligible impact on the shape of the 
differential distributions. Thus, they can be safely neglected here.

Further, both \Sherpa/\Photons and \PHOTOS share a common soft limit, showing 
the compatibility of the inherent soft resummation of \Sherpa/\Photons and the 
superimposed soft limit correction in \PHOTOS. \BLOR exhibits an (almost) 
constant off-set of a few percent owing to the lack of resummed contributions 
modeled in \Sherpa/\Photons by virtue of the YFS form factor.

\myfigure{tbp}{
  \includegraphics[width=0.48\textwidth]{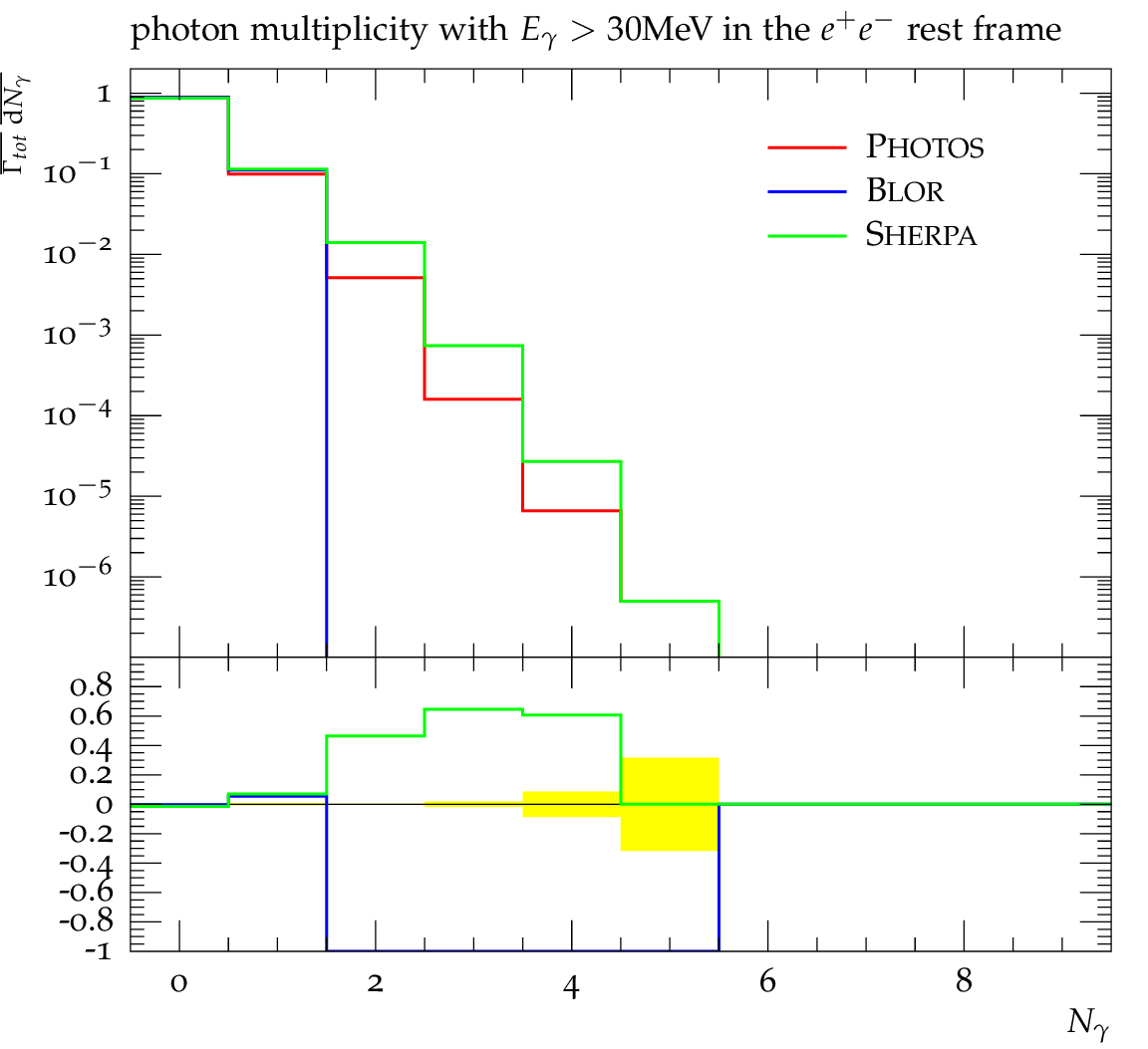} 
  \hspace*{0.02\textwidth}
  \includegraphics[width=0.48\textwidth]{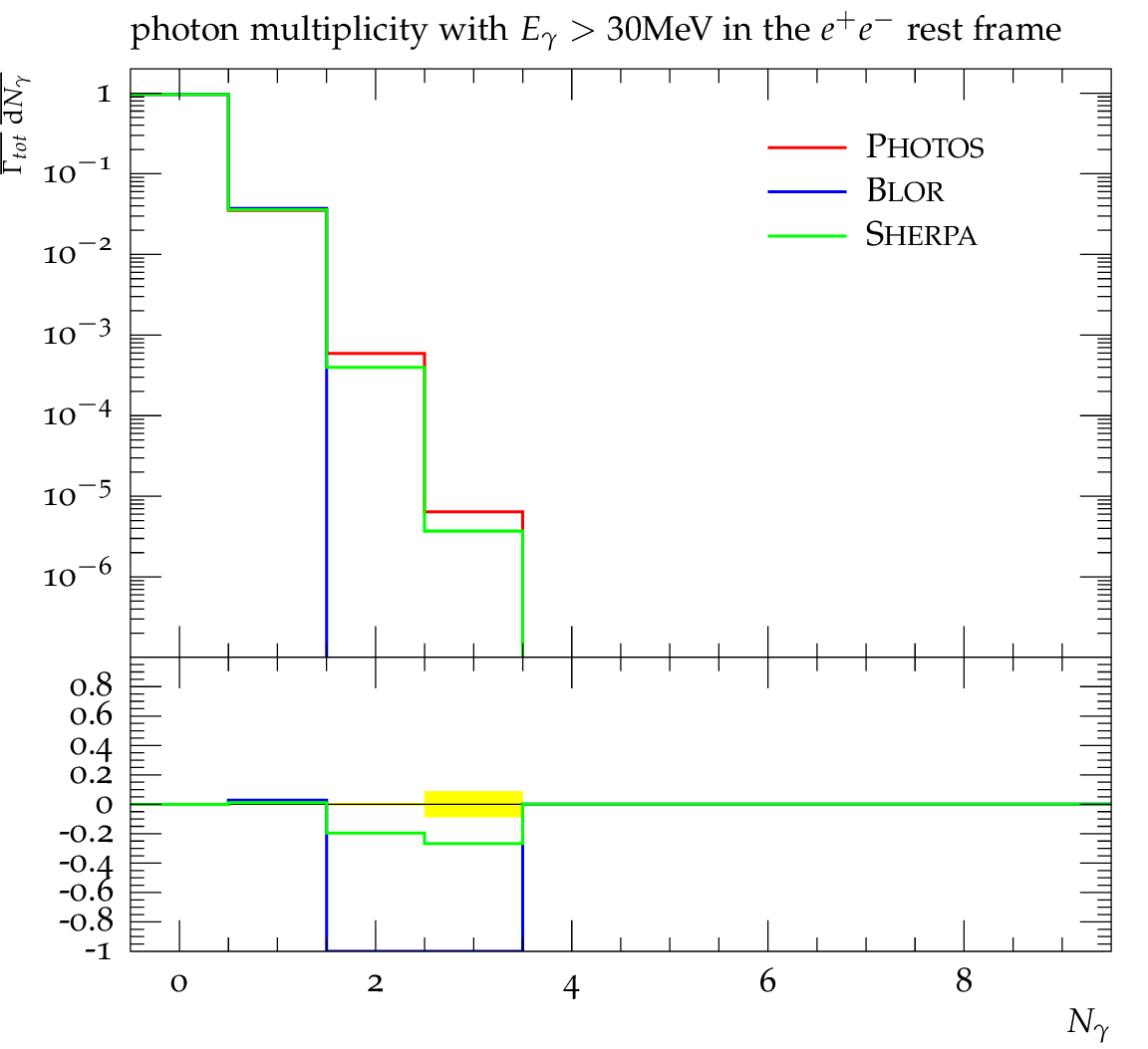}}
  {Multiplicities of photons with at least 30MeV in in the $e^+e^-$ rest frame 
   $B^0\to D^-\,e^+\,\nu_e$ on the right hand side and 
   $B^0\to D^-\,\mu^+\,\nu_\mu$ on the left hand side. In the ratio plot 
   \protect\PHOTOS was chosen as the reference. \label{Fig:B0_D-_l_nu_NGamma}}

\myfigure{tbp}{
  \includegraphics[width=0.48\textwidth]{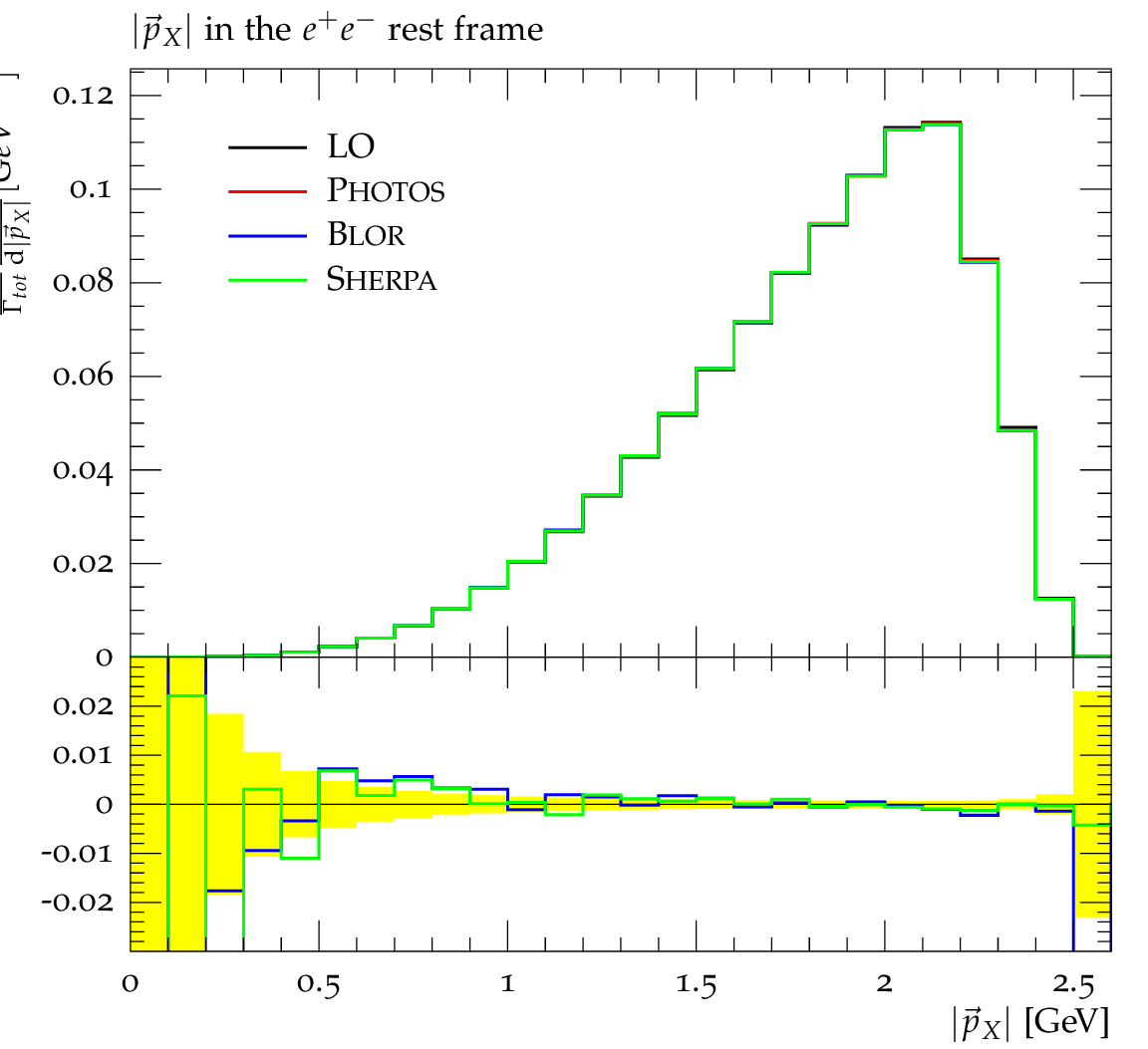} 
  \hspace*{0.02\textwidth}
  \includegraphics[width=0.48\textwidth]{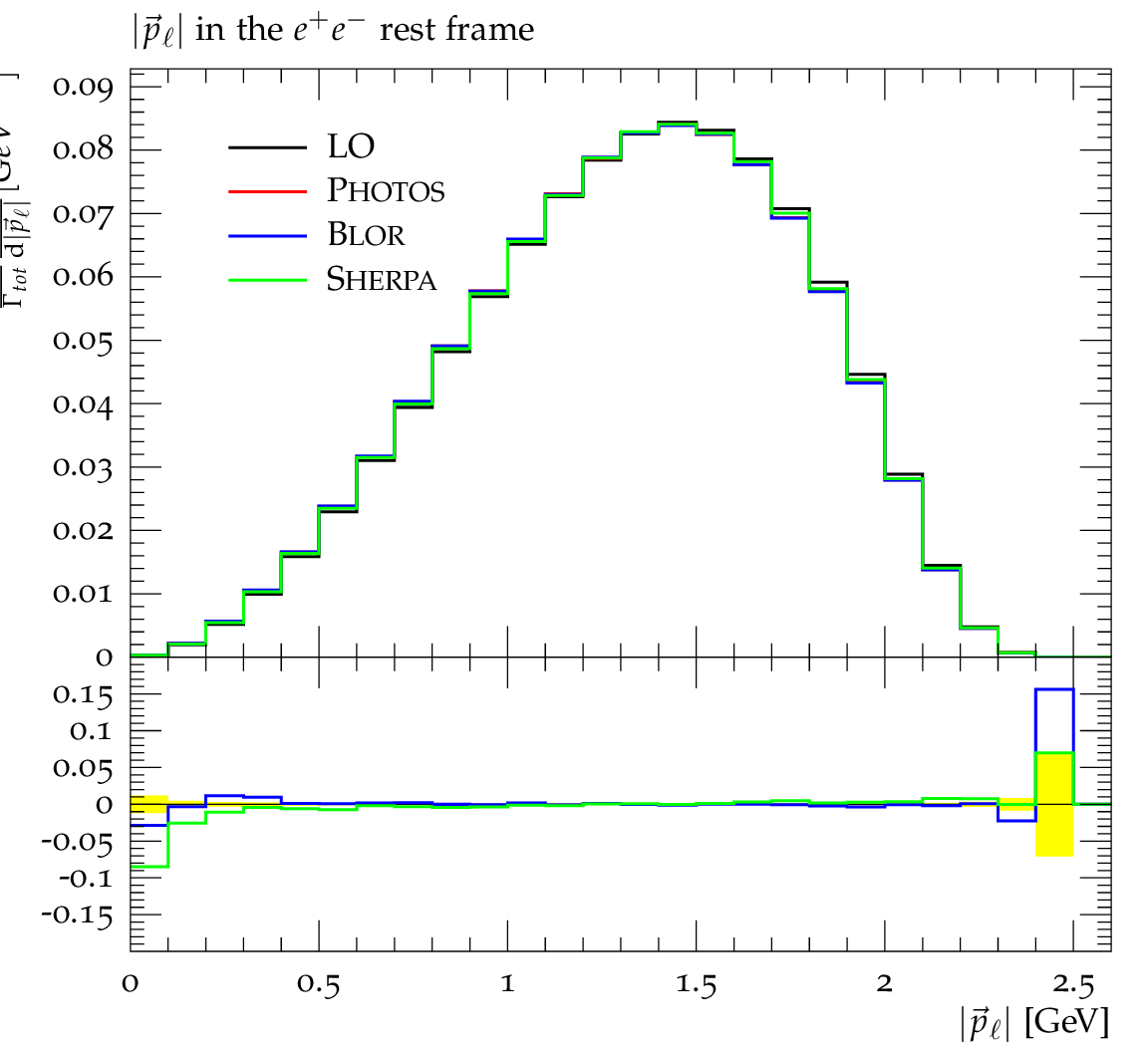}}
  {Lepton and Meson momentum spectrum in the $e^+e^-$ rest frame in the decay 
   $B^0 \to D^- \, \mu^+ \, \nu_\mu$. The \protect\PHOTOS prediction is taken 
   as the reference in the ratio plot.\label{Fig:B0_D-_mu_nu}}

\myfigure{tbp}{
  \includegraphics[width=0.48\textwidth]{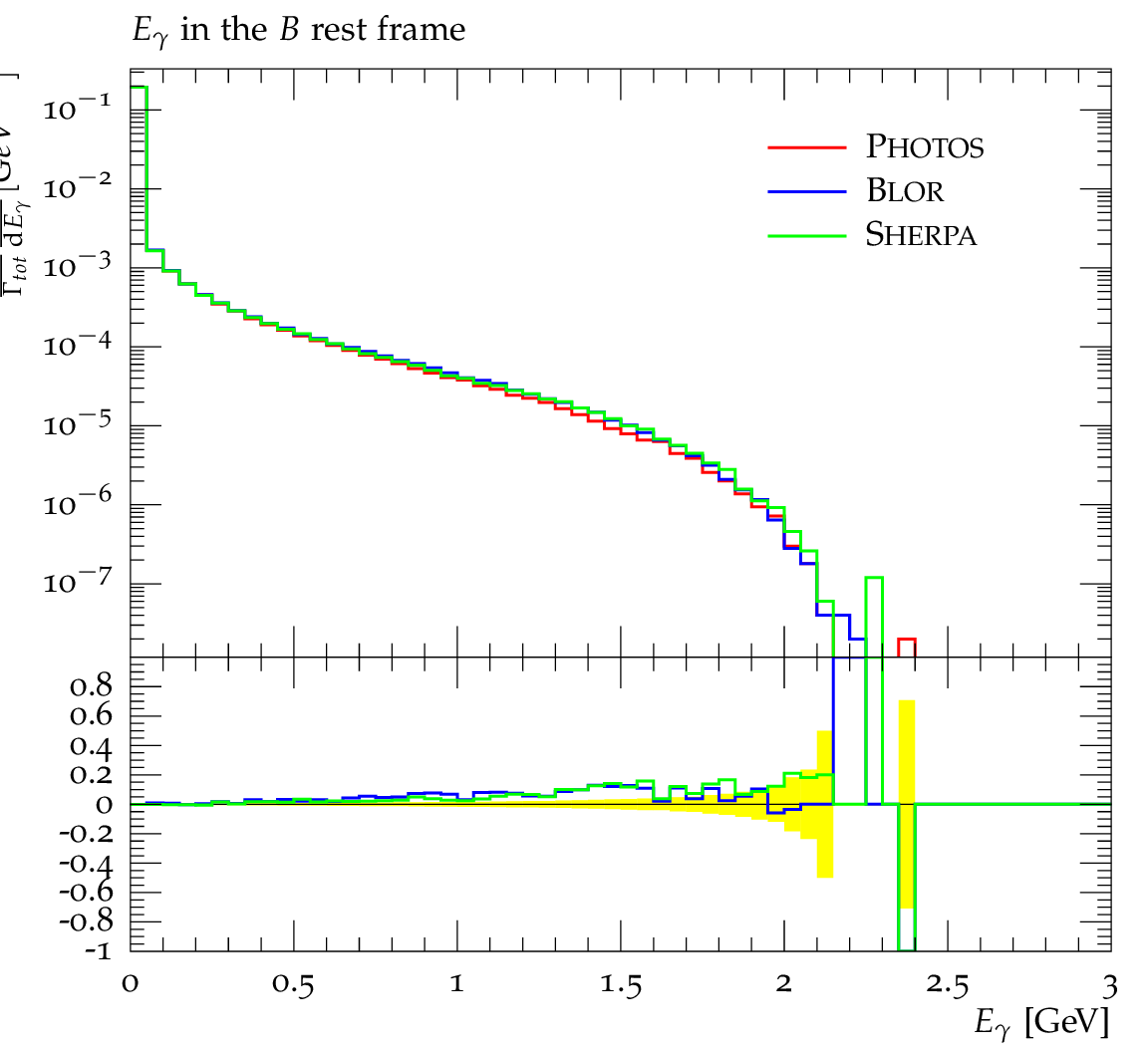} 
  \hspace*{0.02\textwidth}
  \includegraphics[width=0.48\textwidth]{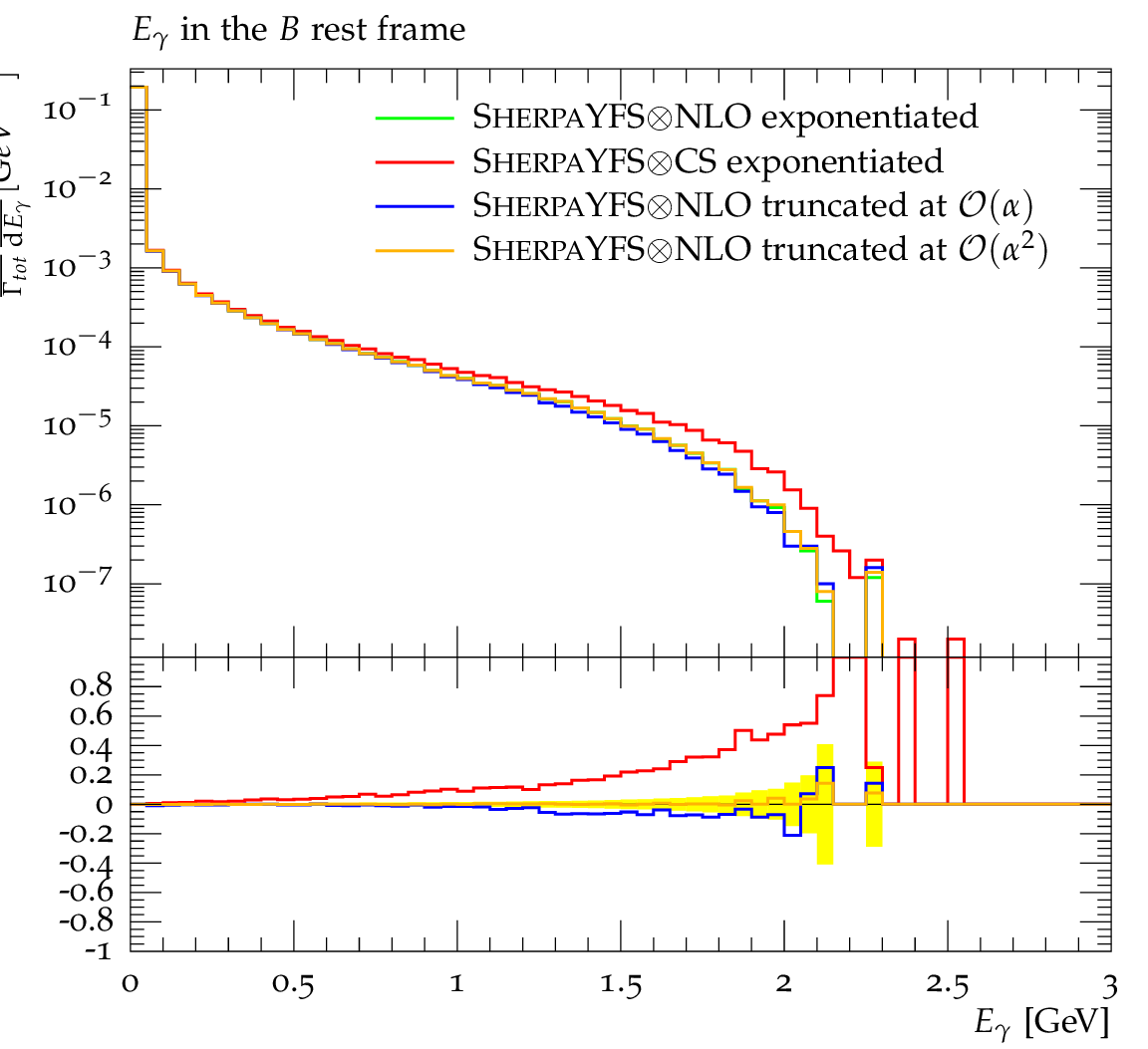}
  }
  {Total radiative energy loss, i.e.~the sum of all photons radiated, in the 
   decay $B^0 \to D^- \, \mu^+ \, \nu_\mu$ in the $B$ rest frame. The labels are 
   identical to those in Fig.~\ref{Fig:B0_D-_e_nu_different_photons}.
   \label{Fig:B0_D-_mu_nu_different_photons}}

\myfigure{tbp}{
  \includegraphics[width=0.48\textwidth]{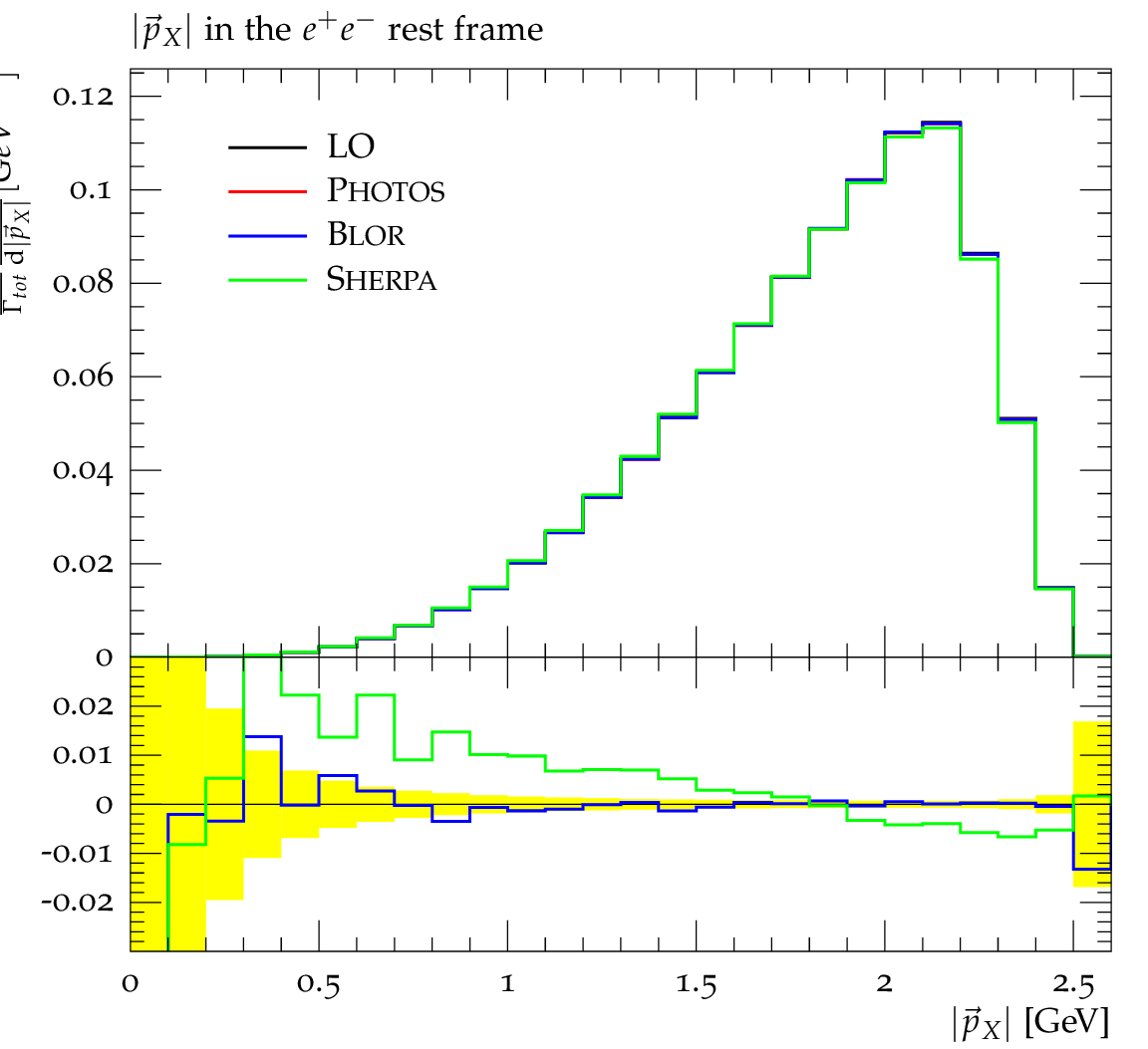} 
  \hspace*{0.02\textwidth}
  \includegraphics[width=0.48\textwidth]{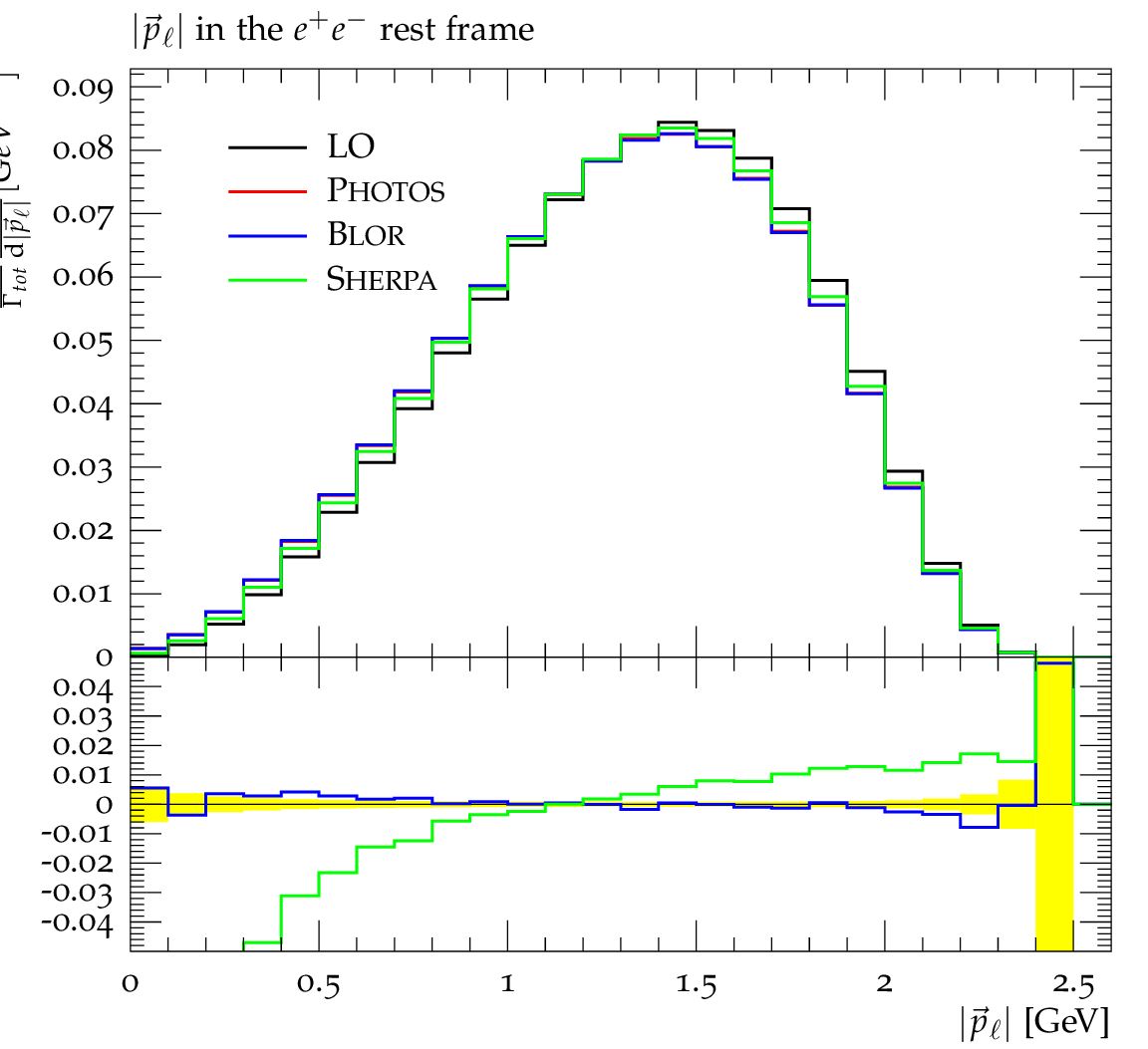}}
  {Lepton and Meson momentum spectrum in the $e^+e^-$ rest frame in the decay 
   $B^+ \to \bar D^0 \, e^+ \, \nu_e$. The \protect\PHOTOS prediction is taken 
   as the reference in the ratio plot.\label{Fig:B+_D0_e_nu}}

\myfigure{tbp}{
  \includegraphics[width=0.48\textwidth]{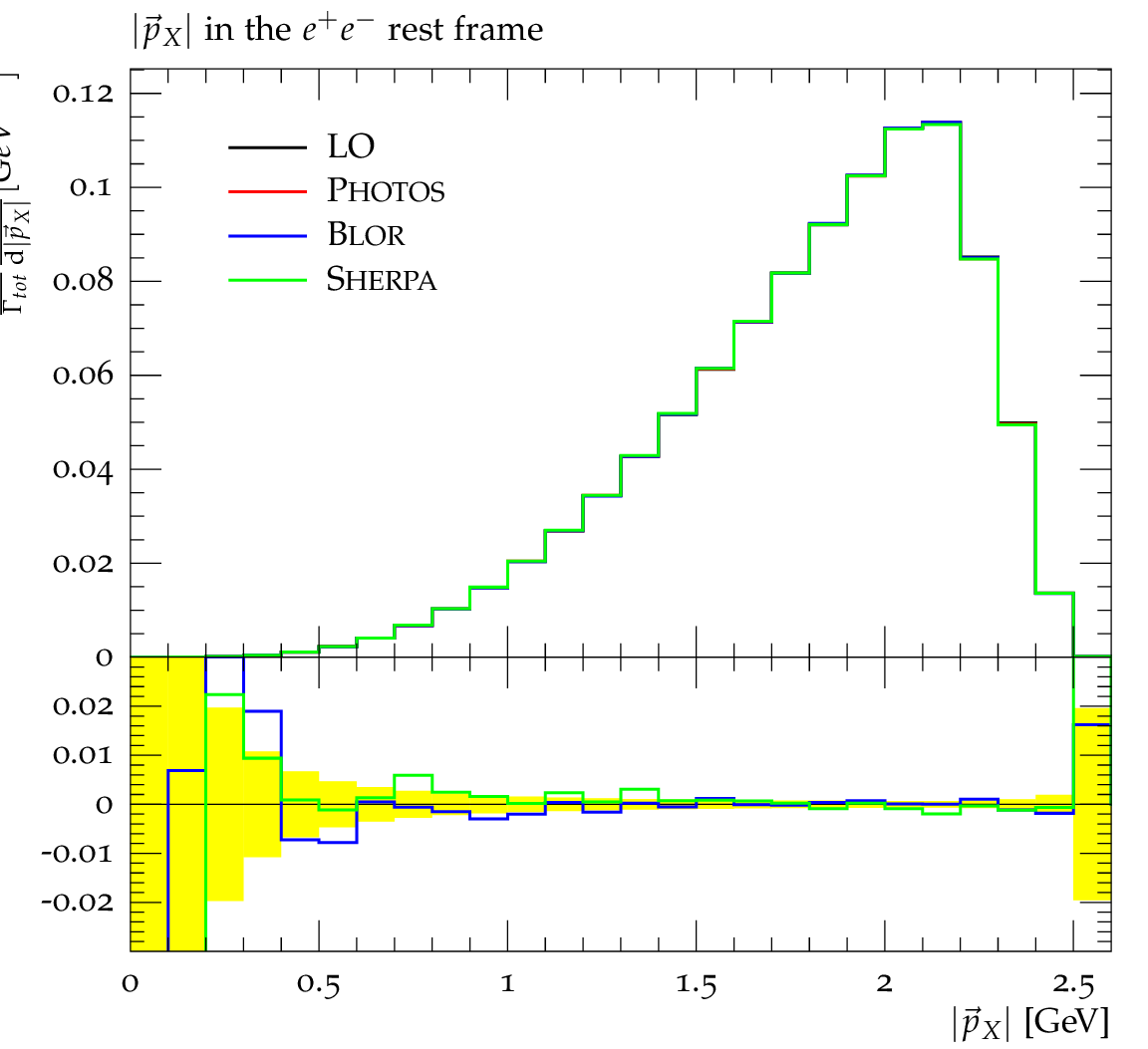} 
  \hspace*{0.02\textwidth}
  \includegraphics[width=0.48\textwidth]{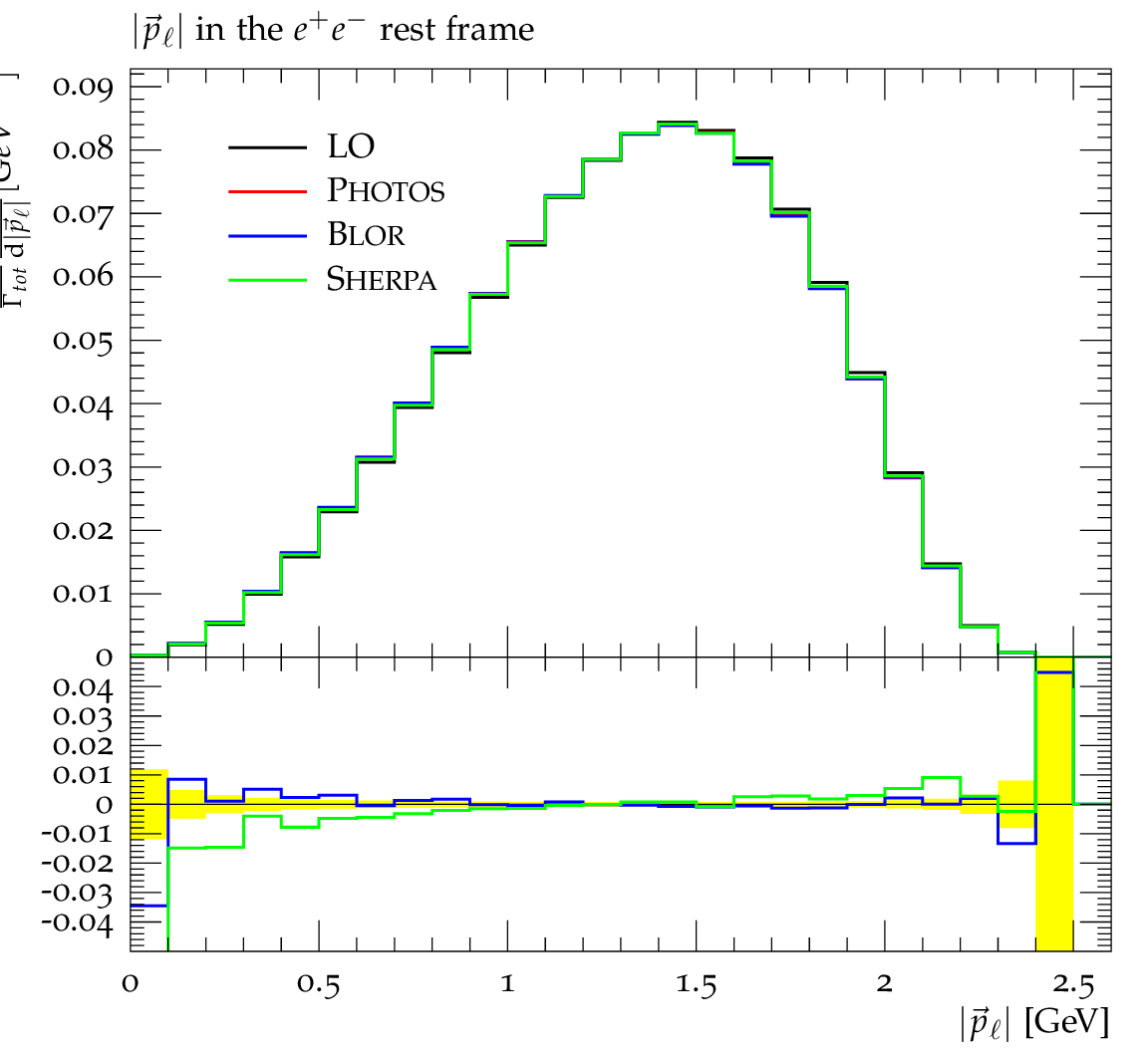}}
  {Lepton and Meson momentum spectrum in the $e^+e^-$ rest frame in the decay 
   $B^+ \to \bar D^0 \, \mu^+ \, \nu_\mu$. The \protect\PHOTOS prediction is 
   taken as the reference in the ratio plot.\label{Fig:B+_D0_mu_nu}}

The lepton and meson spectra for $B^0\to D^-\,\mu^+\,\nu_\mu$ are shown in 
Fig.~\ref{Fig:B0_D-_mu_nu}. Figs.~\ref{Fig:B+_D0_e_nu} and 
\ref{Fig:B+_D0_mu_nu} show strong isospin rotated processes 
$B^+\to\bar D^0\,e^+\,\nu_e$ and $B^+\to\bar D^0\,\mu^+\,\nu_\mu$, 
respectively. Thus, the radiating dipole is spanned between the 
initial state $B^+$ and the lepton. Radiation off the initial state meson is 
suppressed by its much larger mass, as compared to the $D^-$. Thus, 
multi-photon emission is also strongly suppressed. In these cases 
\Sherpa/\Photons predict slightly smaller radiative corrections in the 
electron decay channel than either \PHOTOS or \BLOR. The differences are of the 
order of five percent; note that the scale was enlarged in the reference plot to better 
highlith the differences. These, to the largest extent, root in 
differences in the modeling of emission off the initial state charged meson.

\subsubsection{Decays $B \to D_0^* \, \ell \, \nu_\ell$}
\label{Sec:BDstarLN}


\myfigure{tbp}{
  \includegraphics[width=0.48\textwidth]{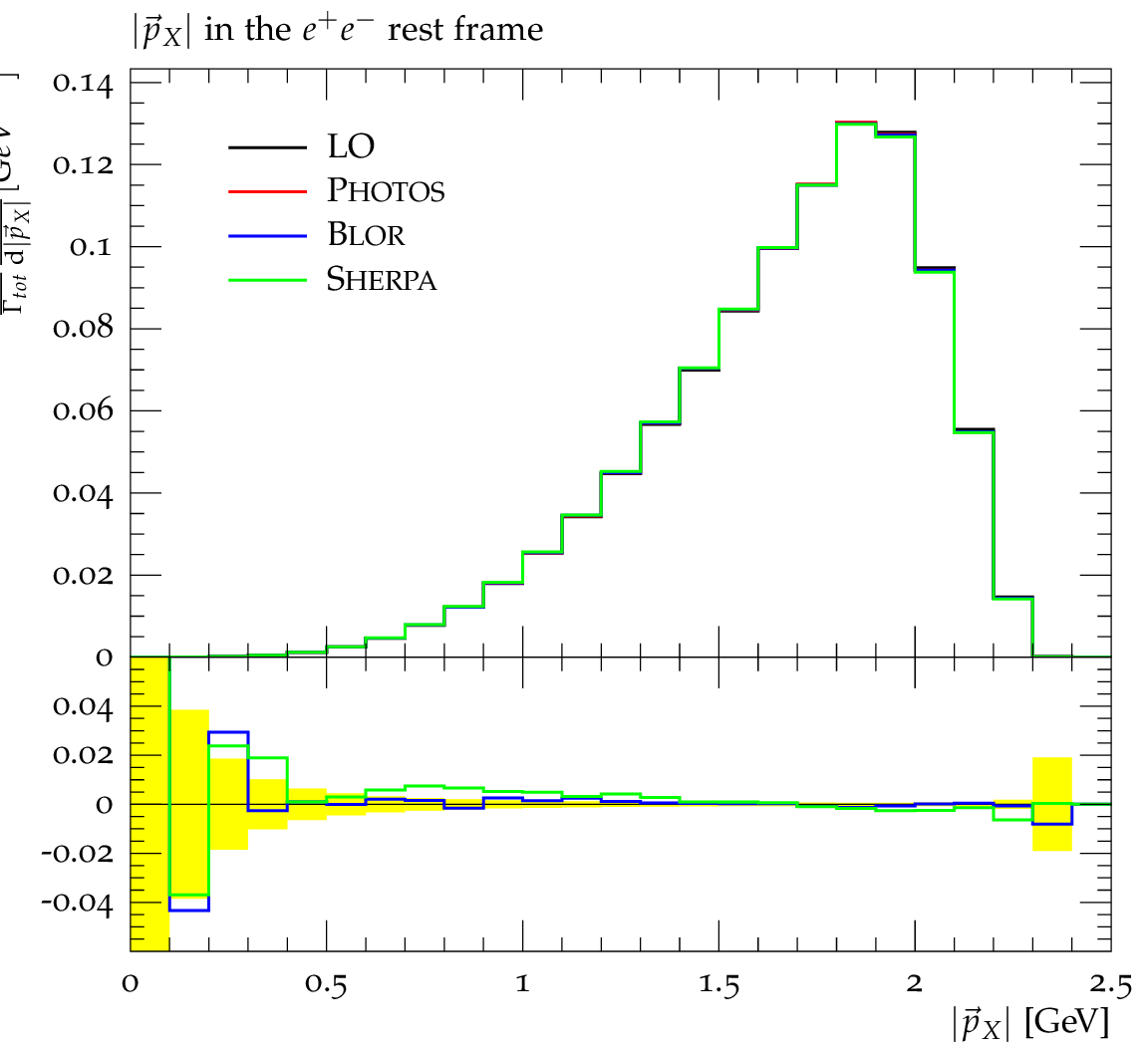} 
  \hspace*{0.02\textwidth}
  \includegraphics[width=0.48\textwidth]{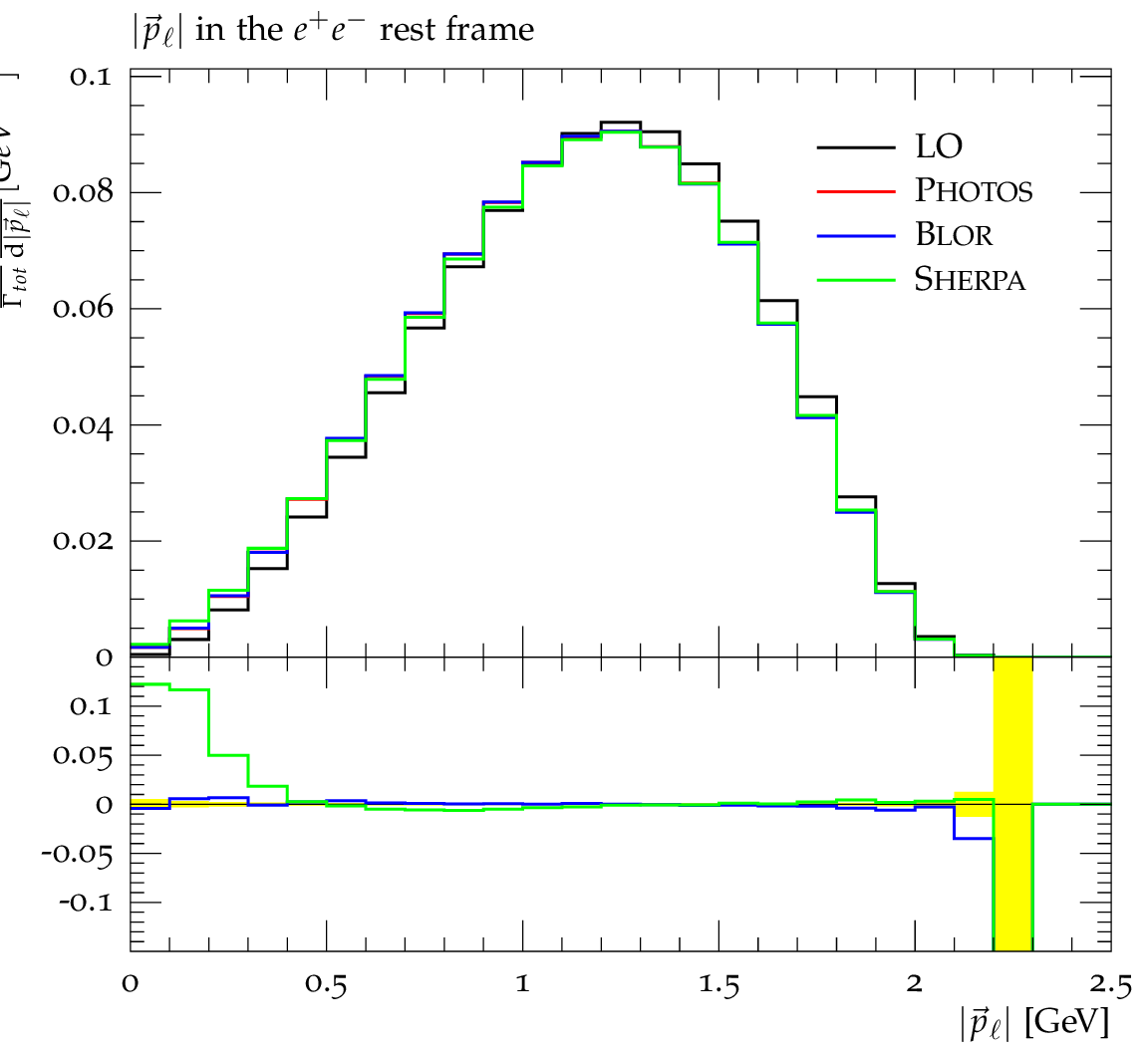}}
  {Lepton and Meson momentum spectrum in the $e^+e^-$ rest frame in the decay 
   $B^0 \to  D_0^{*\,-}\, e^+ \, \nu_e$. The \protect\PHOTOS prediction is 
   taken as the reference in the ratio plot.\label{Fig:B0_D0star-_e_nu}}

\myfigure{tbp}{
  \includegraphics[width=0.48\textwidth]{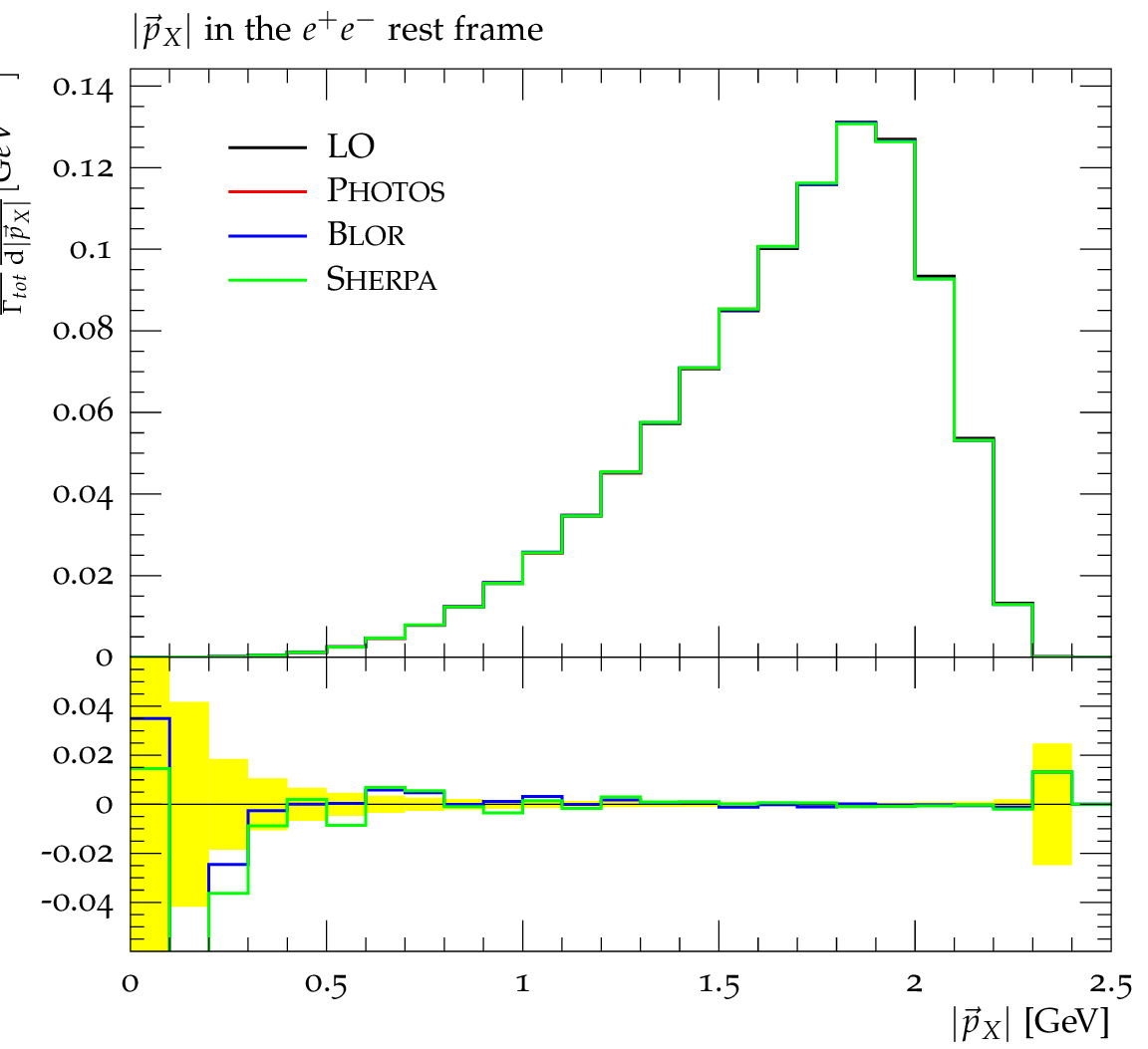} 
  \hspace*{0.02\textwidth}
  \includegraphics[width=0.48\textwidth]{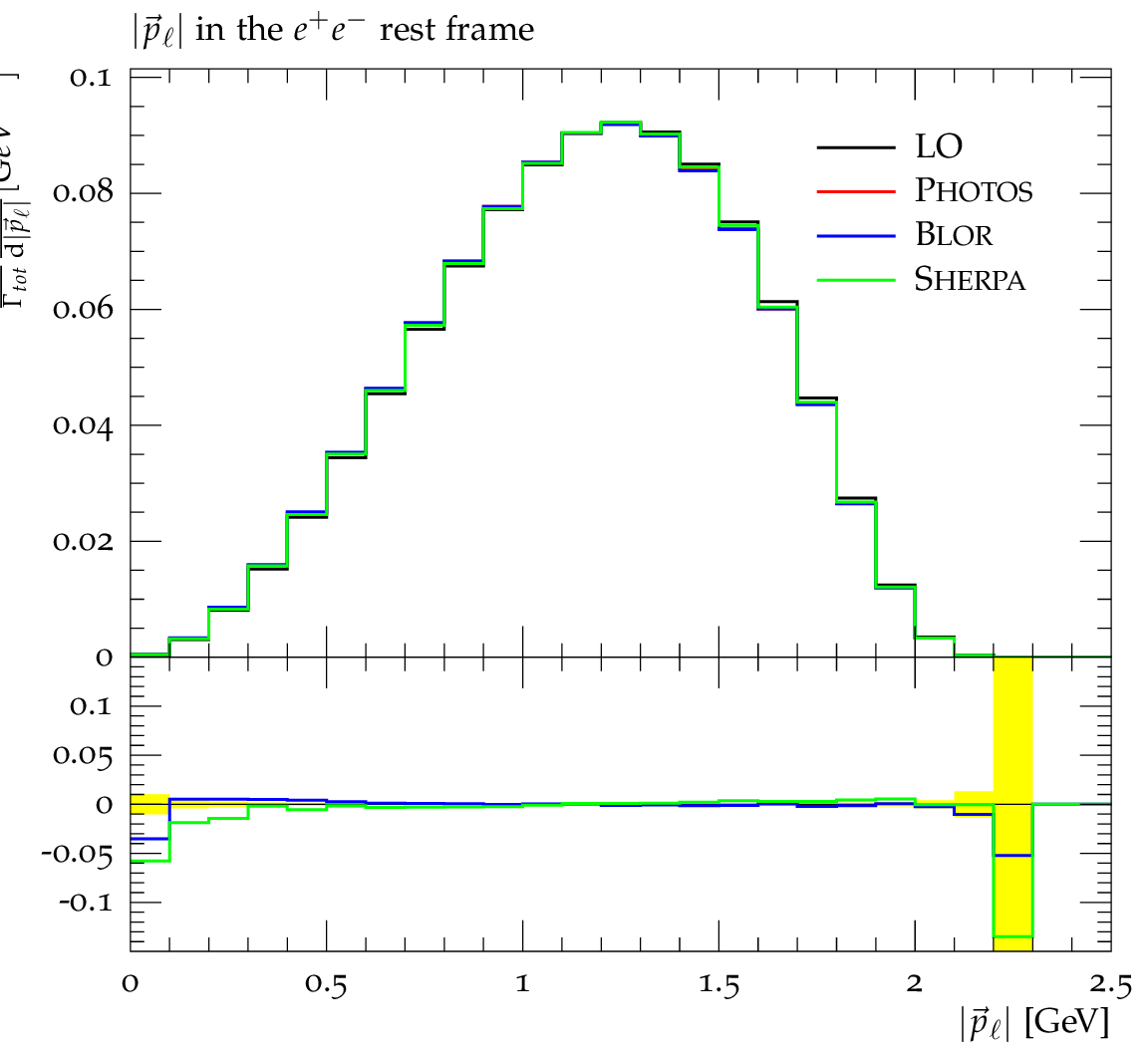}}
  {Lepton and Meson momentum spectrum in the $e^+e^-$ rest frame in the decay 
   $B^0 \to D_0^{*\,-} \, \mu^+ \, \nu_\mu$. The \protect\PHOTOS prediction is 
   taken as the reference in the ratio plot.\label{Fig:B0_D0star-_mu_nu}}

\myfigure{tbp}{
  \includegraphics[width=0.48\textwidth]{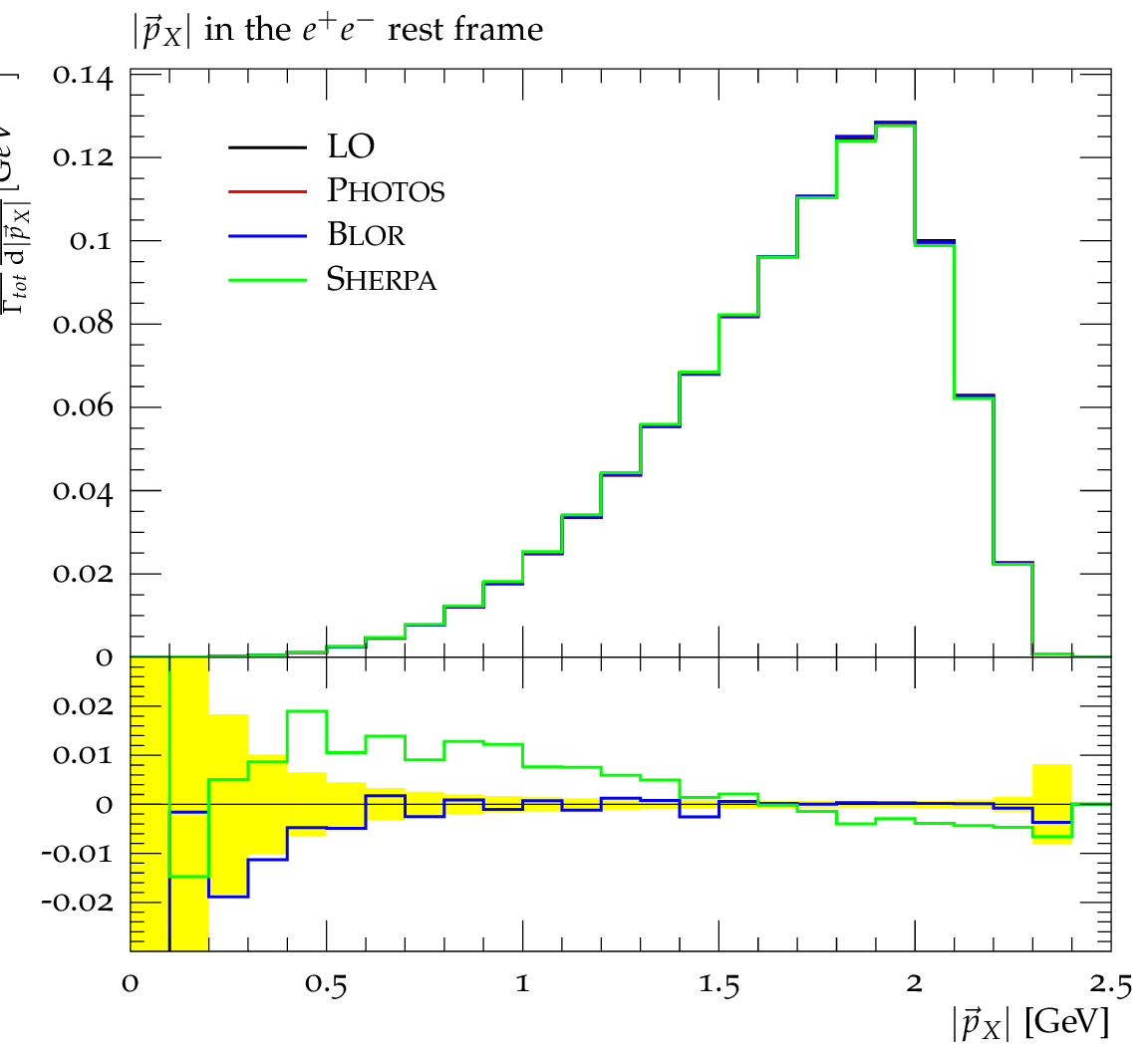} 
  \hspace*{0.02\textwidth}
  \includegraphics[width=0.48\textwidth]{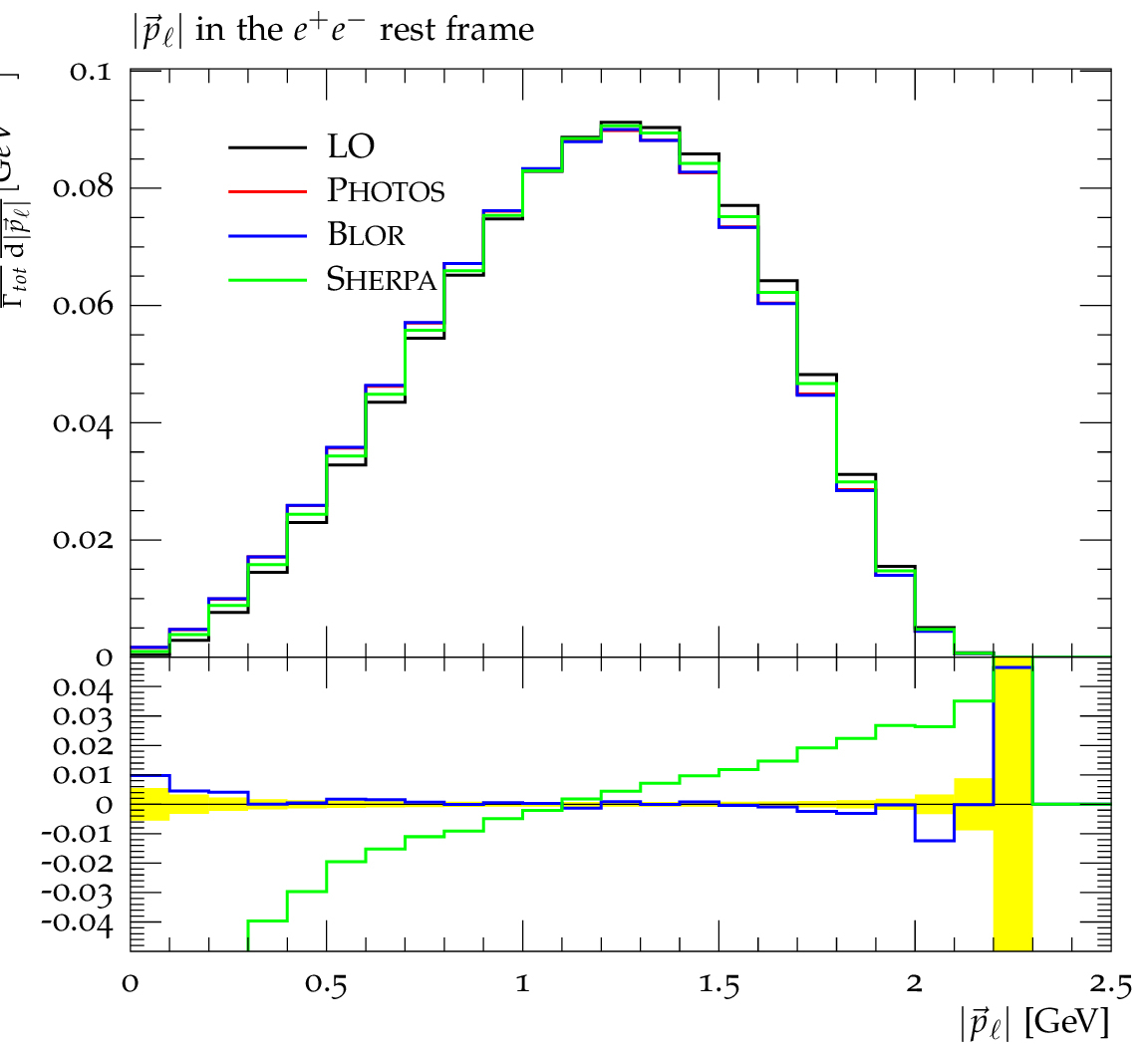}}
  {Lepton and Meson momentum spectrum in the $e^+e^-$ rest frame in the decay 
   $B^+ \to \bar D_0^{* \, 0} \, e^+ \, \nu_e$. The \protect\PHOTOS prediction 
   is taken as the reference in the ratio plot.\label{Fig:B+_D0star0_e_nu}}

\myfigure{tbp}{
  \includegraphics[width=0.48\textwidth]{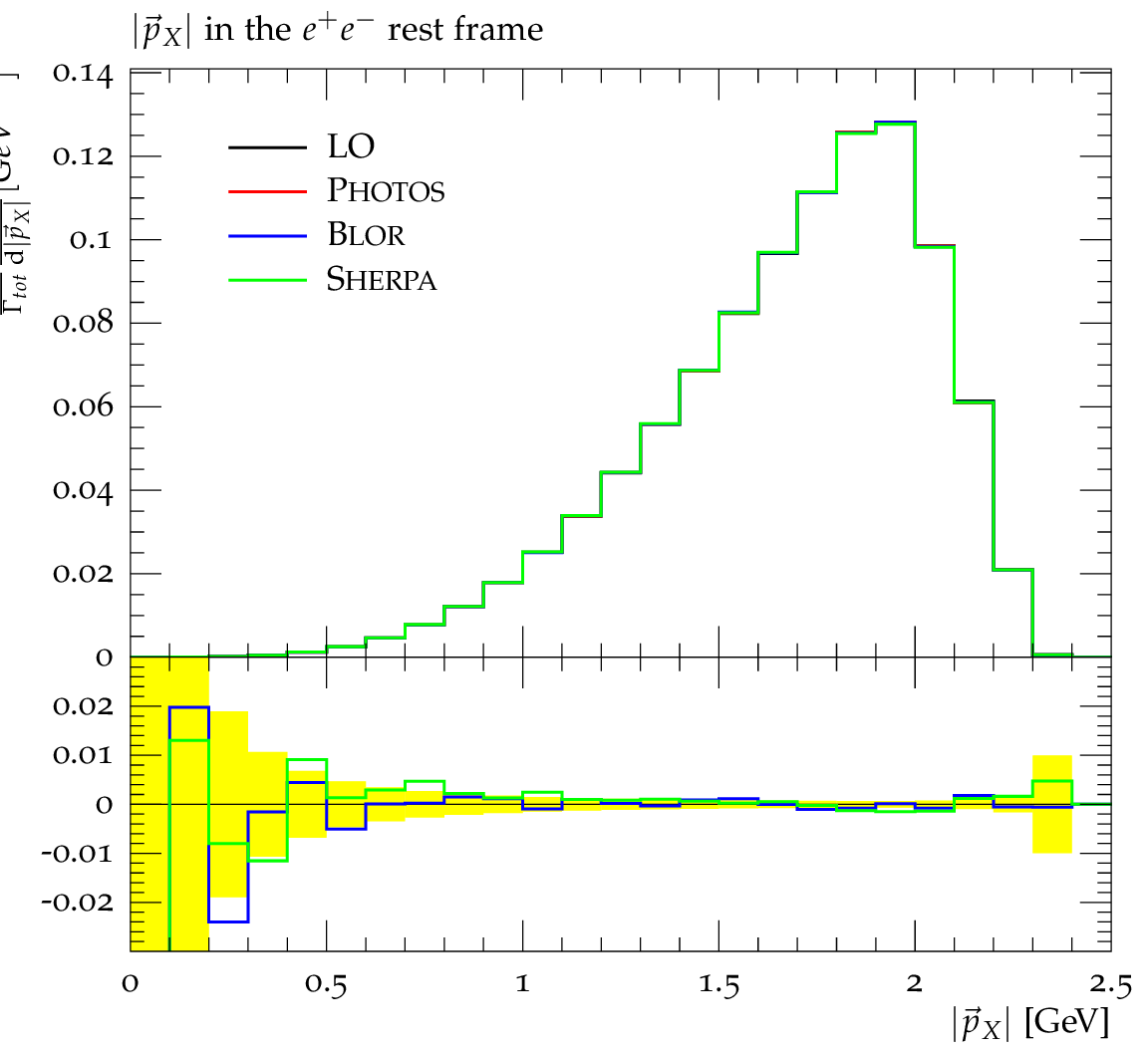} 
  \hspace*{0.02\textwidth}
  \includegraphics[width=0.48\textwidth]{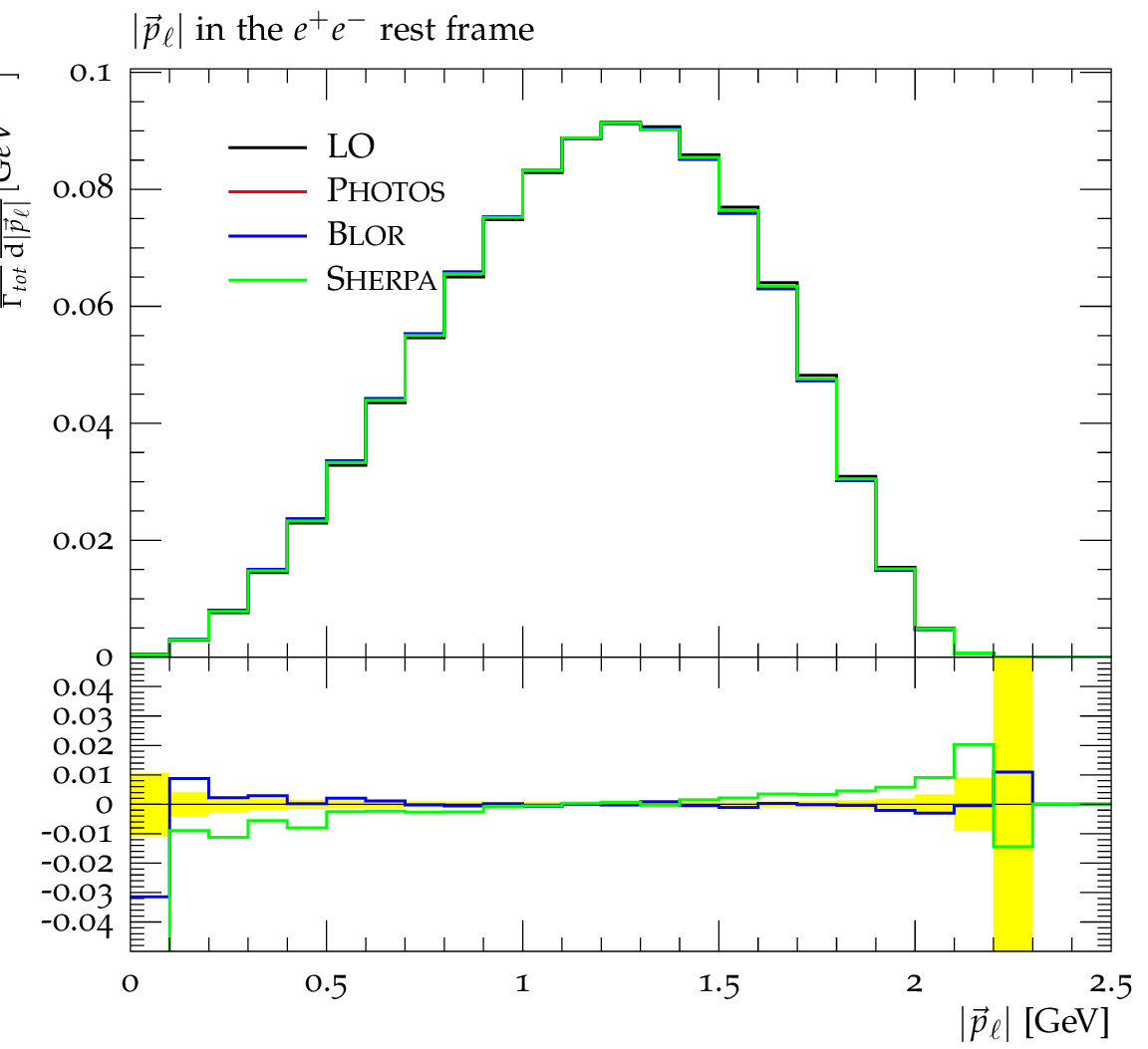}}
  {Lepton and Meson momentum spectrum in the $e^+e^-$ rest frame in the decay 
   $B^+ \to \bar D_0^{* \, 0} \, \mu^+ \, \nu_\mu$. The \protect\PHOTOS 
   prediction is taken as the reference in the ratio plot.\label{Fig:B+_D0star0_mu_nu}}

The final state lepton and meson momentum spectra in the decays 
$B^{0,+} \to \bar D_0^{* \, -,0} \, \ell^+ \, \nu_\ell$, with $\ell=e,\mu$,
are shown in Figs~\ref{Fig:B0_D0star-_e_nu}, \ref{Fig:B0_D0star-_mu_nu}, 
and \ref{Fig:B+_D0star0_e_nu}, \ref{Fig:B+_D0star0_mu_nu}. The $B\to D_0^*$ 
transition current is modeled using Leibovich-Ligeti-Stewart-Wise-parametrised 
form factors, cf.~App.~\ref{App:llswformfactorsintroduction}. Except for 
differences in the form factor parametrisations due to the $D_0^*$ meson being 
a scalar instead of a pseudo-scalar and its higher mass, the effects of higher 
order corrections are comparable to the case of Sec.~\ref{Sec:BDLN}.


\subsubsection{Decays $B \to \pi \, \ell \, \nu_\ell$}
\label{Sec:BpiLN}

\myfigure{tbp}{
  \includegraphics[width=0.48\textwidth]{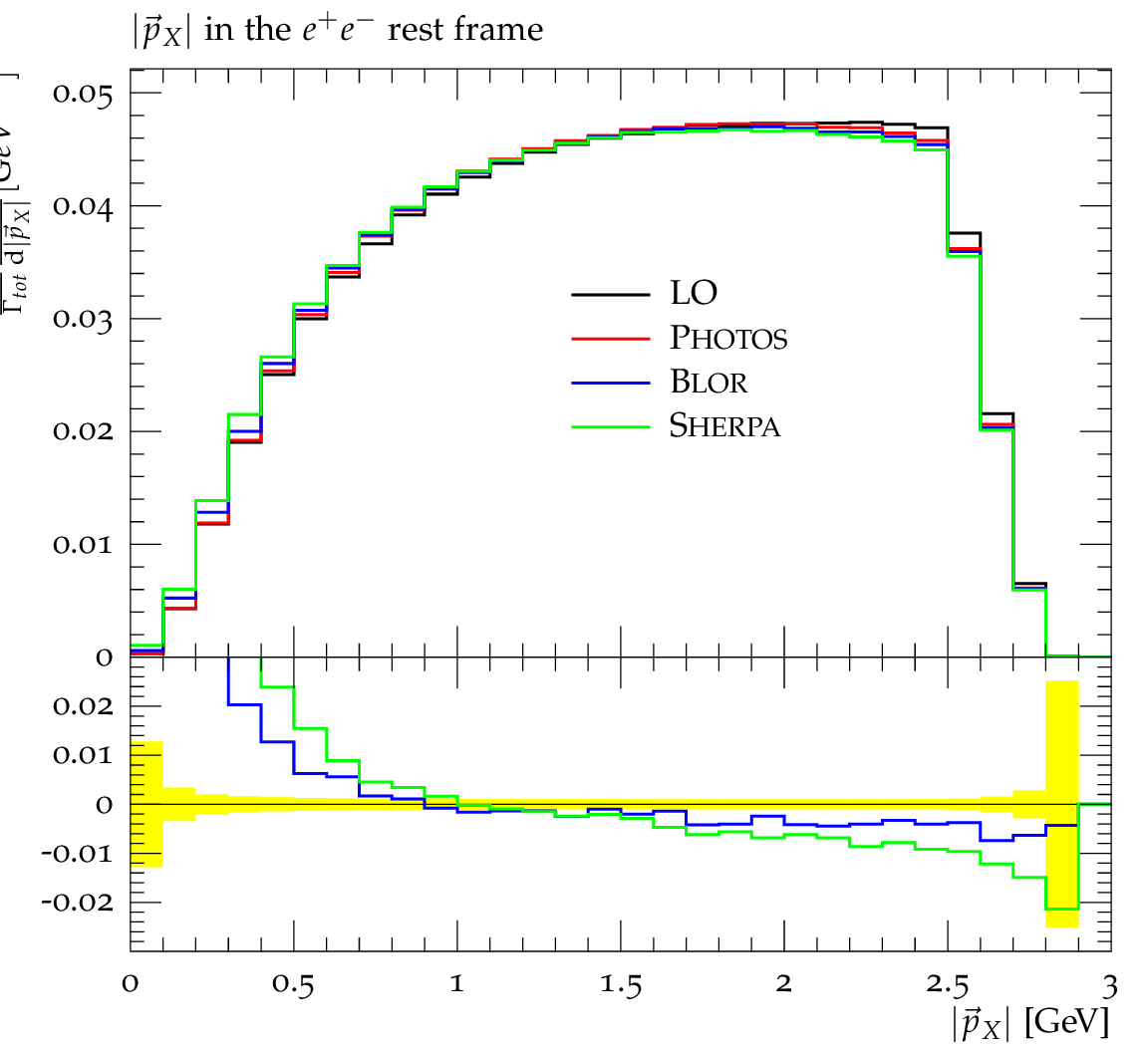} 
  \hspace*{0.02\textwidth}
  \includegraphics[width=0.48\textwidth]{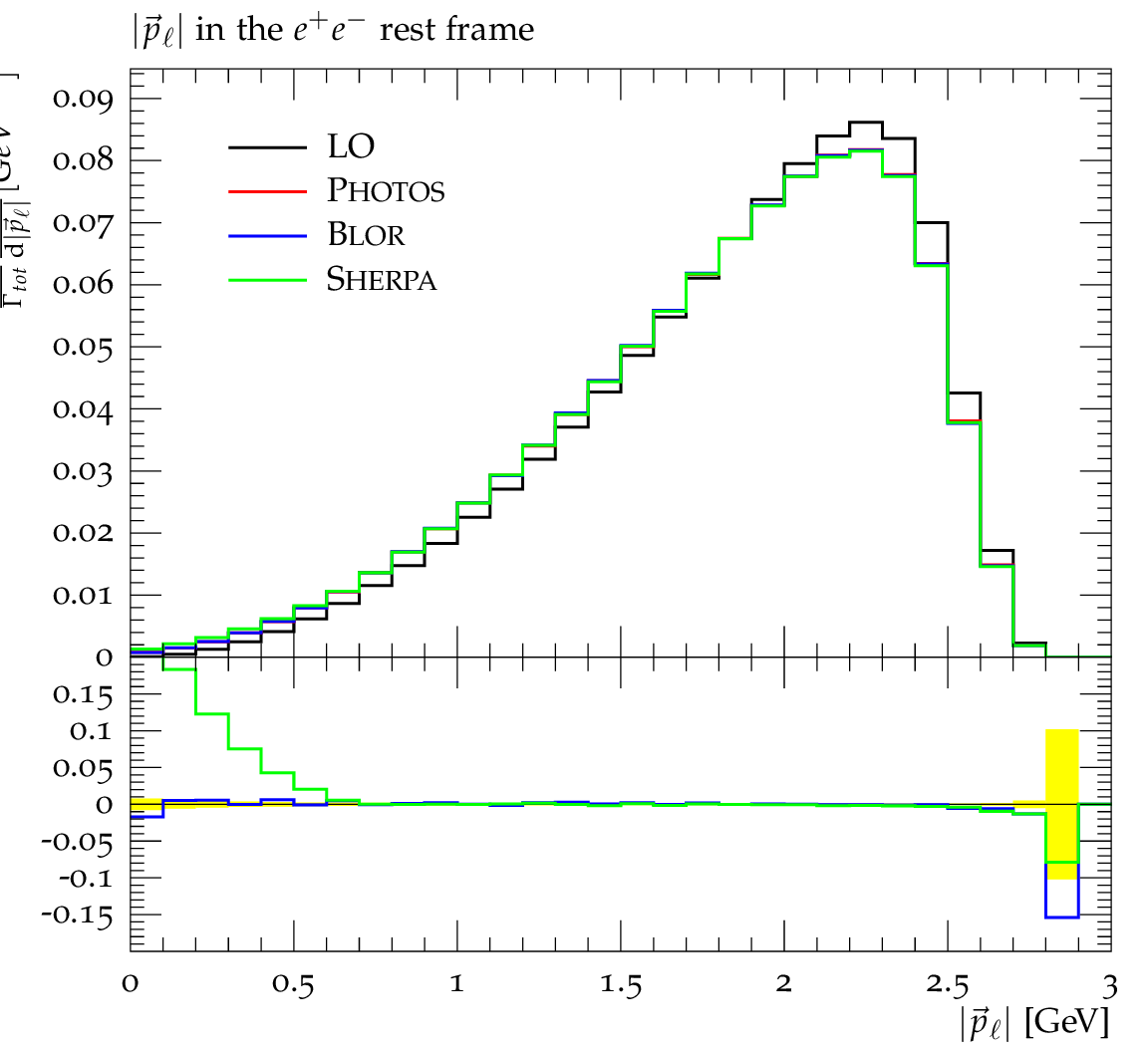}}
  {Lepton and Meson momentum spectrum in the $e^+e^-$ rest frame in the decay 
   $B^0 \to  \pi^-\, e^+ \, \nu_e$. Both matrix-element-corrected multi-photon 
   radiation and the IB terms for $t\neq t'$ exhibit a strong influence here. 
   The \protect\PHOTOS prediction is taken as the reference in the ratio plot.
   \label{Fig:B0_pi-_e_nu}}

\myfigure{tbp}{
  \includegraphics[width=0.48\textwidth]{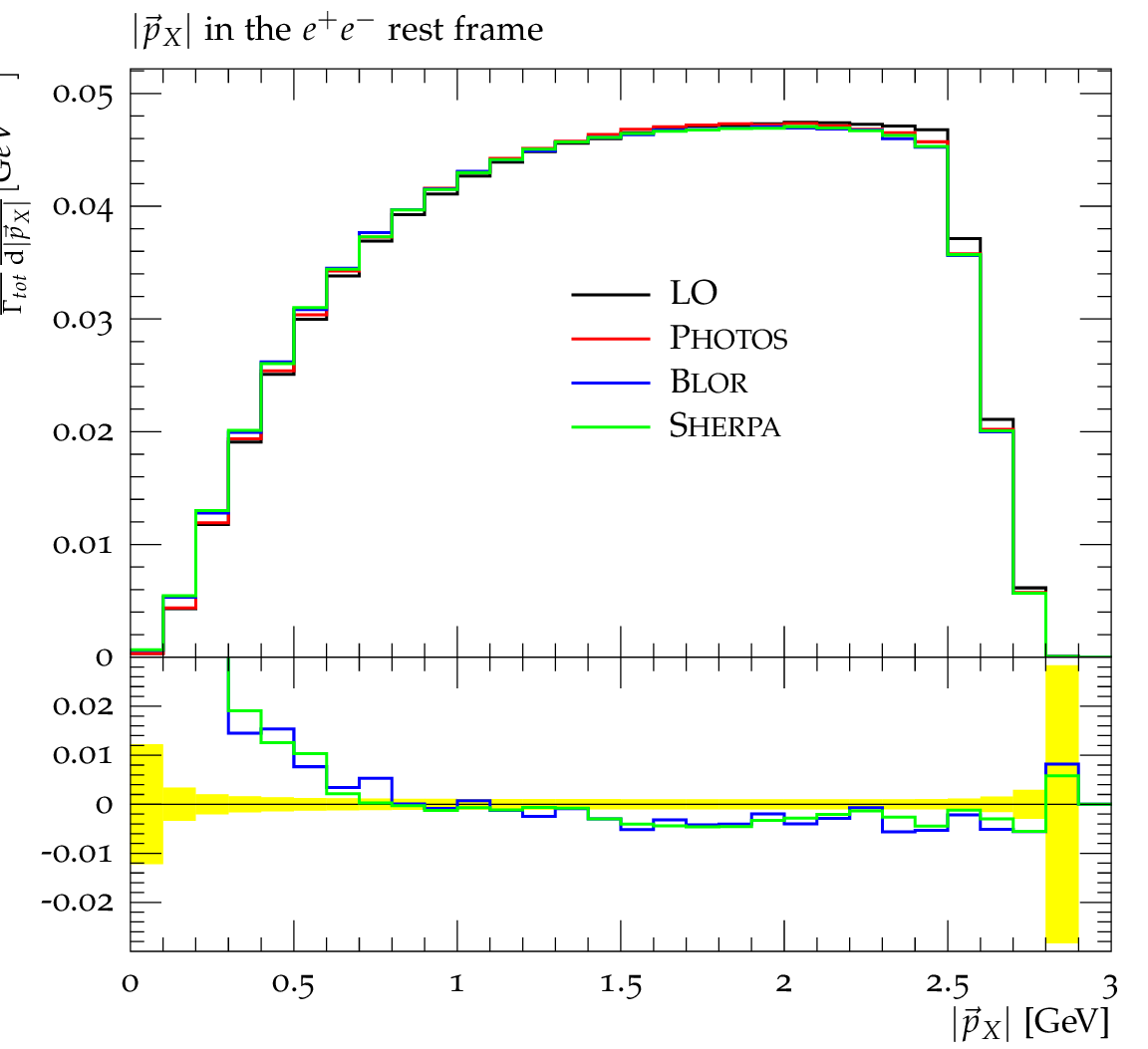} 
  \hspace*{0.02\textwidth}
  \includegraphics[width=0.48\textwidth]{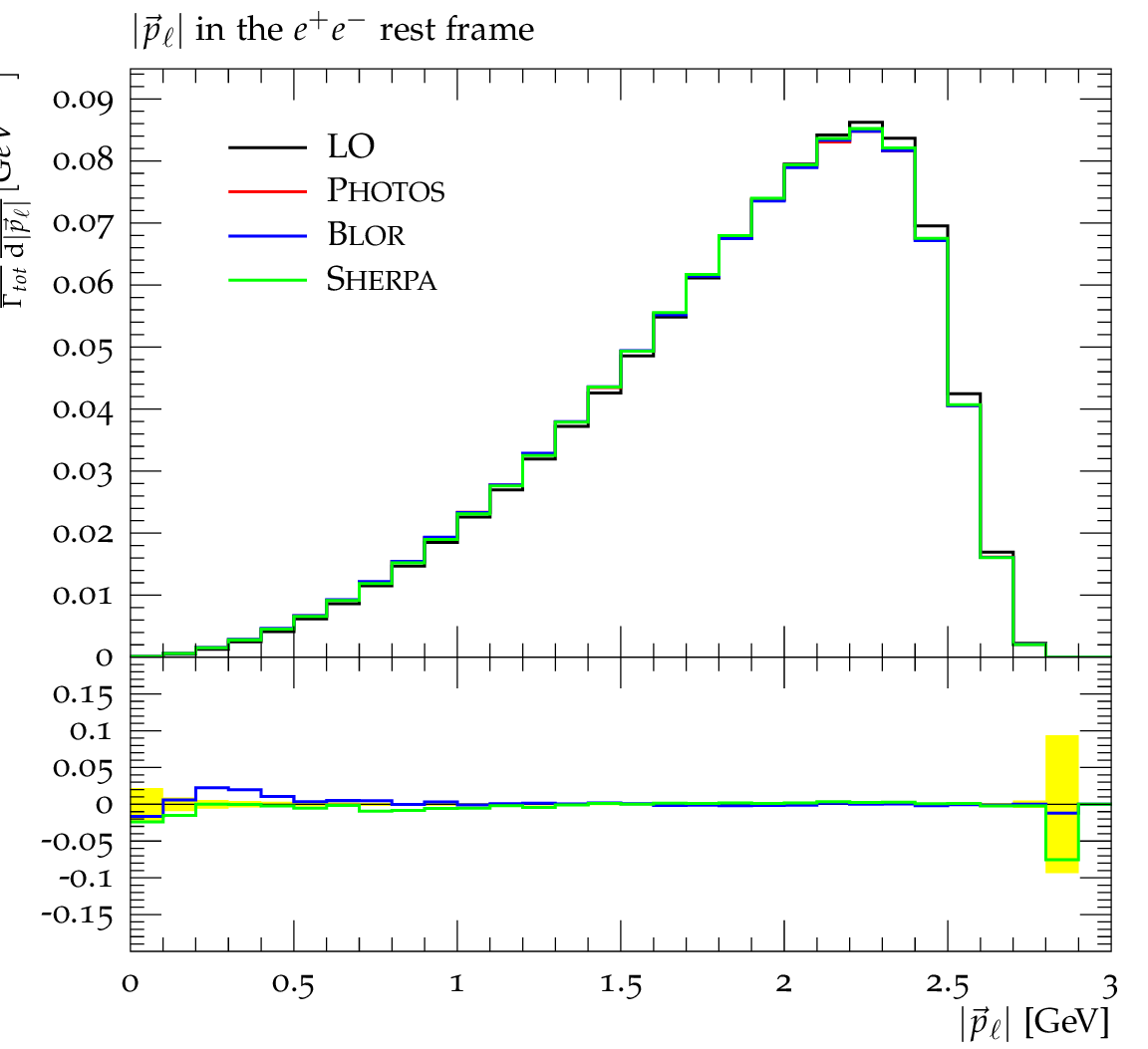}}
  {Lepton and Meson momentum spectrum in the $e^+e^-$ rest frame in the decay 
   $B^0 \to \pi^- \, \mu^+ \, \nu_\mu$. The IB terms for $t\neq t'$ exhibit a 
   strong influence here. The \protect\PHOTOS prediction is taken as the 
   reference in the ratio plot.\label{Fig:B0_pi-_mu_nu}}

\myfigure{tbp}{
  \includegraphics[width=0.48\textwidth]{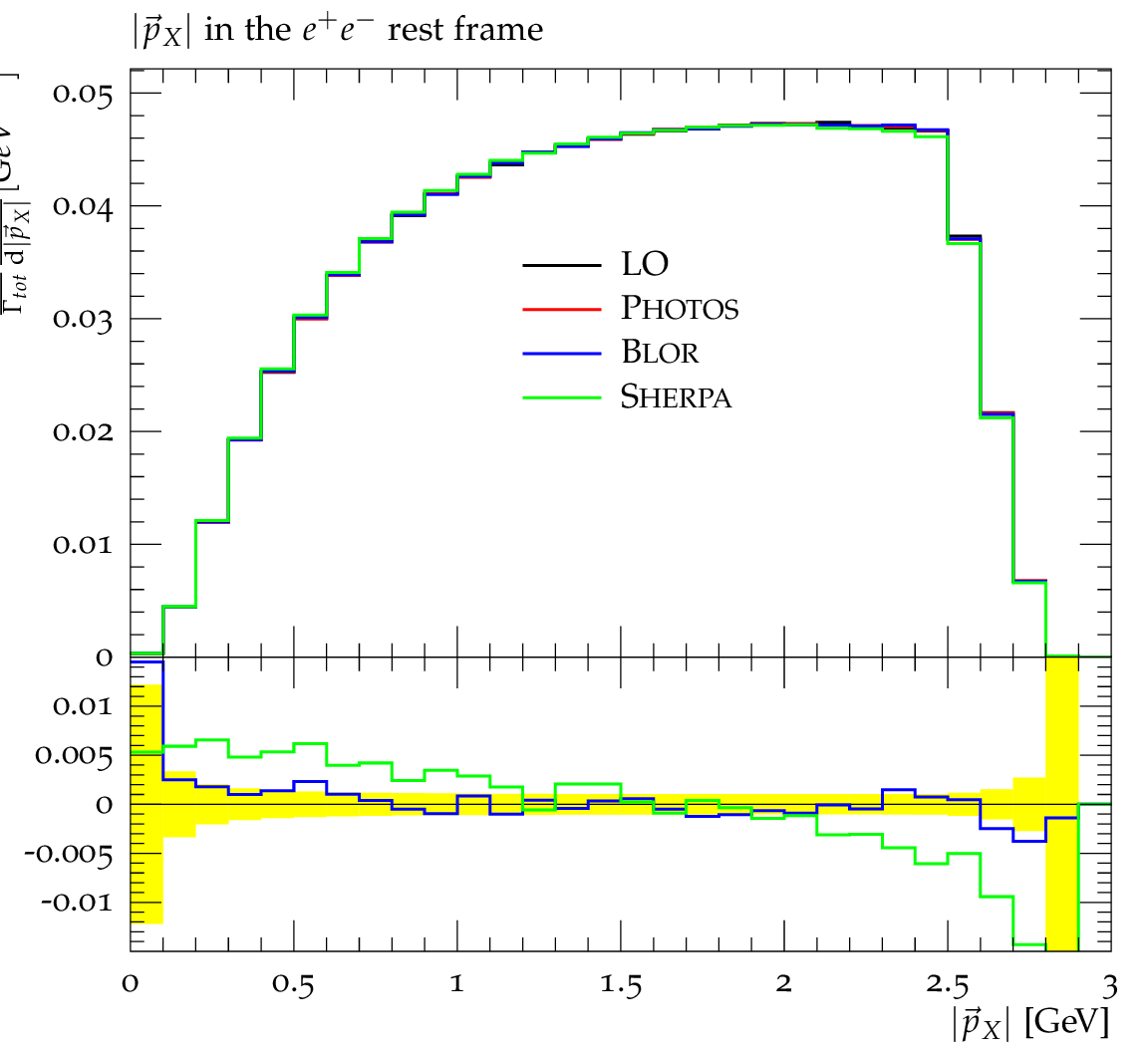} 
  \hspace*{0.02\textwidth}
  \includegraphics[width=0.48\textwidth]{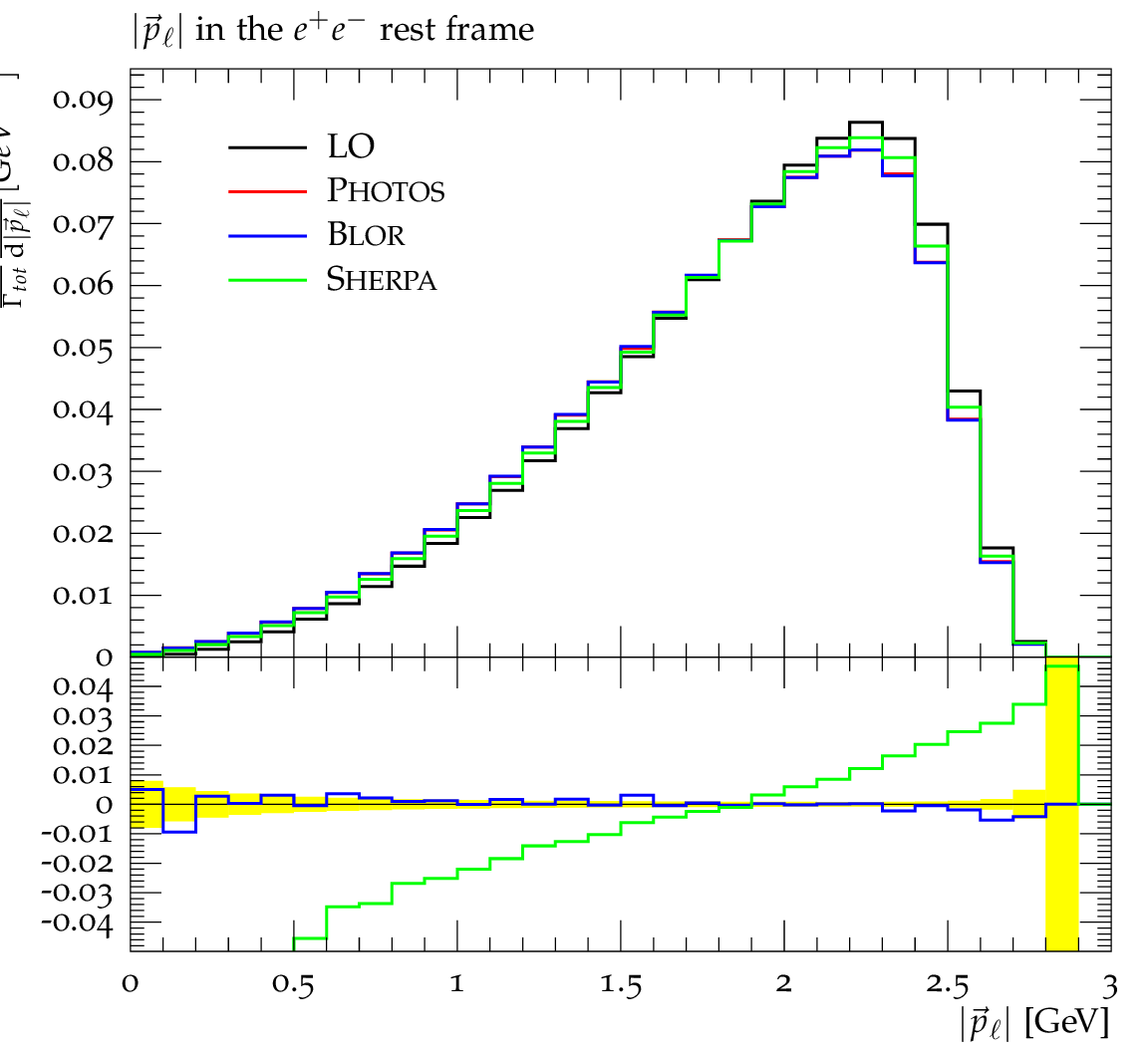}}
  {Lepton and Meson momentum spectrum in the $e^+e^-$ rest frame in the decay 
   $B^+ \to \pi^0 \, e^+ \, \nu_e$. The \protect\PHOTOS prediction is taken as 
   the reference in the ratio plot.\label{Fig:B+_pi_e_nu}}

\myfigure{tbp}{
  \includegraphics[width=0.48\textwidth]{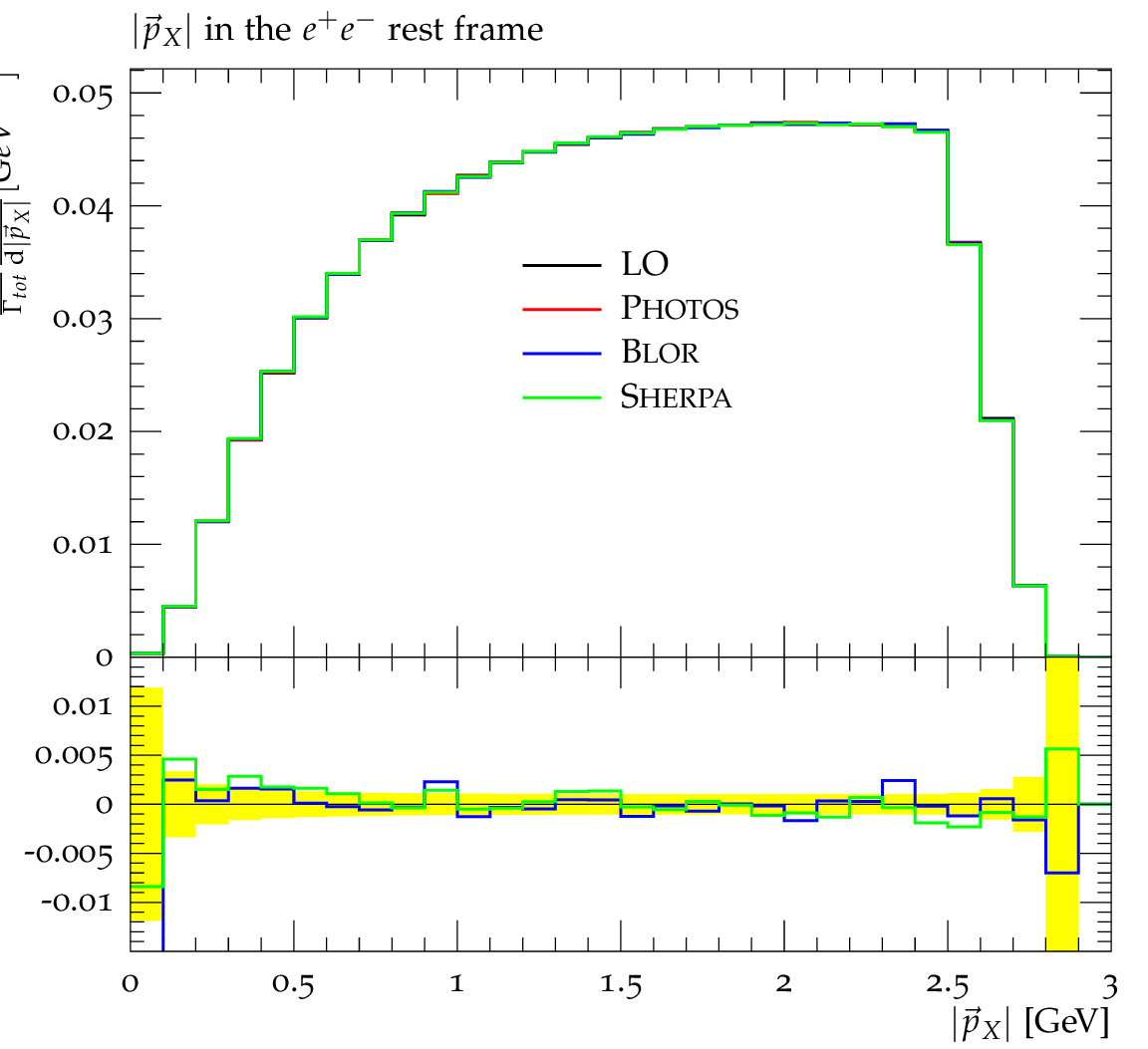} 
  \hspace*{0.02\textwidth}
  \includegraphics[width=0.48\textwidth]{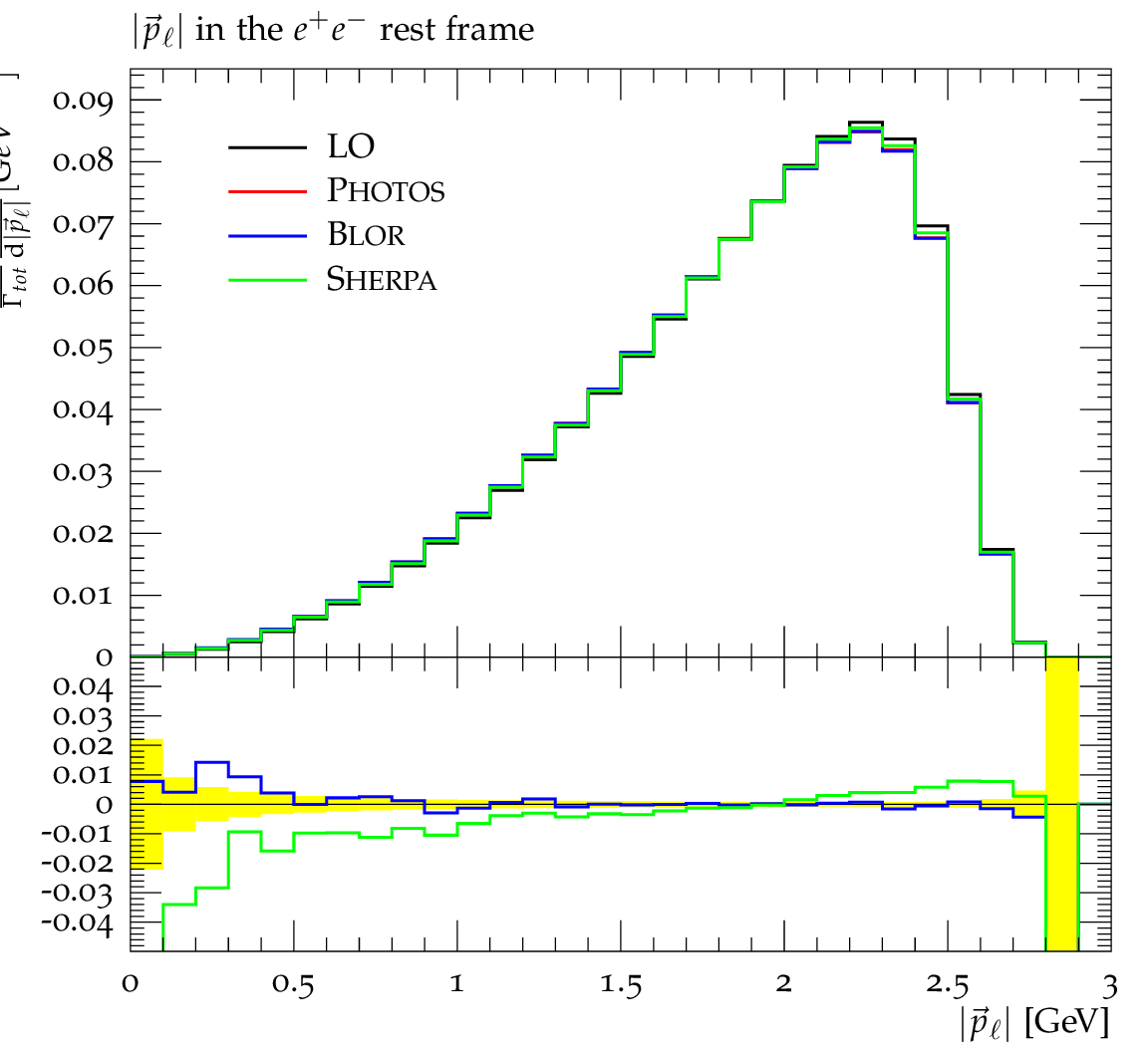}}
  {Lepton and Meson momentum spectrum in the $e^+e^-$ rest frame in the decay 
   $B^+ \to \pi^0 \, \mu^+ \, \nu_\mu$. The \protect\PHOTOS prediction is taken 
   as the reference in the ratio plot.\label{Fig:B+_pi_mu_nu}}

Figs.~\ref{Fig:B0_pi-_e_nu}, \ref{Fig:B0_pi-_mu_nu} and \ref{Fig:B+_pi_e_nu}, 
\ref{Fig:B+_pi_mu_nu} show the decay channels into charged and neutral pions. 
The $B\to\pi$ transition current is modeled using the Ball-Zwicky form factor 
model described in App.~\ref{App:bzformfactorsintroduction}.
Here, because of the comparably small mass of the charged pion effects due 
IB corrections for $t\neq t'$ become important. The structure-dependent 
contributions still have negligible impact on the differential distributions, 
as is shown in Sec.~\ref{Sec:result_SD_terms}. Again, in the 
electron channel of the decay into a charged pion the correct treatment of 
hard multi-photon radiation, assuming they are sufficiently well described in 
the {\sc Qed}-enhanced phenomenological model, leads to comparably large 
deviations. 

Nonetheless, it should be noted, as was also discussed earlier, that the 
matching procedure employed in this study runs into conceptual problems when 
applied to a $B\to\pi$ transition due to large difference between the hadronic 
mass scale $\Lambda=m_\pi$ and the maximal energy of an emitted photon, 
$E_\gamma^\text{max}=2.6379\GeV$ ($B^0\!\to\pi^-e^+\nu_e$). Consequently, a 
considerable fraction of the real emission phase space wherein the photon is 
able to resolve the pion is described by the effective theory only. Thus, the 
results obtained here should be considered with caution. However, they still 
are an improvement over the leading logarithmic corrections employed in 
standard analyses.

 \subsection{Influence of explicit short-distance terms}
\label{Sec:result_SD_terms}

\myfigure{tbp}{
  \includegraphics[width=0.48\textwidth]{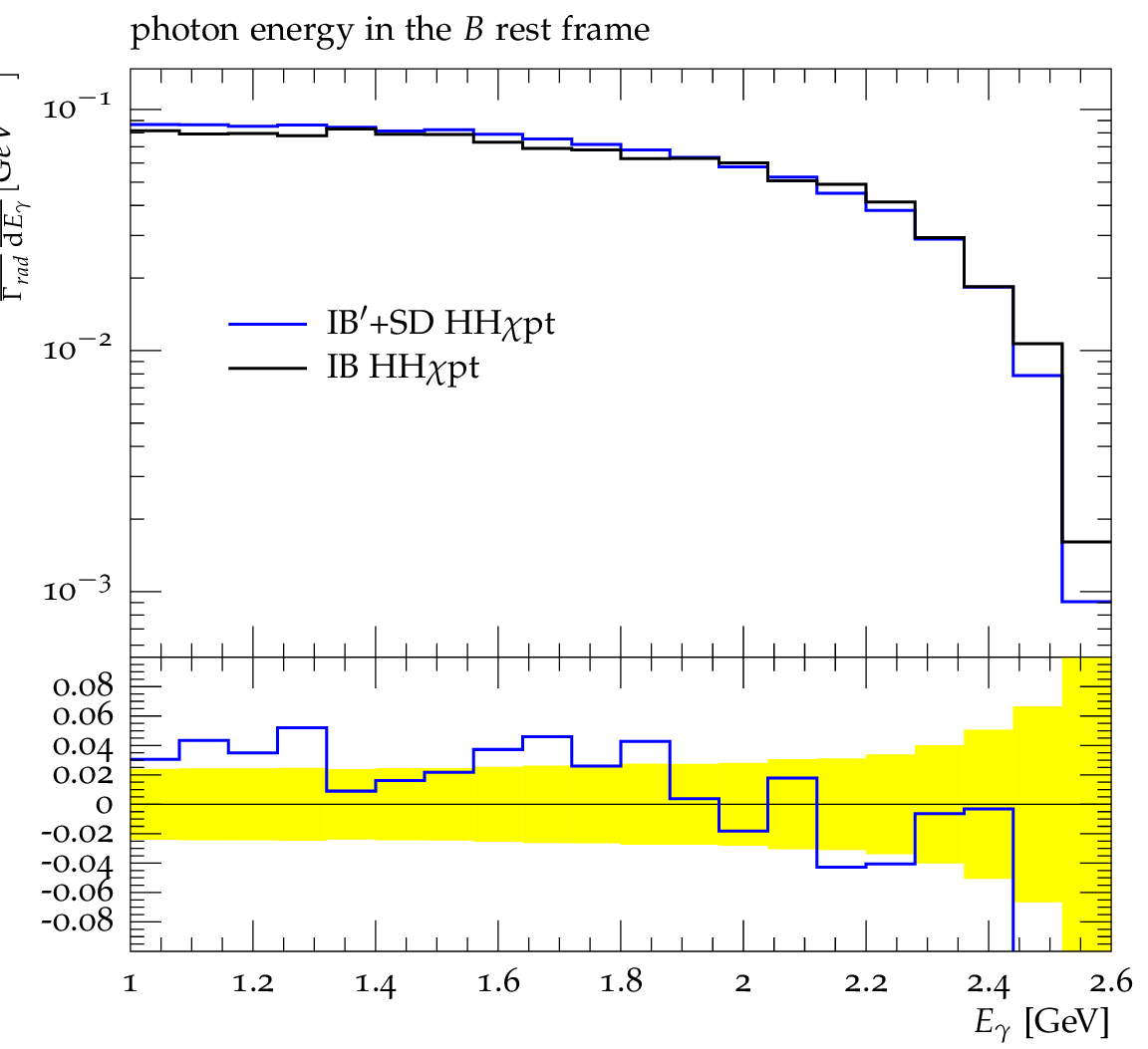}
  }
  {The photon energy spectrum in the decay $B^0\to \pi^-\,e^+\,\nu_e\,\gamma$ 
   ($E_\gamma>1\GeV$) is shown. The complete IB+SD result of 
   \cite{Cirigliano:2005ms} (blue) is compared against the prediction of the IB 
   terms only according to Sec.~\ref{Sec:theory_SD_terms} (black) in the 
   HH$\chi$pt form factor model. In the ratio plot the latter is taken as the 
   reference. \label{Fig:SD_pi_terms}}

\myfigure{tbp}{
  \includegraphics[width=0.48\textwidth]{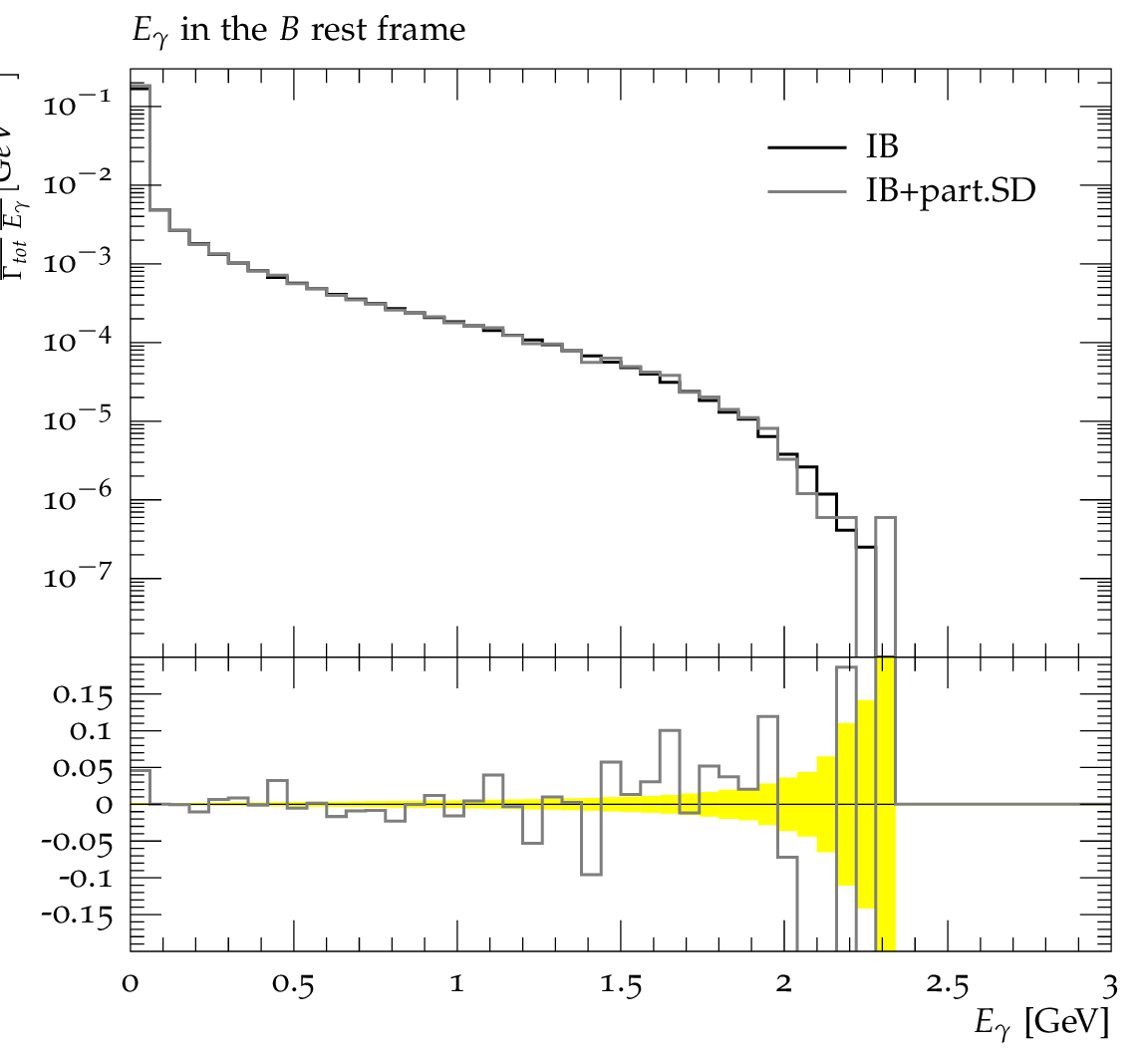}
  \hspace*{0.02\textwidth}
  \includegraphics[width=0.48\textwidth]{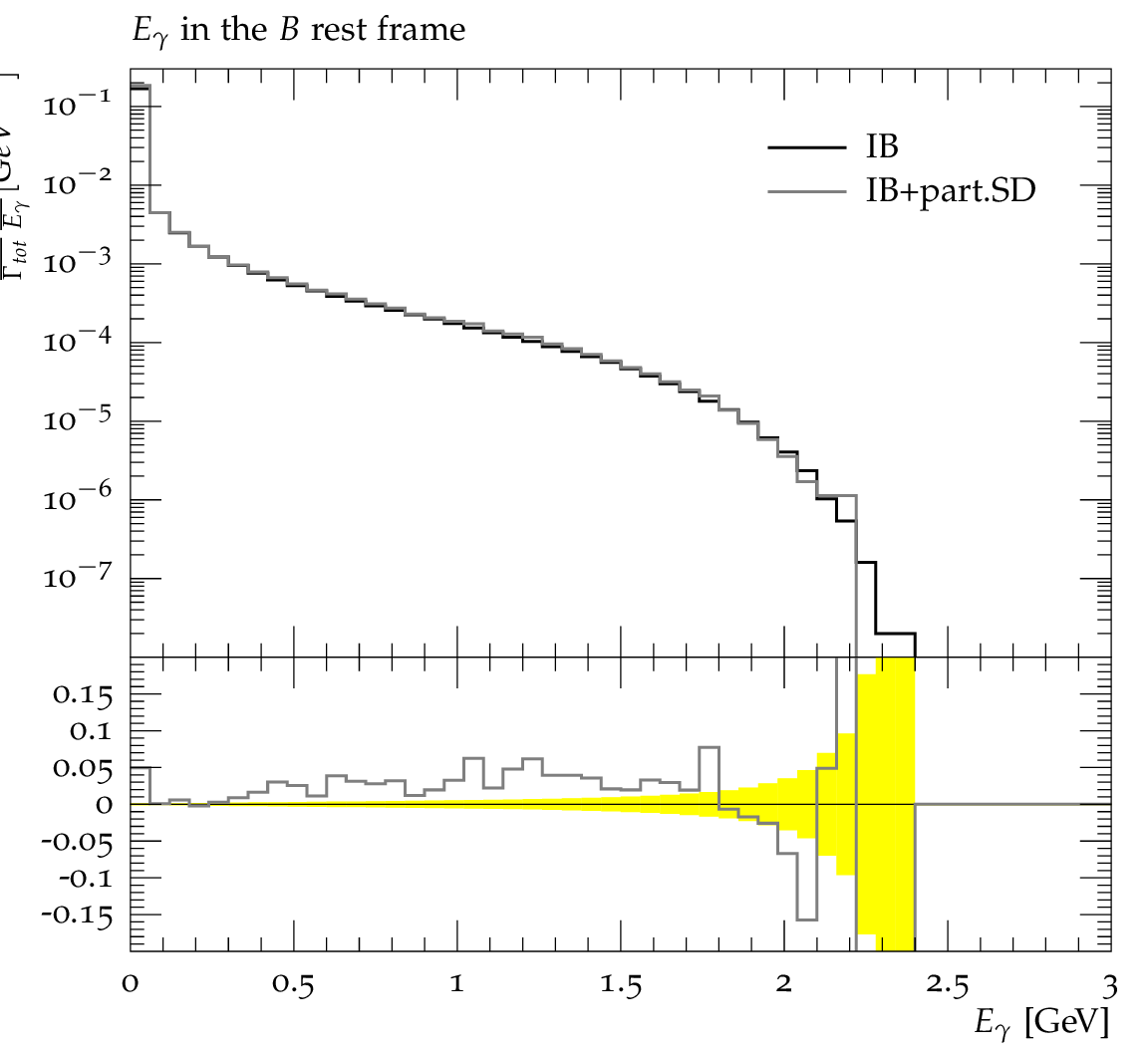}
  }
  {The photon energy spectra in the decays $B^0\to D^-\,e^+\,\nu_e\,\gamma$ on 
   the left and $B^+\to D^0\,e^+\,\nu_e\,\gamma$ on the right are shown. The 
   result including partial SD terms arising due to intermediate excited 
   mesons, $D^{*-}$ and $D^{*0}$ \cite{Becirevic:2009fy}, and the associated 
   coupling $X^*\to X\gamma$ (grey) is compared against the prediction of the 
   IB terms only according to Sec.~\ref{Sec:theory_SD_terms} (black) in the 
   HQET form factor model. In the ratio plot the latter is taken as the 
   reference. \label{Fig:SD_D_terms}}

\mytable{tbp}{
  \begin{tabular}{llll}
     & $\delta_\text{sd}+\delta_\text{ld(IB)}$ & $\delta_\text{sd}+\delta_\text{ld(IB+part.SD)}$ & $\sigma_\text{SD}$ \\
    \hline
    $B^0\to D^-\,\ell^+\,\nu_\ell(\gamma)$  &  $0.02223(6)$      &     $0.02225(7)$           &   $0.00002$  \\
    $B^+\to D^0\,\ell^+\,\nu_\ell(\gamma)$  &  $0.01463(5)$      &     $0.01627(6)$           &   $0.00158$ \\
    \hline
  \end{tabular}
  \vspace*{2mm}
  }
  {The effect of including the partial SD terms arising due to intermediate excited 
   $D^{*-}$ and $D^{*0}$ mesons is shown. Because of the unknown size of the 
   full SD contributions this single term, argued to be dominating, is used 
   as an estimate on the systematic uncertainty $\sigma_\text{SD}$ associated 
   to the IB-only result of Sec.~\ref{Sec:res_rates}.\label{Tab:SD_D_terms}}

In section the influence of explicitly calculated structure dependent terms, 
as introduced in Sec.~\ref{Sec:theory_SD_terms}, is investigated. The analysis 
is performed for the decay channel $B^0\to\pi^-\,e^+\,\nu_e$ with the results 
of \cite{Cirigliano:2005ms} and the decay channels 
$B^{0,+}\to D^{-,0}\,e^+\,\nu_e$ with the results of \cite{Becirevic:2009fy}. 
The same conclusions also apply to the muon channels.

For the charged pion channel, \cite{Cirigliano:2005ms} employs a form factor 
model of the heavy-hadron chiral perturbation theory 
(HH$\chi$pt) \cite{Wise:1992hn,Burdman:1992gh,Yan:1992gz,Wise:1993wa}, valid in 
the region of $E_\gamma>1\GeV$. 
Despite the mismatch of the form factor model used in the present study the 
result depicted in Fig.~\ref{Fig:SD_pi_terms} shows the structure-dependent contributions have 
little influence on the photon energy spectrum in this region. Due to their finiteness 
in the limit $k\to 0$, they are expected to behave similarly for 
$E_\gamma<1\GeV$. 
For the Ball-Zwicky form factor model used in this study, 
cf.~App.~\ref{App:bzformfactorsintroduction}, the IB and SD correction are 
expected to behave similarly.

In the $D$ meson channels, \cite{Becirevic:2009fy} uses lattice results for 
the trilinear couplings of an exited $D^*$ meson to a photon and its ground 
state: $g_{D^{*+}D^+\gamma}=-0.1(7)$ and $g_{D^{*0}D^0\gamma}=2.7(1.2)$. The 
effects manifest themselves as corrections to the total decay widths and are summarized
in Tab.~\ref{Tab:SD_D_terms}: they prove minor in 
the case of $D^-$, and sizable in the case of $D^0$. The case is similar for the 
radiative spectra: a slight change in the shape of the radiative energy loss 
for the $D^0$ channel on the scale of less than 5\%, while no such change 
occurs in the $D^-$ channel. These particular SD corrections, however, only 
form a single term in one class of SD correction. Note that higher order charm 
resonances, i.e. through $D^{**}D \g$ processes, do not contribute to real 
corrections at {\sc Nlo} of the studied decay modes because of angular momentum
and spin conservation.  Therefore, the lowest order are believed to be one of the 
dominant terms, they are taken as an estimate for the error associated to neglecting
all SD contributions. 

Note, that the samples containing (partial) SD terms have larger statistical 
uncertainties.

\section{Conclusions}\label{Sec:Conclusions}
 
In this paper, electroweak correction to semileptonic $B$ decays were studied. 
A long-distance calculation in the {\sc Qed} improved effective Lagrangian was 
matched to the partonic short-distance result of \cite{Sirlin:1977sv} for 
(pseudo)scalar final state mesons. Structure dependent terms, e.g.~due to 
non-local photon-charged meson interactions, intermediate resonant meson 
propagators or modifications to the effective weak meson decay due to off-shell 
currents, were not taken into account for the computed central values. This was 
done because they are only known for a very limited set of processes, and there 
usually only partially. 

The results achieved with this method, detailed in Sec.~\ref{Sec:Results}, give 
more reliable predictions for total and differential decay rates, 
accompanied by quantifiable errors. The improved predictions of the total 
decay rate were applied on two selected measurements of 
$V_\text{cb}$~\cite{Aubert:2009ac} and $V_\text{ub}$~\cite{Aubert:2006px} 
resulting in small corrections to their respective central values. These 
exemplifications, however, are a mere reweighting of their stated results and 
correction factors. To fully assess the impact of the corrections to the 
leading order decay presented in this paper the form factors of the 
phenomenological models will have to be refitted with the results presented 
in Sec.~\ref{Sec:res_diff}. Here, special attention is again drawn to the 
large deviations near the endpoints of the kinematic distributions arising 
when both fixed next-to-leading order and resummed leading-logarithmic calculations are 
combined. Finally, the parts of the analyses relying on Monte Carlo estimates 
of the radiation pattern need to be corrected for the improved description 
presented in this publication.

It should again be emphasised that the results presented for the decay of a 
$B$ meson into a pion should be considered with care. The prescription of 
matching long- and short-distance corrections runs into conceptual problems 
for this particular process. This is due to the large hierarchy of the scales 
of the pion mass (the scale where a photon is able to resolve a pion) and the 
maximally allowed photon energy ($E_\gamma^\text{max}\sim 2.5\GeV$, 
cf.~Sec.~\ref{Sec:BpiLN}). Consequently, the prediction for the total decay 
rate may receive significant corrections when a more elaborate matching 
scheme is used. Nonetheless, the differential distributions are unaffected.

Similarly, structure-dependent contributions, where known, have been shown to 
have negligible influence on the differential distributions while their impact 
on the total decay rates can be sizeable. Despite this fact, these 
structure-dependent contributions were not included in the predictions of the 
central values of the total decay rates, but only have been used to estimate 
their potential error. This treatment is justified since in all cases 
considered here they are only known partially.

\section*{Acknowledgements}
 The authors would like to thank Heiko Lacker, Dominik St\"ockinger, and 
S\'ebastien Descotes-Genon for useful discussions. MS thanks Frank Siegert for 
help in introducing new form factor models into \Sherpa/\Hadrons. The work of 
MS was supported by the DFG Graduate College 1504 and the MCnet Marie Curie 
Research Training Network (contract number MRTN-CT-2006-035606).

All histograms were plotted using tools from Rivet \cite{Buckley:2010ar}.

\appendix
\section{Form factor models of exclusive semileptonic $B$ meson decays reviewed.}
\label{App:form_factors}

\subsection{Form factors for $B \to D \, \ell \, \nu$}
\label{App:hqetformfactorsintroduction}

\mytable{h!}
  {\begin{tabular}{cl}
     Parameter & Value \\ \hline
     $\mathcal{G}(1)$ & 0.98 \\
     $\rho_D^2$ & 1.19 
   \end{tabular}\vspace*{2mm}}
  {Parameter values used for the transition current $\bra D | V^\mu  | B \ket$
   taken from Heavy Quark Effective Theory.
   \label{Tab:HQET_vals}}

The vector current describing the semileptonic $B \to D \, \ell \, \nu$ decay 
is given by 
\bea\label{heavyquarkcurrent}
 \bra D | V^\mu  | B \ket
& = & \sqrt{m_B m_D} 
      \left(h_+(w)\bve{v_B+v_D}^\mu+h_-(w)\bve{v_B-v_D}^\mu\right) \,,
\eea
with the heavy quark form factors $h_\pm$ parametrised \cite{Caprini:1997mu}
\bea
 h_+(w) 
& = &  \mathcal{G}(1) \times \left[ 1 - 8 \rho_D^2 \, z
                                      + (51 \rho_D^2 - 10) \, z^2 
                                      - (252 \rho_D^2 - 84) \, z^3) \right] \,, \\
 h_-(w) 
& = & 0 \,.
\eea
It is $z=\frac{\sqrt{w+1}-\sqrt{2}}{\sqrt{w+1}+\sqrt{2}}$ with 
$w=\frac{m_B^2+m_D^2-t}{2m_Bm_D}$, $\rho_D^2$ the form factor slope, 
and $\mathcal{G}(1)$ the normalisation at $w=1$. The values used are given in 
Tab.~\ref{Tab:HQET_vals}.

\subsection{Form factors for $B \to \pi \, \ell \, \nu$}
\label{App:bzformfactorsintroduction}

\mytable{h!}
  {\begin{minipage}{0.45\textwidth}
    \begin{tabular}{cl}
      Parameter & Value \\ \hline
      $m_{f_{+1}}^2$ & 28.40 $\GeV^2$ \\
      $m_{f_{+2}}^2$ & 40.73 $\GeV^2$ \\
      $m_{f_0}^2$ & 33.81 $\GeV^2$
    \end{tabular}
  \end{minipage}
  \begin{minipage}{0.03\textwidth}
  \end{minipage}
  \begin{minipage}{0.45\textwidth}
    \begin{tabular}{cl}
      Parameter & Value \\ \hline
      $r_{f_{+1}}$ & 0.744 \\
      $r_{f_{+2}}$ & -0.486 \\
      $r_{f_0}$ & 0.258
    \end{tabular}
  \end{minipage}\vspace*{2mm}}
  {Parameter values used for the transition current $\bra \pi | V^\mu  | B \ket$
   taken from the pole parametrisation in \cite{Ball:2004ye}.
   \label{Tab:BZ_vals}}

The vector current describing the semileptonic $B \to \pi \, \ell \, \nu$ decay 
is given by 
\bea
 \bra \pi | V^\mu  | B \ket 
& = &  \left(\bve{p_B + p_\pi}^\mu 
              - \frac{m_B^2 - m_\pi^2}{t} \bve{p_B - p_\pi}^\mu \right) f_+(t) 
      +\left(\frac{m_B^2 - m_\pi^2}{t} \bve{p_B - p_\pi}^\mu \right) f_0(t) \,, \nnb \\
 \eea
with form factors parametrised as \cite{Ball:2004ye} 
\bea
 f_+(t) 
& = &  \frac{r_{f_{+1}} }{ 1 - \frac{t}{m_{f_{+1}}^2}}  
      +\frac{r_{f_{+2}} }{ 1 - \frac{t}{m_{f_{+2}}^2}} \,, \\
 f_0(t) 
& = &  \frac{r_{f_{0}} }{ 1 - \frac{t}{m_{f_{0}}^2 }}  \,.
\eea
$r_{f_{+1}}$, $r_{f_{+2}}$ , and $r_{f_{0}}$ are normalisations and
$m_{f_{+1}}$, $m_{f_{+2}}$, and $m_{f_{0}}$ pole masses. Their values are 
listed in Tab.~\ref{Tab:BZ_vals}.

\subsection{Form factors for $B \to D^*_0 \, \ell \, \nu$ }
\label{App:llswformfactorsintroduction}

\mytable{h!}
  {\begin{minipage}{0.45\textwidth}
    \begin{tabular}{cl}
      Parameter & Value \\ \hline
      $\epsilon_c$ & 0.3571 $\GeV^{-1}$ \\
      $\epsilon_b$ & 0.1042 $\GeV^{-1}$ \\
      $\zeta'$ & -1.0
    \end{tabular}
  \end{minipage}
  \begin{minipage}{0.03\textwidth}
  \end{minipage}
  \begin{minipage}{0.45\textwidth}
    \begin{tabular}{cl}
      Parameter & Value \\ \hline
      $\bar\Lambda$ & 0.4 GeV \\
      $\bar\Lambda^*$ & 0.75 GeV \\
      $\zeta(1)$ & 1.0
    \end{tabular}
  \end{minipage}\vspace*{2mm}}
  {Parameter values used for the transition current $\bra D_0^* | A^\mu | B \ket$
   taken from the pole parametrisation in \cite{Leibovich:1997em,Leibovich:1997tu}.
   \label{Tab:LLSW_vals}}

The axial-vector current describing the semileptonic 
$B \to D^*_0 \, \ell \, \nu$ decay is given by 
\bea\label{llswheavyquarkcurrent}
  \bra {D^*_0} | A^\mu  | B \ket  
& = & \sqrt{m_B m_{D^*_0}} \left( g_+(w) \bve{v_B + v_{D^*_0}}^\mu 
                                 +g_-(w) \bve{v_B - v_{D^*_0}}^\mu \right) \,.
\eea
with the form factors $g_\pm$ parametrised as 
\cite{Leibovich:1997em,Leibovich:1997tu}
\bea\label{llswformfactexp}
  g_{+}(w) 
& = & \epsilon_c \left[ 2 (w - 1) \zeta_1(w) 
                       -3 \zeta(w) \frac{w \bar \Lambda^*
                       -\bar\Lambda}{w+1} \right]\nnb\\
&&{} -\epsilon_b \left[ \frac{\bar\Lambda^*(2w+1)-\bar\Lambda(w+2)}{w+1}\zeta(w)
                       -2(w-1)\zeta_1(w)\right]\,, \\
  g_{-}(w)
& = & \zeta(w)  \,.
\eea
with 
\bea
 \zeta(w)
& = & \zeta(1) \times \left[ 1 + \zeta'(w-1) \right] \,, \\
 \zeta_1(w) 
& = & \bar\Lambda \, \zeta(w) \,.
\eea
where $ \zeta'$ denotes the form factor slope. The parameters are defined as 
$\epsilon_c\equiv\frac{1}{2m_c}$, $\epsilon_b\equiv\frac{1}{2m_b}$, 
$\bar\Lambda\equiv m_D-m_c$, and $\bar\Lambda^*\equiv m_{D_0^*}-m_c$.
Their values are listed in Tab.~\ref{Tab:LLSW_vals}.
As can be seen from the hadronic current of eq.~(\ref{llswheavyquarkcurrent}) 
the role of vector and axial-vector terms are reversed in decays to scalars as 
opposed to decays to pseudo-scalars. Thus, in the discussion of IB and SD terms 
the role of $V_{\mu\nu}$ and $A_{\mu\nu}$ are reversed. In particular, 
$V_{\mu\nu}^\text{IB}=0$ and $A_{\mu\nu}^\text{IB}\neq 0$.

\section{Next-to-leading order matrix elements}
\label{App:NLO_ME}

This appendix presents details on the real emission matrix elements with 
special focus on the inner bremsstrahlungs (IB) vertex emission terms. 
The corresponding virtual emission matrix elements can be found in 
\cite{Bernlochner:2010yd}, Sec.~4.3. They are calculated in $D = 4 - 2\epsilon$ 
dimensions and are UV regularised using the Pauli-Villars prescription 
\cite{Pauli:1949zm} of introducing an unphysical heavy photon of mass 
$\Lambda$, the matching scale to the short distance result.
\vspace*{2mm}

The summed matrix element of the Feynman graphs a to c in Fig.~\ref{Fig:nlofg1} 
for $B^+ \to \bar X^0 \, \ell^+ \, \nu_\ell\, \g$ is
 \bea\label{Eq:nloem1}
  \mathcal{M}_1^\half 
  & = &   i\,e\,\frac{\GF}{\sqrt{2}} V_{\text{xb}}\,H_\mu(p_B,p_X;t)\;\;
            \bar u_\nu\, \pr\gamma^\mu\,
            \frac{p_\ell\cdot\epsilon^*+\half\ds{k}\ds{\epsilon}^*}
                 {p_\ell \cdot k}\,v_\ell \nnb \\
  &&{}  - i\,e\,\frac{\GF}{\sqrt{2}} V_{\text{xb}}\,
            \frac{p_B\cdot\epsilon^*}{p_B\cdot k}\,H_\mu(p_B-k,p_X;t')\;\;
            \bar u_\nu\, \pr\gamma^\mu\,v_\ell \nnb \\
  &&{}  + \mathcal{M}_{1,\text{vertex emission}}^\half \,.
  \eea
Similarly, the summed matrix element of the Feynman graphs a to c in 
Fig.~\ref{Fig:nlofg2} for $B^0 \to \bar X^- \, \ell^+ \, \nu_\ell\, \g$ is
\bea\label{Eq:nloem2}
 \mathcal{M}^{\half}_1
  & = &   i\,e\,\frac{\GF}{\sqrt{2}} V_{\text{xb}}\,H_\mu(p_B,p_X;t)\;\;
            \bar u_\nu\, \pr\gamma^\mu\,
            \frac{p_\ell\cdot\epsilon^*+\half\ds{k}\ds{\epsilon}^*}
                 {p_\ell \cdot k}\,v_\ell \nnb \\
  &&{}  - i\,e\,\frac{\GF}{\sqrt{2}} V_{\text{xb}}\,
            \frac{p_X\cdot\epsilon^*}{p_X\cdot k}\,H_\mu(p_B,p_X+k;t')\;\;
            \bar u_\nu\, \pr\gamma^\mu\,v_\ell \nnb \\
  &&{}  + \mathcal{M}_{1,\text{vertex emission}}^\half \,.
\eea
The real emission changes the definition of the four momentum transfer squared, 
depending on the emission leg:
\bea
 t  \; = \; \bve{p_B - p_X}^2 \; = \; \bve{p_\ell + p_\nu + k}^2 \,,
 &\qquad&
 t' \; = \; \bve{p_B - p_X - k}^2 \; = \; \bve{p_\ell + p_\nu}^2 \,,
\eea
and it is 
\bea
 H_\mu (p_1,p_2;t) 
 & = & \bve{p_1+p_2}_\mu \, f_+(t) + \bve{p_1-p_2}_\mu \, f_-(t)\,,
\eea
as defined in eq.~(\ref{Eq:hadcurtreelev}). The emission terms off the external 
mesons and leptons in eqs.~(\ref{Eq:nloem1}) and (\ref{Eq:nloem2}) are, 
however, not gauge invariant by themselves. The vertex emission terms are thus 
needed to restore gauge invariance. Assuming $t = t'$, 
$f_\pm(t)=f_\pm(t^\prime)$, the vertex emission terms of the 
constant-form-factor {\sc Qed} invariant Lagrangian of 
eq.~(\ref{Eq:QED_W_Lagrangian}) are recovered:
\bea
\mathcal{M}_{1,\text{vertex emission}}^\half
& = &{} -i\,e\,\frac{\GF}{\sqrt{2}} V_{\text{xb}}\,\left(f_+(t)+f_-(t)\right)\;\;
       \bar u_\nu\, \pr\ds{\epsilon}^* v_\ell
      \qquad (B^+\to\bar X^0\,\ell^+\nu_\ell)\\
\mathcal{M}_{1,\text{vertex emission}}^\half
& = &{} +i\,e\,\frac{\GF}{\sqrt{2}} V_{\text{xb}}\,\left(f_+(t)-f_-(t)\right)\;\;
       \bar u_\nu\, \pr\ds{\epsilon}^* v_\ell
      \qquad (B^0\to\bar X^-\,\ell^+\nu_\ell)\,.
\eea
This is a reasonable approximation for heavy meson processes. Here, hard 
photons are emitted predominantly collinear to the charged lepton. Light meson 
processes, however, also radiate a considerable fraction of their hard 
radiation in the direction of the light meson and, hence, give rise to 
non-negligible corrections for $t\neq t^\prime$. There are two ways to obtain 
these corrections.\\
\noindent{\bf A:} Supposing the form factors $f_\pm$ can be expanded around 
$t$, the hadronic current reads 
\bea \label{Eq:hadcurexp}
 H_\mu (t') 
 & = & H_\mu(t) + k' \, \frac{\ud H_\mu}{\ud t} \Big|_{k=0} 
                + k'^2 \, \half \frac{\ud^2 H_\mu}{\ud t'^2} \Big|_{k=0}
                + \order(k^3) \, ,
\eea
with $t' = t+k'$ and $k' = {}-2 k \cdot \bve{p_B - p_X}$. Introducing 
eq.~(\ref{Eq:hadcurexp}) into eqs.~(\ref{Eq:nloem1}) and (\ref{Eq:nloem2}) and 
employing Ward's identity \cite{Ward:1950xp} to obtain the gauge restoring 
terms, results in
\bea\label{nloem11} 
 \lefteqn{\mathcal{M}_{1,\text{vertex emission}}^{\half,\;B^+\to\bar X^0\ell^+\nu_\ell}}\nnb\\
 & \hspace*{-2mm}= & i\,e\,\frac{\GF}{\sqrt{2}} V_{\text{xb}}\;\;
       \bar u_\nu\, \pr\gamma^\mu v_\ell
       \left(\frac{p_B\cdot\epsilon^*}{p_B\cdot k}\,k_\alpha
              -\epsilon^*_\alpha\right)
       \left(\delta^\alpha_{\ph{\alpha}\mu}\left(f_+(t)+f_-(t)\right)
              - 2\bve{p_B-p_X}^\alpha
                  \frac{\ud H_\mu}{\ud t'} \Big|_{k=0} \right) \nnb \\
 &&{}  +\order(k^2) \,,
\eea
and
\bea\label{nloem22}
 \lefteqn{\mathcal{M}_{1,\text{vertex emission}}^{\half,\;B^0\to\bar X^-\ell^+\nu_\ell}}\nnb\\
 & \hspace*{-2mm}= &\hspace{-3mm}{}-i\,e\,\frac{\GF}{\sqrt{2}} V_{\text{xb}}\;\;
       \bar u_\nu\,\pr\gamma^\mu v_\ell
       \left(\frac{p_X\cdot\epsilon^*}{p_X\cdot k}\,k_\alpha
              -\epsilon^*_\alpha\right)
       \left(\delta^\alpha_{\ph{\alpha}\mu}\left(f_+(t)-f_-(t)\right)
              - 2\bve{p_B-p_X}^\alpha
                  \frac{\ud H_\mu}{\ud t'} \Big|_{k=0} \right)  \nnb \\ 
 &&{}  + \order(k^2) \, , 
\eea 
respectively. Neglecting higher order terms in the expansion of 
eq.~(\ref{Eq:hadcurexp}) results in Low's matrix element \cite{Low:1958sn} 
for these processes. This 
approach, by implying the existence of a Taylor-series representation of the 
form factors $f_\pm(t)$, yields a consistent result both for the interaction 
terms of the phenomenological Lagrangian and the Feynman rules.
However, the exact functional form of the form factors has to be known.
Further, by including higher order corrections in $k$ structure 
dependent contributions to the matrix element are introduced: the isolated 
terms that restore gauge invariance are not unique, and undesired ambiguities 
are apparent. The impact of such terms were studied for $K_{l3}$ decays in 
\cite{Fischbach:1970qw,Fearing:1970zz}, finding negligible impact on the 
next-to-leading order decay rate. This result however can't be extrapolated
to $B$ meson decays, due to the wide range of possible excited intermediate states.  

\noindent{\bf B:} A result independent of the functional form of the form 
factors, and thus not relying on their differentiability, can be derived 
similarly. Instead of their argument, the form factors themselves are 
decomposed
\bea
 f_\pm(t') & = & f_\pm(t) + Z_\pm(t,t') \,.
\eea
Now, the missing terms for achieving gauge invariance of the real emission 
amplitude are determined as
\bea\label{nloem111} 
 \lefteqn{\mathcal{M}_{1,\text{vertex emission}}^{\half,\;B^+\to\bar X^0\ell^+\nu_\ell}}\nnb\\
 & \hspace*{-2mm}= & i\,e\,\frac{\GF}{\sqrt{2}} V_{\text{xb}}\;\;
       \bar u_\nu\, \pr\gamma^\mu v_\ell\;\;
       \Big({}-\left(f_+(t')+f_-(t')\right)\,\epsilon_\mu^* \nnb\\
 &&\hspace*{43mm}
             +(p_B+p_X)_\mu\;\epsilon^*\!\!\cdot Z_+(t,t')
             +(p_B-p_X)_\mu\;\epsilon^*\!\!\cdot Z_-(t,t')\Big)
\eea
and
\bea\label{nloem222}
 \lefteqn{\mathcal{M}_{1,\text{vertex emission}}^{\half,\;B^0\to\bar X^+\ell^-\nu_\ell}}\nnb\\
 & = & i\,e\,\frac{\GF}{\sqrt{2}} V_{\text{xb}}\;\;
       \bar u_\nu\,\pr\gamma^\mu v_\ell\;\;
       \Big(\left(f_+(t')-f_-(t')\right)\,\epsilon_\mu^* \nnb\\
 &&\hspace*{43mm}
             +(p_B+p_X)_\mu\;\epsilon^*\!\!\cdot Z_+(t,t')
             +(p_B-p_X)_\mu\;\epsilon^*\!\!\cdot Z_-(t,t')\Big)\,,
\eea 
respectively. $Z_\pm(t,t')=\frac{k\cdot n}{k\cdot n}Z_\pm(t,t')=
k_\alpha\frac{n^\alpha}{k\cdot n}\left(f_\pm(t')-f_\pm(t)\right)\equiv
k_\alpha Z_\pm^\alpha(t,t')$, 
$k\cdot n \neq 0$ but otherwise arbitrary. Through the definition of $n$, 
however, again ambiguities are introduced into this generic result which, 
again, are assumed to be negligible in this paper. 
The emission terms constructed this way constitute the inner-bremstrahlungs 
part of non-local emission term $V_{\mu\nu}^{\text{IB}}$ of 
Sec.~\ref{Sec:theory_SD_terms}.

\bibliographystyle{amsunsrt_mod}
\bibliography{journal}

\begin{thebibliography}{10}

\bibitem{Cabibbo:1963yz}
N.~Cabibbo, \emph{{Unitary Symmetry and Leptonic Decays}}, Phys. Rev. Lett.
  \textbf{10} (1963),
  \href{http://www-spires.dur.ac.uk/spires/find/hep/www?j=PRLTA,10,531}{531--5%
33}. \relax
 \relax
\bibitem{Kobayashi:1973fv}
M.~Kobayashi and T.~Maskawa, \emph{{CP Violation in the Renormalizable Theory
  of Weak Interaction}}, Prog. Theor. Phys. \textbf{49} (1973),
  \href{http://www.slac.stanford.edu/spires/find/hep/www?j=PTPKA,49,652}{652--%
657}. \relax
 \relax
\bibitem{Charles:2004jd}
J.~Charles et~al., CKMfitter Group collaboration, \emph{{CP violation and the
  CKM matrix: Assessing the impact of the asymmetric $B$ factories}}, Eur.
  Phys. J. \textbf{C41} (2005),
  \href{http://www.slac.stanford.edu/spires/find/hep/www?eprint=HEP-PH/0406184%
}{1--131},  [\href{http://arXiv.org/pdf/hep-ph/0406184}{{\tt hep-ph/0406184}}].
  \relax
 \relax
\bibitem{Bona:2005eu}
M.~Bona et~al., UTfit collaboration, \emph{{The UTfit collaboration report on
  the status of the unitarity triangle beyond the standard model. I: Model-
  independent analysis and minimal flavour violation}}, JHEP \textbf{03}
  (2006),
  \href{http://www.slac.stanford.edu/spires/find/hep/www?eprint=hep-ph/0509219%
}{080},  [\href{http://arXiv.org/pdf/hep-ph/0509219}{{\tt hep-ph/0509219}}].
  \relax
 \relax
\bibitem{Low:1958sn}
F.~E. Low, \emph{{Bremsstrahlung of very low-energy quanta in elementary
  particle collisions}}, Phys. Rev. \textbf{110} (1958),
  \href{http://www-spires.dur.ac.uk/spires/find/hep/www?j=PHRVA,110,974}{974--%
977}. \relax
 \relax
\bibitem{Yennie:1961ad}
D.~R. Yennie, S.~C. Frautschi and H.~Suura, \emph{{The Infrared Divergence
  Phenomena and High-Energy Processes}}, Ann. Phys. \textbf{13} (1961),
  \href{http://www.slac.stanford.edu/spires/find/hep/www?j=APNYA,13,379}{379--%
452}. \relax
 \relax
\bibitem{Altarelli:1977zs}
G.~Altarelli and G.~Parisi, \emph{{Asymptotic freedom in parton language}},
  Nucl. Phys. \textbf{B126} (1977),
  \href{http://www.slac.stanford.edu/spires/find/hep/www?j=NUPHA,B126,298}{298%
--318}. \relax
 \relax
\bibitem{Sirlin:1977sv}
A.~Sirlin, \emph{{Current Algebra Formulation of Radiative Corrections in Gauge
  Theories and the Universality of the Weak Interactions}}, Rev. Mod. Phys.
  \textbf{50} (1978),
  \href{http://www.slac.stanford.edu/spires/find/hep/www?j=RMPHA,50,573}{573}.
  \relax
 \relax
\bibitem{Sirlin:1981ie}
A.~Sirlin, \emph{{Large m(W), m(Z) Behavior of the O(alpha) Corrections to
  Semileptonic Processes Mediated by W}}, Nucl. Phys. \textbf{B196} (1982),
  \href{http://www.slac.stanford.edu/spires/find/hep/www?j=NUPHA,B196,83}{83}.
  \relax
 \relax
\bibitem{Eidelman:2004wy}
S.~Eidelman et~al., Particle Data Group collaboration, \emph{{Review of
  particle physics}}, Phys. Lett. \textbf{B592} (2004),
  \href{http://www-spires.dur.ac.uk/spires/find/hep/www?j=PHLTA,B592,1}{1}.
  \relax
 \relax
\bibitem{Sher:2003fb}
A.~Sher et~al., \emph{{New, high statistics measurement of the $K^+ \to \pi^0 e^+ \nu$
  (Ke3) branching ratio}}, Phys. Rev. Lett. \textbf{91} (2003),
  \href{http://www-spires.dur.ac.uk/spires/find/hep/www?eprint=HEP-EX/0305042}%
{261802},  [\href{http://arXiv.org/pdf/hep-ex/0305042}{{\tt hep-ex/0305042}}].
  \relax
 \relax
\bibitem{Alexopoulos:2004sw}
T.~Alexopoulos et~al., KTeV collaboration, \emph{{A Determination of the CKM
  Parameter $|V_{us}|$}}, Phys. Rev. Lett. \textbf{93} (2004),
  \href{http://www.slac.stanford.edu/spires/find/hep/www?eprint=hep-ex/0406001%
}{181802},  [\href{http://arXiv.org/pdf/hep-ex/0406001}{{\tt hep-ex/0406001}}].
  \relax
 \relax
\bibitem{Alexopoulos:2004sx}
T.~Alexopoulos et~al., KTeV collaboration, \emph{{Measurements of KL Branching
  Fractions and the CP Violation Parameter $|\eta^\pm|$}}, Phys. Rev. \textbf{D70}
  (2004),
  \href{http://www.slac.stanford.edu/spires/find/hep/www?eprint=hep-ex/0406002%
}{092006},  [\href{http://arXiv.org/pdf/hep-ex/0406002}{{\tt hep-ex/0406002}}].
  \relax
 \relax
\bibitem{Alexopoulos:2004sy}
T.~Alexopoulos et~al., KTeV collaboration, \emph{{Measurements of Semileptonic
  KL Decay Form Factors}}, Phys. Rev. \textbf{D70} (2004),
  \href{http://www.slac.stanford.edu/spires/find/hep/www?eprint=hep-ex/0406003%
}{092007},  [\href{http://arXiv.org/pdf/hep-ex/0406003}{{\tt hep-ex/0406003}}].
  \relax
 \relax
\bibitem{Barberio:1990ms}
E.~Barberio, B.~van Eijk and Z.~W{\c a}s, \emph{{PHOTOS: A Universal Monte
  Carlo for QED radiative corrections in decays}}, Comput. Phys. Commun.
  \textbf{66} (1991),
  \href{http://www.slac.stanford.edu/spires/find/hep/www?j=CPHCB,66,115}{115--%
128}. \relax
 \relax
\bibitem{Barberio:1993qi}
E.~Barberio and Z.~W{\c a}s, \emph{{PHOTOS - a universal monte carlo for QED
  radiative corrections: version 2.0}}, Comput. Phys. Commun. \textbf{79}
  (1994),
  \href{http://www.slac.stanford.edu/spires/find/hep/www?j=CPHCB,79,291}{291--%
308}. \relax
 \relax
\bibitem{Andre:2004fs}
T.~C. Andre, \emph{{Radiative corrections in K0l3 decays}}, Nucl. Phys. Proc.
  Suppl. \textbf{142} (2005),
  \href{http://www.slac.stanford.edu/spires/find/hep/www?j=NUPHZ,142,58}{58--6%
1}, UMI-31-49380. \relax
 \relax
\bibitem{Bernlochner:2010yd}
\href{http://www.slac.stanford.edu/spires/find/hep/www?eprint=arXiv:1003.1620}%
{F.~U. Bernlochner and H.~Lacker}, \emph{{Radiative corrections in exclusive
  semileptonic $B$-meson decays to (pseudo)scalar final state mesons}},
  \href{http://arXiv.org/pdf/1003.1620}{{\tt arXiv:1003.1620}} [hep-ph]. \relax
 \relax
\bibitem{Gleisberg:2008ta}
T.~Gleisberg, S.~H{\"o}che, F.~Krauss, M.~Sch\"{o}nherr, S.~Schumann,
  F.~Siegert and J.~Winter, \emph{{Event generation with \Sherpa 1.1}}, JHEP
  \textbf{02} (2009),
  \href{http://www.slac.stanford.edu/spires/find/hep/www?eprint=0811.4622}{007%
},  [\href{http://arXiv.org/pdf/0811.4622}{{\tt arXiv:0811.4622}} [hep-ph]].
  \relax
 \relax
\bibitem{Schonherr:2008av}
M.~Sch\"{o}nherr and F.~Krauss, \emph{Soft photon radiation in particle decays
  in \Sherpa}, JHEP \textbf{12} (2008),
  \href{http://www.slac.stanford.edu/spires/find/hep/www?eprint=arXiv:0810.507%
1}{018},  [\href{http://arXiv.org/pdf/0810.5071}{{\tt arXiv:0810.5071}}
  [hep-ph]]. \relax
 \relax
\bibitem{DescotesGenon:2005pw}
S.~Descotes-Genon and B.~Moussallam, \emph{{Radiative corrections in weak
  semi-leptonic processes at low energy: A two-step matching determination}},
  Eur. Phys. J. \textbf{C42} (2005),
  \href{http://www.slac.stanford.edu/spires/find/hep/www?eprint=HEP-PH/0505077%
}{403--417},  [\href{http://arXiv.org/pdf/hep-ph/0505077}{{\tt
  hep-ph/0505077}}]. \relax
 \relax
\bibitem{Pauli:1949zm}
W.~Pauli and F.~Villars, \emph{{On the Invariant regularization in relativistic
  quantum theory}}, Rev. Mod. Phys. \textbf{21} (1949),
  \href{http://www.slac.stanford.edu/spires/find/hep/www?j=RMPHA,21,434}{434--%
444}. \relax
 \relax
\bibitem{Kinoshita:1962ur}
T.~Kinoshita, \emph{{Mass singularities of Feynman amplitudes}}, J. Math. Phys.
  \textbf{3} (1962),
  \href{http://www.slac.stanford.edu/spires/find/hep/www?j=JMAPA,3,650}{650--6%
77}. \relax
 \relax
\bibitem{Lee:1964is}
T.~D. Lee and M.~Nauenberg, \emph{{Degenerate Systems and Mass Singularities}},
  Phys. Rev. \textbf{133} (1964),
  \href{http://www.slac.stanford.edu/spires/find/hep/www?j=PHRVA,133,B1549}{B1%
549--B1562}. \relax
 \relax
\bibitem{Burnett:1967km}
T.~H. Burnett and N.~M. Kroll, \emph{{Extension of the low soft photon
  theorem}}, Phys. Rev. Lett. \textbf{20} (1968),
  \href{http://www-spires.slac.stanford.edu/spires/find/hep/www?j=PRLTA,20,86}%
{86}. \relax
 \relax
\bibitem{Becirevic:2009xp}
\href{http://www.slac.stanford.edu/spires/find/hep/www?eprint=0903.2407}{D.~Be%
cirevic and B.~Haas}, \emph{{$D^*\to D \pi$ and $D^*\to D \gamma$ decays: Axial coupling
  and Magnetic moment of $D^*$ meson}},
  \href{http://arXiv.org/pdf/0903.2407}{{\tt arXiv:0903.2407}} [hep-lat].
  \relax
 \relax
\bibitem{Gasser:2004ds}
J.~Gasser, B.~Kubis, N.~Paver and M.~Verbeni, \emph{{Radiative K(e3) decays
  revisited}}, Eur. Phys. J. \textbf{C40} (2005),
  \href{http://www.slac.stanford.edu/spires/find/hep/www?eprint=hep-ph/0412130%
}{205--227},  [\href{http://arXiv.org/pdf/hep-ph/0412130}{{\tt
  hep-ph/0412130}}]. \relax
 \relax
\bibitem{Becirevic:2009fy}
\href{http://www-spires.dur.ac.uk/spires/find/hep/www?eprint=arXiv:0910.5031}{%
D.~Becirevic and N.~Kosnik}, \emph{{Soft photons in semileptonic $B \to D$
  decays}},  \href{http://arXiv.org/pdf/0910.5031}{{\tt arXiv:0910.5031}}
  [hep-ph]. \relax
 \relax
\bibitem{Bijnens:1992en}
J.~Bijnens, G.~Ecker and J.~Gasser, \emph{{Radiative semileptonic kaon
  decays}}, Nucl. Phys. \textbf{B396} (1993),
  \href{http://www.slac.stanford.edu/spires/find/hep/www?eprint=hep-ph/9209261%
}{81--118},  [\href{http://arXiv.org/pdf/hep-ph/9209261}{{\tt
  hep-ph/9209261}}]. \relax
 \relax
\bibitem{Poblaguev:1999ys}
A.~A. Poblaguev, \emph{{What can be learned from an experimental study of
  radiative K(l3) decay?}}, Phys. Atom. Nucl. \textbf{62} (1999),
  \href{http://www.slac.stanford.edu/spires/find/hep/www?j=PANUE,62,975}{975--%
979}. \relax
 \relax
\bibitem{Cirigliano:2005ms}
V.~Cirigliano and D.~Pirjol, \emph{{Factorization in exclusive semileptonic
  radiative B decays}}, Phys. Rev. \textbf{D72} (2005),
  \href{http://www.slac.stanford.edu/spires/find/hep/www?eprint=HEP-PH/0508095%
}{094021},  [\href{http://arXiv.org/pdf/hep-ph/0508095}{{\tt hep-ph/0508095}}].
  \relax
 \relax
\bibitem{Fearing:1970zz}
H.~W. Fearing, E.~Fischbach and J.~Smith, \emph{{Current algebra, anti-k0-l-3
  form-factors, and radiative anti-k0-l-3 decay}}, Phys. Rev. \textbf{D2}
  (1970),
  \href{http://www.slac.stanford.edu/spires/find/hep/www?j=PHRVA,D2,542}{542--%
560}. \relax
 \relax
\bibitem{Krauss:2010xx}
F.~Krauss, T.~Laubrich and F.~Siegert, \emph{{Simulation of hadron decays in
  \Sherpa}}, in preparation. \relax
 \relax
\bibitem{Catani:1996vz}
S.~Catani and M.~H. Seymour, \emph{{A general algorithm for calculating jet
  cross sections in NLO QCD}}, Nucl. Phys. \textbf{B485} (1997),
  \href{http://www.slac.stanford.edu/spires/find/hep/www?eprint=hep-ph/9605323%
}{291--419},  [\href{http://arXiv.org/pdf/hep-ph/9605323}{{\tt
  hep-ph/9605323}}]. \relax
 \relax
\bibitem{Dittmaier:1999mb}
S.~Dittmaier, \emph{{A general approach to photon radiation off fermions}},
  Nucl. Phys. \textbf{B565} (2000),
  \href{http://www.slac.stanford.edu/spires/find/hep/www?eprint=hep-ph/9904440%
}{69--122},  [\href{http://arXiv.org/pdf/hep-ph/9904440}{{\tt
  hep-ph/9904440}}]. \relax
 \relax
\bibitem{Catani:2002hc}
S.~Catani, S.~Dittmaier, M.~H. Seymour and Z.~Trocsanyi, \emph{{The dipole
  formalism for next-to-leading order QCD calculations with massive partons}},
  Nucl. Phys. \textbf{B627} (2002),
  \href{http://www.slac.stanford.edu/spires/find/hep/www?eprint=hep-ph/0201036%
}{189--265},  [\href{http://arXiv.org/pdf/hep-ph/0201036}{{\tt
  hep-ph/0201036}}]. \relax
 \relax
\bibitem{Golonka:2005pn}
P.~Golonka and Z.~W{\c a}s, \emph{{PHOTOS Monte Carlo: A Precision tool for QED
  corrections in $Z$ and $W$ decays}}, Eur. Phys. J. \textbf{C45} (2006),
  \href{http://www.slac.stanford.edu/spires/find/hep/www?eprint=hep-ph/0506026%
}{97--107},  [\href{http://arXiv.org/pdf/hep-ph/0506026}{{\tt
  hep-ph/0506026}}]. \relax
 \relax
\bibitem{Aubert:2009ac}
B.~Aubert et~al., BABAR collaboration, \emph{Measurement of $|V_{cb}|$ and the
  Form-Factor Slope in $\bar{B} \to D \ell^- \bar{\nu}$ Decays in Events Tagged by a Fully
  Reconstructed B Meson}, Phys. Rev. Lett. \textbf{104} (2010),
  \href{http://www.slac.stanford.edu/spires/find/hep/www?eprint=0904.4063}{011%
802},  [\href{http://arXiv.org/pdf/0904.4063}{{\tt arXiv:0904.4063}} [hep-ex]].
  \relax
 \relax
\bibitem{Aubert:2006px}
B.~Aubert et~al., BABAR collaboration, \emph{{Measurement of the $B^0 \to \pi^-
  l^+ \nu$ Form-Factor Shape and Branching Fraction, and Determination of
  $|V_{ub}|$ with a Loose Neutrino Reconstruction Technique}}, Phys. Rev. Lett.
  \textbf{98} (2007),
  \href{http://www.slac.stanford.edu/spires/find/hep/www?eprint=hep-ex/0612020%
}{091801},  [\href{http://arXiv.org/pdf/hep-ex/0612020}{{\tt hep-ex/0612020}}].
  \relax
 \relax
\bibitem{Ball:2004ye}
P.~Ball and R.~Zwicky, \emph{{New results on $B \to \pi, K, \eta$ decay form
  factors from light-cone sum rules}}, Phys. Rev. \textbf{D71} (2005),
  \href{http://www.slac.stanford.edu/spires/find/hep/www?eprint=hep-ph/0406232%
}{014015},  [\href{http://arXiv.org/pdf/hep-ph/0406232}{{\tt hep-ph/0406232}}].
  \relax
 \relax
\bibitem{Aubert:2008yv}
B.~Aubert et~al., BABAR collaboration, \emph{{Measurements of the Semileptonic
  Decays $\bar B\to D\ell\bar\nu$ and $\bar B\to D^*\ell\bar\nu$ Using a Global
  Fit to $DX\ell\bar\nu$ Final States}}, Phys. Rev. \textbf{D79} (2009),
  \href{http://www-spires.dur.ac.uk/spires/find/hep/www?eprint=arXiv:0809.0828%
}{012002},  [\href{http://arXiv.org/pdf/0809.0828}{{\tt arXiv:0809.0828}}
  [hep-ex]]. \relax
 \relax
\bibitem{Wise:1992hn}
M.~B. Wise, \emph{{Chiral perturbation theory for hadrons containing a heavy
  quark}}, Phys. Rev. \textbf{D45} (1992),
  \href{http://www.slac.stanford.edu/spires/find/hep/www?j=PHRVA,D45,2188}{218%
8--2191}. \relax
 \relax
\bibitem{Burdman:1992gh}
G.~Burdman and J.~F. Donoghue, \emph{{Union of chiral and heavy quark
  symmetries}}, Phys. Lett. \textbf{B280} (1992),
  \href{http://www-spires.dur.ac.uk/spires/find/hep/www?j=PHLTA,B280,287}{287-%
-291}. \relax
 \relax
\bibitem{Yan:1992gz}
T.-M. Yan et~al., \emph{{Heavy quark symmetry and chiral dynamics}}, Phys. Rev.
  \textbf{D46} (1992),
  \href{http://www.slac.stanford.edu/spires/find/hep/www?j=PHRVA,D46,1148}{114%
8--1164}. \relax
 \relax
\bibitem{Wise:1993wa}
\href{http://www.slac.stanford.edu/spires/find/hep/www?eprint=hep-ph/9306277}{%
M.~B. Wise}, \emph{{Combining chiral and heavy quark symmetry}},
  \href{http://arXiv.org/pdf/hep-ph/9306277}{{\tt hep-ph/9306277}}. \relax
 \relax
\bibitem{Buckley:2010ar}
\href{http://www.slac.stanford.edu/spires/find/hep/www?eprint=1003.0694}{A.~Bu%
ckley, J.~Butterworth, L.~Lonnblad, H.~Hoeth, J.~Monk et~al.}, \emph{{Rivet
  user manual}},  \href{http://arXiv.org/pdf/1003.0694}{{\tt arXiv:1003.0694}}
  [hep-ph]. \relax
 \relax
\bibitem{Caprini:1997mu}
I.~Caprini, L.~Lellouch and M.~Neubert, \emph{{Dispersive bounds on the shape
  of $\bar{B} \to D^{(*)} l \bar{\nu}$ form factors}}, Nucl. Phys.
  \textbf{B530} (1998),
  \href{http://www.slac.stanford.edu/spires/find/hep/www?eprint=hep-ph/9712417%
}{153--181},  [\href{http://arXiv.org/pdf/hep-ph/9712417}{{\tt
  hep-ph/9712417}}]. \relax
 \relax
\bibitem{Leibovich:1997em}
A.~K. Leibovich, Z.~Ligeti, I.~W. Stewart and M.~B. Wise, \emph{{Semileptonic B
  decays to excited charmed mesons}}, Phys. Rev. \textbf{D57} (1998),
  \href{http://www.slac.stanford.edu/spires/find/hep/www?eprint=hep-ph/9705467%
}{308--330},  [\href{http://arXiv.org/pdf/hep-ph/9705467}{{\tt
  hep-ph/9705467}}]. \relax
 \relax
\bibitem{Leibovich:1997tu}
A.~K. Leibovich, Z.~Ligeti, I.~W. Stewart and M.~B. Wise, \emph{{Model
  independent results for B $\to$ D1(2420) l anti-nu and B $\to$ D*2(2460) l
  anti-nu at order Lambda(QCD)/m(c,b)}}, Phys. Rev. Lett. \textbf{78} (1997),
  \href{http://www.slac.stanford.edu/spires/find/hep/www?eprint=hep-ph/9703213%
}{3995--3998},  [\href{http://arXiv.org/pdf/hep-ph/9703213}{{\tt
  hep-ph/9703213}}]. \relax
 \relax
\bibitem{Ward:1950xp}
J.~C. Ward, \emph{{An Identity in Quantum Electrodynamics}}, Phys. Rev.
  \textbf{78} (1950),
  \href{http://www.slac.stanford.edu/spires/find/hep/www?j=PHRVA,78,182}{182}.
  \relax
 \relax
\bibitem{Fischbach:1970qw}
E.~Fischbach and J.~Smith, \emph{{Current algebra, k-l-3+ form-factors, and
  radiative k-l-3+ decay}}, Phys. Rev. \textbf{184} (1969),
  \href{http://www.slac.stanford.edu/spires/find/hep/www?j=PHRVA,184,1645}{164%
5--1660}. \relax
 \relax
\end{thebibliography}

\end{fmffile}
\end{document}